\documentclass[11pt, oneside]{article}   	
\usepackage{jheppub}
\usepackage{amsmath}
\usepackage{amssymb}
\usepackage{amsfonts}
\usepackage{amsthm}
\usepackage{amsbsy}
\usepackage{array}
\usepackage{mathtools}
\usepackage[usenames,dvipsnames]{xcolor}
\usepackage{subcaption}
\usepackage{dsdshorthand}
\usepackage{graphicx}
\usepackage[vcentermath]{youngtab}
\usepackage{tikzit}

\tikzstyle{small circle}=[shape=circle, fill=white, draw=black]
\tikzstyle{medium circle}=[fill=white, draw=black, shape=circle, minimum width=1cm, minimum height=1cm]
\tikzstyle{twoPoint}=[fill=white, draw=black, shape=circle, twopt]
\tikzstyle{threePoint}=[fill=white, draw=black, shape=circle, threept]
\tikzstyle{m vert rect}=[fill=white, draw=black, shape=rectangle, minimum height=2cm, minimum width=.5cm]
\tikzstyle{s box}=[fill=white, draw=black, shape=rectangle]
\tikzstyle{celestialThreePoint}=[fill=white, draw=red, shape=circle, threept]
\tikzstyle{large circle}=[fill=white, draw=black, shape=circle, minimum height=1.5cm, minimum width=1.5cm]
\tikzstyle{green circle of inversion}=[draw={black!20!green}, shape=circle, minimum size=2cm, dashed, tikzit draw={black!20!green}]
\tikzstyle{large green circle}=[draw={black!20!green}, shape=circle, dashed, minimum size=3cm, semithick]
\tikzstyle{small triangle}=[fill=white, draw=black, regular polygon, regular polygon sides=3, inner sep=1.5pt, rotate=90]
\tikzstyle{small blue triangle}=[fill=white, draw={black!20!blue}, regular polygon, regular polygon sides=3, inner sep=1.5pt, rotate=-90, tikzit draw=blue]
\tikzstyle{small green circle}=[draw={black!20!green}, shape=circle, tikzit draw={black!20!green}, dashed, minimum size=.5cm]
\tikzstyle{celestialTwoPoint}=[fill=red, draw=red, shape=circle, inner sep=1pt, minimum size=1pt]

\tikzstyle{arrow}=[->]
\tikzstyle{op}=[-, spinning]
\tikzstyle{dashedOp}=[-, scalar]
\tikzstyle{celestial}=[-, color=red, scalar, tikzit draw=red]
\tikzstyle{minkowski}=[-, spinning, color={black!20!blue}, tikzit draw=blue]

\input{harmonic_analysis.tikzdefs}

\tikzstyle{pole}=[fill=black, draw=black, shape=circle, inner sep=0pt, minimum size=2pt]

\tikzstyle{Arrow}=[->, draw=blue]
\tikzstyle{anti arrow}=[<-, draw=blue]
\tikzstyle{axis}=[->]
\tikzstyle{dashed_arrow}=[->, draw=red]
\tikzstyle{dashed_only}=[-, draw=red]
\tikzstyle{dashed_anti}=[<-, draw=red]

\tikzstyle{point}=[fill=black, draw=black, shape=circle, inner sep=0pt, minimum size=3pt]
\tikzstyle{lr_point}=[fill=red, draw=red, shape=circle, inner sep=0pt, minimum size=2pt]
\tikzstyle{sh_point}=[fill={rgb,255: red,128; green,128; blue,128}, draw={rgb,255: red,128; green,128; blue,128}, shape=circle, inner sep=0pt, minimum size=3pt]
\tikzstyle{sh_lr_point}=[fill={red!40}, draw={red!40}, shape=circle, inner sep=0pt, minimum size=2pt, tikzit fill={rgb,255: red,255; green,128; blue,0}, tikzit draw={rgb,255: red,255; green,128; blue,0}]

\tikzstyle{dashed_st}=[-, dashed]
\tikzstyle{arrow}=[->]
\tikzstyle{dashed_grey}=[-, draw=black, dotted]
\tikzstyle{blue_line}=[draw=blue, ->]
\tikzstyle{energy}=[draw={rgb,255: red,255; green,128; blue,0}, decoration={{snake,amplitude=1pt,segment length=6pt,post length=1pt}}, decorate, ->]
\tikzstyle{blue_line_0}=[-, draw=blue]
\tikzstyle{redline}=[-, draw=red]

\newcommand\wL{\mathbf{L}}

\renewcommand\vol{\mathop{\mathrm{vol}}}
\newcommand{\EEEC}{\mathrm{EEEC}}

\definecolor{energycolor}{RGB}{230,50,10}

\tikzset{
  energy/.style={->,
  energycolor,
  decoration={
      snake,
      amplitude=1pt,
      segment length=6pt,
      post length=1pt
    },
  decorate
  }
}

\makeatletter
\def\@fpheader{\ }
\makeatother

\title{Three-point energy correlators and the celestial block expansion}
\author{Cyuan-Han Chang and David Simmons-Duffin}
\affiliation{Walter Burke Institute for Theoretical Physics, Caltech, Pasadena, California 91125, USA}
\emailAdd{cchang7@caltech.edu}
\emailAdd{dsd@caltech.edu}

\date{}
\abstract{We study the three-point energy correlator (EEEC), defined as a matrix element of a product of three energy detectors at different locations on the celestial sphere. Lorentz symmetry implies that the EEEC can be decomposed into special functions called  celestial blocks. We compute three-point celestial blocks in an expansion around the collinear limit, where the three detectors approach each other on the celestial sphere. The leading term is a traditional $d{-}2$-dimensional four-point conformal block, and thus the collinear EEEC behaves like a conformally-invariant four-point function in $d{-}2$ dimensions. We obtain the coefficients of the conformal block decomposition for the collinear EEEC at leading nontrivial order in weakly-coupled $\cN=4$ SYM and QCD. These data allow us to make certain all-orders predictions for the collinear EEEC in various kinematic limits, including the OPE limit and the double lightcone limit. We also study Ward identities satisfied by the EEEC and compute contact terms in the EEEC in weakly-coupled $\cN=4$ SYM. Finally, we study the celestial block expansion of the EEEC in planar $\cN=4$ SYM at strong coupling, determining celestial block coefficients to leading and first subleading order at large $\l$.
}

\preprint{CALT-TH 2022-005}

\begin{document}

\maketitle
\pagenumbering{roman}
\setcounter{page}{2}
\newpage
\pagenumbering{arabic}
\setcounter{page}{1}

\section{Introduction}

Energy correlators \cite{Basham:1978zq,Basham:1978bw,Basham:1977iq} are natural Lorentzian observables with numerous applications in collider physics, conformal field theory, and string theory, see e.g.\ \cite{DELPHI:1990sof,OPAL:1990reb,SLD:1994idb,SLD:1994yoe,Belitsky:2001ij,Hofman:2008ar,Zhiboedov:2013opa,Larkoski:2013eya,Belitsky:2013xxa,Belitsky:2013bja,Belitsky:2013ofa,Faulkner:2016mzt,Hartman:2016lgu,Cordova:2017dhq,Cordova:2017zej,Cordova:2018ygx,Dixon:2018qgp,Dixon:2019uzg,Luo:2019nig,Henn:2019gkr,Chen:2019bpb,Gao:2020vyx,Chen:2020vvp,Ebert:2020sfi,Korchemsky:2021okt,Korchemsky:2021htm,Chen:2021gdk,Komiske:2022enw,Holguin:2022epo}. They are given by an expectation value of a product of energy flux operators $\cE(\vec n_i)$ \cite{Sveshnikov:1995vi} that measure the flux of energy at locations $\vec n_i \in S^{d-2}$ on the celestial sphere:
\be
\<\Psi|\cE(\vec n_1)\cdots \cE(\vec n_k)|\Psi\>.
\ee
Energy correlators are examples of more general ``event shapes," which are expectation values of products of detectors at different locations on the celestial sphere.

The kinematics of energy correlators and event shapes exhibit many features of a (fictitious) Euclidean $d{-}2$-dimensional CFT on the celestial sphere. In particular, the Lorentz group $\SO(d-1,1)$ acts as the conformal group on the celestial sphere, so event shapes exhibit conformal symmetry.

However, other aspects of event shapes are different from $d{-}2$-dimensional CFT. Event shapes are not necessarily computed (in an obvious way) by a local path integral on the celestial sphere. Consequently, structures like radial quantization and a $d{-}2$-dimensional operator product expansion (OPE) can't obviously be used to analyze them. Nevertheless, it was argued by Hofman and Maldacena \cite{Hofman:2008ar} that a kind of OPE should exist between energy flux operators $\cE(\vec n_1)\x \cE(\vec n_2)$ in the limit $\vec n_1\to \vec n_2$, i.e.\ as the corresponding points on the celestial sphere approach each other. In \cite{Kologlu:2019mfz,Chang:2020qpj}, the OPE of two energy flux operators $\cE(\vec n_1)\x \cE(\vec n_2)$ was explicitly constructed in a general nonperturbative CFT$_d$, and it was shown that the objects appearing are the light-ray operators $\mathbb{O}_J(\vec n)$ of \cite{Kravchuk:2018htv}. This leads to a useful expansion for two-point energy correlators in special functions called ``celestial blocks," which re-sum the contributions of light-ray operators and their descendants on the celestial sphere.

If it were possible to iterate the light-ray OPE, we would obtain a simple and beautiful procedure for evaluating higher-point energy correlators. However, the arguments of \cite{Kologlu:2019mfz,Chang:2020qpj} do not extend in a simple way to describe the OPE of an energy flux operator and a more general light-ray operator $\cE(\vec n_1)\x \mathbb{O}_J(\vec n_2)$, or to describe an OPE of general light-ray operators $\mathbb{O}_{J_1}(\vec n_1) \x \mathbb{O}_{J_2}(\vec n_2)$. Perturbative studies of these more complicated OPEs were undertaken recently in \cite{Chen:2021gdk}. Finding an appropriate nonperturbative generalization of the light-ray OPE is an important problem. However, we will not solve it in this work. Instead, we assume that a general light-ray OPE exists and study some of its consequences for higher-point energy correlators.

One consequence is that higher-point energy correlators should admit an expansion in a discrete sum of multi-point celestial blocks. Mathematically, harmonic analysis with respect to the Lorentz group \cite{Dobrev:1977qv} guarantees that energy correlators can be expanded in an integral of celestial ``partial waves." However, going from a partial wave expansion to a celestial block expansion requires a dynamical assumption about poles in partial wave coefficients. We check this assumption by studying the celestial block expansion of three-point energy correlators (EEEC) at both weak coupling (in QCD and $\cN=4$ SYM) and strong coupling ($\cN=4$ SYM). In all cases, we find that a discrete celestial block expansion exists, and that the quantum numbers of objects appearing can be understood from symmetries. At weak coupling, we use the recent perturbative expressions for the EEEC in \cite{Chen:2019bpb}, and at strong coupling, we study Hofman and Maldacena's famous result for the EEEC \cite{Hofman:2008ar}.

A particularly interesting limit of the EEEC is the collinear limit \cite{Chen:2019bpb,Chen:2021gdk}, where all three operators approach each other on the celestial sphere $|\vec n_{ij}|\to 0$ with $|\vec n_{ij}|/|\vec n_{kl}|$ fixed, where $\vec n_{ij}=\vec n_i-\vec n_j$. Physically, the collinear limit is obtained by simultaneously boosting all three detectors. By Lorentz-invariance, this is equivalent to boosting the state $|\Psi\>$ in the opposite direction, causing its momentum $p$ to approach the null cone. However, a point on the null cone encodes a point on the celestial sphere $(1,\vec n_4)=p/p^0$, so the kinematics of the EEEC in the collinear limit are the same as for a conformal four-point function in CFT$_{d-2}$. In particular, celestial blocks have an expansion in the collinear limit, where the leading term is the usual four-point conformal block. This observation was made for the leading term in  \cite{Chen:2021gdk}, and we will extend it to a systematic expansion around the collinear limit. Furthermore, event shapes inherit crossing symmetry from the $d$-dimensional bulk theory. This allows us to apply techniques from the analytic bootstrap for CFT four-point functions to the collinear EEEC, including the lightcone bootstrap \cite{Fitzpatrick:2012yx,Komargodski:2012ek,Alday:2015eya,Alday:2015ewa,Alday:2016njk,Simmons-Duffin:2016wlq} and Lorentzian inversion formula \cite{Caron-Huot:2017vep,Simmons-Duffin:2017nub}. (Interestingly, the Lorentzian inversion formula requires analytically continuing to Lorentzian signature on the celestial sphere, which is $(d-2,2)$ signature from the point of view of the bulk theory.)

This paper is organized as follows. In section~\ref{sec:blockdecomposition}, we study implications of Lorentz symmetry for event shapes. We explain the form that the celestial block expansion should take for 2- and 3-point event shapes, and study the expansion of 3-point celestial blocks around the collinear limit. We furthermore explore general constraints of celestial crossing symmetry for the collinear EEEC using lightcone bootstrap methods. In section~\ref{sec:EEEC_LO}, we study recent leading-order weak-coupling results for the collinear EEEC in QCD and $\cN=4$ SYM from the point of view of the celestial block expansion, using the Lorentzian inversion formula to extract celestial block coefficients. In section~\ref{sec:allorderEEEC}, we describe some predictions for higher orders in the weak coupling expansion that follow from a discrete celestial block expansion. In section~\ref{sec:Wardidentities_contactterm}, we discuss consequences of Ward identities, in particular using them to determine the leading nontrivial contact terms in the EEEC in weakly-coupled $\cN=4$ SYM. In section~\ref{sec:strongcoupling}, we study the EEEC in strongly-coupled $\cN=4$ SYM for general configurations on the celestial sphere --- not just the collinear limit. We explain how the corresponding celestial OPE data can be obtained from a three-point celestial inversion formula, and then apply the inversion formula to results from \cite{Hofman:2008ar} to obtain simple analytic formulas for the full EEEC celestial OPE data at $O(1/\l)$. Finally, we conclude in section~\ref{sec:discussion}.
\\
\\
{\bf{Note Added:}} This paper will appear simultaneously with a paper by Hao Chen, Ian Moult, Joshua Sandor, and Hua Xing Zhu, that also studies three-point correlators of light-ray operators from the perspective of the light-ray OPE. We thank these authors for coordinating submission.

\section{Lorentz symmetry and event shapes}\label{sec:blockdecomposition}

Because the Lorentz group $\SO(d-1,1)$ is also the conformal group on the celestial sphere $S^{d-2}$, event shapes can be decomposed into ``celestial blocks," which are natural objects from the point of view of $d{-}2$ dimensional CFT. We will be particularly interested in three-point event shapes. In the ``collinear" limit where the three detectors are close to each other, the kinematics of a three-point event shape become the same as a CFT four-point function, and celestial blocks become four-point conformal blocks. We will begin by reviewing event shapes in CFT. We then discuss celestial blocks for two-point event shapes, before introducing three-point celestial blocks and their collinear limit.

\subsection{Review: event shapes and the light transform}\label{sec:eventshapeintro}

An event shape can be thought of as a weighted cross section, or alternatively as the expectation value of an operator at future null infinity. For example, consider the three-point energy correlator (EEEC), conventionally defined by
\be\label{eq:EEEC_def}
&\mathrm{EEEC}(\z_{12},\z_{13},\z_{23})\nn \\
&=\sum_{i,j,k}\int d\s\frac{E_iE_jE_k}{Q^3}\de\p{\z_{12}-\frac{1-\cos\th_{ij}}{2}}\de\p{\z_{13}-\frac{1-\cos\th_{ik}}{2}}\de\p{\z_{23}-\frac{1-\cos\th_{jk}}{2}}.
\ee
Here, $d\s$ is the phase space measure multiplied by the squared amplitude for some state $|\cO(p)\>$ to create outgoing particles, and the sum $\sum_{i,j,k}$ runs over triplets of outgoing particles. The definition (\ref{eq:EEEC_def}) is convenient for perturbative calculations and deriving Ward identities (see section \ref{sec:Wardidentities_contactterm}). However, it obscures some features like IR safety, and furthermore requires the existence of asymptotic states.

An alternative definition of the EEEC, that works in any nonperturbative QFT, is \cite{Belitsky:2013xxa}
\be\label{eq:EEEC_def_CFT}
\mathrm{EEEC}(\z_{12},\z_{13},\z_{23})=&\int d\O_{\vec{n}_1}d\O_{\vec{n}_2}d\O_{\vec{n}_3}\de(\z_{12}-\tfrac{1-\vec{n}_1\.\vec{n}_2}{2})\de(\z_{13}-\tfrac{1-\vec{n}_1\.\vec{n}_3}{2})\de(\z_{23}-\tfrac{1-\vec{n}_2\.\vec{n}_3}{2}) \nn \\
&\quad\x\frac{\<\cO(p)|\cE(\vec{n}_1)\cE(\vec{n}_2)\cE(\vec{n}_3)|\cO(p)\>}{(-p^2)^{\frac{3}{2}}\<\cO(p)|\cO(p)\>},
\ee
where $\cE(\vec n)$ is an energy detector defined by
\be\label{eq:ANEC_def}
\cE(\vec n)=\lim_{r \to \oo}r^{d-2}\int_0^{\oo} dt\ n^{i}T^{0}{}_{i}(t,r\vec n).
\ee
In a CFT, $\cE(\vec n)$ is conformally equivalent to the average null energy operator ANEC --- a null integral of the stress tensor \cite{Hofman:2008ar}. Thus, we often refer to $\cE(\vec n)$ as ANEC operators.
Here, and below, we use the shorthand notation where when a bra and ket have equal momenta, we implicitly strip off an overall momentum-conserving delta function. This is equivalent to Fourier-transforming only one of the operators:
\be
\<\cO(p)|\cdots|\cO(p)\> &\equiv \int d^d x e^{i p\. x} \<0|\cO(x) \cdots \cO(0)|0\>.
\ee

The ANEC operator $\cE(\vec n)$ measures energy flux at a point on the celestial sphere $\vec n \in S^{d-2}$. 
In a CFT, it can be understood in terms of a conformally-invariant integral transform called the light transform \cite{Kravchuk:2018htv}. To describe the light transform, we use index-free notation where we contract indices of an operator with an auxiliary null vector $z$: $\cO(x,z)=\cO^{\mu_1\cdots\mu_J}(x)z_{\mu_1}\cdots z_{\mu_J}$. The light transform of an operator $\cO$ with scaling dimension $\De$ and spin $J$ is 
\be
\wL[\cO](x,z)=\int_{-\oo}^{\oo} d\a (-\a)^{-\De-J}\cO\p{x-\frac{z}{\a},z}.
\ee
Under conformal transformations, $\wL[\cO](x,z)$ transforms like a primary operator at $x$ with quantum numbers $(1-J,1-\De)$.  The ANEC operator defined in \eqref{eq:ANEC_def} can be written as the light transform of the stress-energy tensor placed at spatial infinity:
\be
\cE(\vec n)=2\wL[T](\oo,z=(1,\vec n)).
\ee
In general, an (un-normalized) $n$-point event shape in CFT is the matrix element of a product of $n$ light-transformed operators in a state $|\cO(p)\>$:
\be
\<\cO(p)|\wL[\cO_1](\oo,z_1)\cdots \wL[\cO_n](\oo,z_n)|\cO(p)\>.
\ee 
For the EEEC, we have $\cO_1=\cO_2=\cO_3=T$.

\subsection{Lorentz symmetry and celestial blocks}
\subsubsection{Two-point event shapes}
\label{sec:twopteventshape}
Consider a two-point scalar event shape\footnote{This scalar event shape is only well-defined nonperturbatively if the theory has Regge intercept $J_0<-1$ \cite{Kologlu:2019bco}. In this section, we are only interested in kinematics, so we assume this is the case.}
\be\label{eq:twopteventshape_scalar}
\<\f_4(p)|\wL[\f_1](\oo,z_1)\wL[\f_2](\oo,z_2)|\f_3(p)\>.
\ee
For simplicity, we study event shapes built from scalars $\f_i$ in this section, leaving spinning operators for later.

Let us understand how the product $\wL[\f_1](\oo,z_1)\wL[\f_2](\oo,z_2)$ transforms under the Lorentz group $\SO(d-1,1)$. The Lorentz group is isomorphic to the Euclidean conformal group in $d-2$ dimensions.  From this point of view, the polarization vector $z$ can be thought of as an embedding-space coordinate for the celestial sphere $S^{d-2}$.

As convenient notation, let $\cP_{\de,j}(z,w)$ denote an operator with dimension $\de$ and spin $j$ in a fictitious CFT$_{d-2}$ on the celestial sphere, in the embedding formalism. The embedding space coordinates are null vectors $z,w\in \R^{d-1,1}$ with a gauge redundancy $w\sim w+\a z$. $\cP_{\de,j}(z,w)$ is a homogeneous function of $z$ and $w$ with degrees $-\de$ and $j$, respectively. See \cite{Kologlu:2019mfz} for more details on this notation. We usually refer to the spin $j$ on the celestial sphere as ``transverse spin" to disambiguate it from the Lorentz spin $J$ of a local operator in $d$-dimensions. When $j=0$, we write simply $\cP_\de(z)$.

The light-transformed operator $\wL[\f_i](\oo,z)$ is homogeneous of degree $1-\De_i$ in $z$. Thus, it transforms like a scalar on the celestial sphere with dimension $\de_i=\De_i-1$:
\be
\label{eq:celestialcorrespondence}
\wL[\f_i](\oo,z) &\sim  \cP_{\de_i}(z).
\ee
From this point of view, we can treat correlators of $\wL[\f_i](\oo,z)$ as if they were correlators of $\cP_{\de_i}(z)$ in a fictitious CFT$_{d-2}$.
Note that we do not assert that there exists a {\it local} CFT on $S^{d-2}$. For our purposes, (\ref{eq:celestialcorrespondence}) is convenient notation for keeping track of symmetries.

Using this notation, a product $\wL[\f_1](\oo,z_1)\wL[\f_2](\oo,z_2)$ transforms like a product of scalars $\cP_{\de_1}(z_1)\cP_{\de_2}(z_2)$ in $d{-}2$ dimensions. It is natural to expand such a product in a $d{-}2$ dimensional OPE, where the objects that appear are spin-$j$ traceless symmetric tensors:
\be
\label{eq:formalopesum}
\cP_{\de_1}(z_1)\cP_{\de_2}(z_2) = \sum_{\de,j}r_{\de,j}\cC_{\de,j}(z_1,z_2,\ptl_{z_2},\ptl_{w_2})\cP_{\de,j}(z_2,w_2).
\ee
Here, the dimensions $\de$ and ``OPE coefficients'' $r_{\de,j}$ that appear depend on the theory.  However, the differential operator $\cC_{\de,j}$ is determined by symmetry and is defined by
\be
\cC_{\de,j}(z_1,z_2,\ptl_{z_2},\ptl_{w_2})\<\cP_{\de,j}(z_2,w_2)\cP_{\de,j}(z,w)\>=\<\cP_{\de_1}(z_1)\cP_{\de_2}(z_2)\cP_{\de,j}(z,w)\>,
\ee
where $\<\cP_{\de,j}(z_2,w_2)\cP_{\de,j}(z,w)\>$ and $\<\cP_{\de_1}(z_1)\cP_{\de_2}(z_2)\cP_{\de,j}(z,w)\>$ are standard two- and three-point structures in the embedding space.

The light-ray OPE gives a concrete version of the expansion (\ref{eq:formalopesum}) where the objects on the right-hand side are light-ray operators. Taking $\wL[\f]\wL[\f]$ as an example, the operators appearing are \cite{Kologlu:2019mfz,Chang:2020qpj}
\be\label{eq:phiphi_lightrayOPE}
\wL[\f]\x \wL[\f] \sim \sum_{i}\mathbb{O}_{i,J=-1,j=0} + \sum_{i}\sum_{n=1}^{\oo} \cD_{2n} \mathbb{O}_{i,J=-1+2n,j=0}.
\ee
Here, $\mathbb{O}_{i,J,j}$ denotes a light-ray operator on the $i$-th Regge trajectory with spin $J$ and transverse spin $j$  \cite{Kravchuk:2018htv}. $\cD_{2n}$ is a differential operator that decreases the spin $J$ by $2n$ and increases the transverse spin $j$ by $2n$. Light-ray operators are analytic continuations of light transformed operators $\wL[\cO]$ in $J$, so they satisfy
\be
\mathbb{O}_{i,J,j}=f_{\f\f\cO_{i,J,j}}\wL[\cO_{i,J,j}], \qquad J\in \Z_{\geq 0},\ J~\mathrm{even}.
\ee
The light-ray OPE thus establishes a relation between the scalar two-point event shape, defined as the matrix element of $\wL[\f]\wL[\f]$, and the OPE data of $\f\x\f$ analytically continued to $J=-1,1,3,\dots$. For concrete calculations of two-point event shapes using the light-ray OPE, see \cite{Kologlu:2019mfz,Chang:2020qpj}.

Let us apply the light-ray OPE to the event shape (\ref{eq:twopteventshape_scalar}). As discussed in \cite{Chang:2020qpj}, there is a selection rule for the transverse spin $j$: in a state created by scalar operators $\<\f_4(p)|\cdot|\phi_3(p)\>$, only light-ray operators with $j=0$ can have nonzero matrix elements. Thus, the sum in (\ref{eq:formalopesum}) collapses to just the $j=0$ terms. We find
\be
\<\f_4(p)|\wL[\f_1](\oo,z_1)\wL[\f_2](\oo,z_2)|\f_3(p)\>=\sum_{\de}r_{12\de}\cC_{\de}(z_1,z_2,\ptl_{z_2})\<\f_4(p)|\mathbb{W}_{\de}(z_2)|\f_3(p)\>,
\ee
where $\mathbb{W}_{\de}$ stands for transverse-spin zero light-ray operators $\mathbb{O}_{i,J=-1,j=0}$, and transforms as a scalar primary with dimension $\de$ under $\SO(d-1,1)$.

The form of the matrix element $\<\f_4(p)|\mathbb{W}_{\de}(z_2)|\f_3(p)\>$ is fixed by Lorentz symmetry, homogeneity in $z_2$, and dimensional analysis to be
\be\label{eq:scoeff_defiition}
\<\f_4(p)|\mathbb{W}_{\de}(z_2)|\f_3(p)\>=s_{34\de}(-2p\cdot z_2)^{-\de}(-p^2)^{\frac{\De_3+\De_4+\de-2-d}{2}}.
\ee
Thus, the event shape can be written as
\be
\<\f_4(p)|\wL[\f_1](\oo,z_1)\wL[\f_2](\oo,z_2)|\f_3(p)\>=\sum_{\de}r_{12\de}s_{34\de}(-p^2)^{\frac{\De_3+\De_4+\de-2-d}{2}}\cC_{\de}(z_1,z_2,\ptl_{z_2})(-2p\cdot z_2)^{-\de}.
\ee

The object $\cC_{\de}(z_1,z_2,\ptl_{z_2})(-2p\cdot z_2)^{-\de}$ is called a ``celestial block" and it is completely fixed by $\SO(d-1,1)$ symmetry. It can be computed by solving a Casimir differential equation, similar to the method used by Dolan and Osborn to compute conventional conformal blocks \cite{DO2}. The result is \cite{Kologlu:2019mfz}
\be
\cC_{\de}(z_1,z_2,\ptl_{z_2})(-2p\cdot z_2)^{-\de}=\frac{(-p^2)^{\frac{\de_1+\de_2-\de}{2}}}{(-2p\. z_1)^{\de_1}(-2p\. z_2)^{\de_2}}f_{\de}^{\de_1,\de_2}(\z),
\ee
where $\z$ is a Lorentz-invariant cross ratio
\be\label{eq:crossratio_definition}
\z=\frac{(-p^2)(-2z_1\.z_2)}{(-2p\.z_1)(-2p\.z_2)},
\ee
and the function $f_{\de}^{\de_1,\de_2}(\z)$ is given by
\be
f_{\de}^{\de_1,\de_2}(\z)=\z^{\frac{\de-\de_1-\de_2}{2}}{}_2F_1\p{\tfrac{\de+\de_1-\de_2}{2},\tfrac{\de+\de_2-\de_1}{2},\de+2-\tfrac{d}{2},\z}.
\ee

Combining everything, we obtain a celestial block expansion for the two-point event shape
\be\label{eq:twopteventshape_expansion}
&\<\f_4(p)|\wL[\phi_1](\oo, z_1)\wL[\phi_2](\oo, z_2)|\f_3(p)\>\nn \\
&=\frac{(-p^2)^{\frac{\de_1+\de_2+\de_3+\de_4-d}{2}}}{(-2p\. z_1)^{\de_1}(-2p\. z_2)^{\de_2}}\sum_{\de}r_{12\de}s_{34\de}f_{\de}^{\de_1,\de_2}(\z).
\ee
Note that the form of the celestial block expansion (\ref{eq:twopteventshape_expansion}) is completely dictated by symmetries. The light-ray OPE formula then predicts that the $\de$'s appearing in the expansion (\ref{eq:twopteventshape_expansion}) should be related to dimensions of light-ray operators in the CFT. Furthermore, it makes a prediction for the product of coefficients $r_{12\de}s_{34\de}$:
\be\label{eq:rscoeff_lightrayOPE}
r_{12\de}s_{34\de}=\frac{2^{d+2-\de_3-\de_4}\pi^{\frac{d}{2}+3}e^{i\pi\frac{\de_4-\de_3}{2}}\G(\de-1)}{\G(\tfrac{\de+\de_3-\de_4}{2})\G(\tfrac{\de-\de_3+\de_4}{2})\G(\tfrac{\de_3+\de_4-\de}{2})\G(\tfrac{\de+\de_3+\de_4+2-d}{2})}\p{p^{+}_{\de+1,J=-1}+p^{-}_{\de+1,J=-1}},
\ee
where $p_{\De,J}^{+}$($p_{\De,J}^{-}$) is the product of OPE coefficients of the four-point function $\<\f_4\f_1\f_2\f_3\>$, analytically continued from even(odd) spin.

Even if we do not know the coefficients $r_{12\de}$ and $s_{34\de}$, we can still make some statements about the event shape. We will be particularly interested in the limit where all the detectors are close to each other. For the two-point case, this simply corresponds to $z_1\to z_2$, or $\z\to 0$, and the event shape should behave as
\be
&\<\phi_4(p)|\wL[\phi_1](\oo, z_1)\wL[\phi_2](\oo, z_2)|\phi_3(p)\>\nn \\
&= r_{12\de_{*}}s_{34\de_{*}}\frac{(-p^2)^{\frac{\de_1+\de_2+\de_3+\de_4-d}{2}}\z^{\frac{\de_{*}-\de_1-\de_2}{2}}}{(-2p\.z_1)^{\de_1}(-2p\.z_2)^{\de_2}}+\cdots\nn \\
&= r_{12\de_{*}}s_{34\de_{*}}(-p^2)^{\frac{\de_{*}+\de_3+\de_4-d}{2}}\<\cP_{\de_1}(z_1)\cP_{\de_2}(z_2)\cP_{\de_{*}}(p)\>+\cdots,
\ee
where $\de_{*}$ is the smallest $\de$ appearing in the $\sum_{\de}$ sum.

\subsubsection{Three-point event shapes}
\label{sec:threepteventshape}
Next, consider a three-point event shape
\be\label{eq:threepteventshape_def}
\<\f_5(p)|\wL[\f_1](\oo, z_1)\wL[\f_2](\oo, z_2)\wL[\f_3](\oo, z_3)|\f_4(p)\>.
\ee
We can use the light-ray OPE to decompose the product of a pair of detectors, say $\wL[\f_1]\wL[\f_2]$, as a sum of light-ray operators $\mathbb{O}_i$. However, we do not currently possess a more general light-ray OPE formula that lets us further decompose the product $\mathbb{O}_i \wL[\f_3]$. To make progress, let us use Lorentz symmetry to predict the form that a celestial block expansion for the three-point event shape should have. 

Following the analysis in section \ref{sec:twopteventshape}, we treat $\wL[\f_1](\oo,z_1)\wL[\f_2](\oo,z_2)\wL[\f_3](\oo,z_3)$ as a product of three scalar primary operators $\cP_{\de_1}(z_1)\cP_{\de_2}(z_2)\cP_{\de_3}(z_3)$ in a fictitious $\mathrm{CFT}_{d-2}$. Formally taking consecutive OPEs, we have\footnote{Here, we are following the notation of \cite{Kologlu:2019mfz,Chang:2020qpj}, where $\vec w_3$ denotes a collection of polarization vectors for different rows of the Young diagram of an $\SO(d-2)$ representation. However, after taking the expectation value in a scalar density matrix $\<\f_5(p)|\cdot|\f_4(p)\>$, only scalar representations $\l'$ are allowed, so $\vec w_3$ immediately drops out and can be ignored.}
\be
&\cP_{\de_1}(z_1)\cP_{\de_2}(z_2)\cP_{\de_3}(z_3) \nn \\
&= \sum_{\de,j}r_{12\cP_{\de,j}}\cC_{12\cP_{\de,j}}(z_1,z_2,\ptl_{z_2},\ptl_{w_2})\cP_{\de,j}(z_2,w_2)\cP_{\de_3}(z_3) \nn \\
&=\sum_{\de',\l'}\sum_{\de,j}r_{12\cP_{\de,j}}r'_{\cP_{\de,j}3\cP_{\de',\l'}}\cC_{12\cP_{\de,j}}(z_1,z_2,\ptl_{z_2},\ptl_{w_2})\cC_{\cP_{\de,j}3\cP_{\de',\l'}}(z_2,z_3,\ptl_{z_3},\ptl_{\vec w_3})\cP_{\de',\l'}(z_3,\vec w_3).
\ee
Therefore, the three-point event shape should have the form
\be
&\<\f_5(p)|\wL[\f_1](\oo, z_1)\wL[\f_2](\oo, z_2)\wL[\f_3](\oo, z_3)|\f_4(p)\> \nn \\
&=\sum_{\de'}\sum_{\de,j}r_{12\cP_{\de,j}}r'_{\cP_{\de,j}3\cP_{\de'}}\cC_{12\cP_{\de,j}}(z_1,z_2,\ptl_{z_2},\ptl_{w_2})\cC_{\cP_{\de,j}3\cP_{\de'}}(z_2,z_3,\ptl_{z_3})\<\f_5(p)|\mathbb{W}'_{\de'}(z_3)|\f_4(p)\> \nn \\
&=\sum_{\de'}\sum_{\de,j}r_{12\cP_{\de,j}}r'_{\cP_{\de,j}3\cP_{\de'}}s'_{45\de'}(-p^2)^{\frac{\de_4+\de_5+\de'-1-d}{2}}\nn\\
&\qquad\qquad \x\cC_{12\cP_{\de,j}}(z_1,z_2,\ptl_{z_2},\ptl_{w_2})\cC_{\cP_{\de,j}3\cP_{\de'}}(z_2,z_3,\ptl_{z_3})(-2z_3\.p)^{-\de'},
\label{eq:threepointcelestialblockexpansion}
\ee
where from the second to the third line, we again use the homogeneity of $z_3$ and $p$. In the expansion (\ref{eq:threepointcelestialblockexpansion}), we have three unknown coefficients $r_{12\cP_{\de,j}}$, $r'_{\cP_{\de,j}3\cP_{\de'}}$, and $s'_{45\de'}$.

If an OPE expansion for a three-point event shape exists, symmetries ensure it must take the form (\ref{eq:threepointcelestialblockexpansion}). However, we do not know an argument guaranteeing the existence of such an expansion. Mathematically, the only thing that is guaranteed is that (\ref{eq:threepteventshape_def}) can be decomposed into a double-integral over complex $\de$ and $\de'$ of ``celestial partial waves," defined below in equation (\ref{eq:celestialpartialwavedefinition}). An expansion like (\ref{eq:threepointcelestialblockexpansion}) would arise if we can additionally close the contours to the right, picking up a set of discrete poles, as described in \cite{PhysRevD.13.887} for the conventional conformal block decomposition. In this work, we assume that such a contour maneuver is possible, at least to characterize the leading behavior of the event shape in the collinear limit. In the absence of nonperturbative arguments, it is also important to compare (\ref{eq:threepointcelestialblockexpansion}) to perturbative data, as we do in section~\ref{sec:EEEC_LO}.

The kinematic dependence of (\ref{eq:threepointcelestialblockexpansion}) is accounted for by the object
\be
\label{eq:threeptcelestialblock}
\cC_{12\cP_{\de,j}}(z_1,z_2,\ptl_{z_2},\ptl_{w_2})\cC_{\cP_{\de,j}3\cP_{\de'}}(z_2,z_3,\ptl_{z_3})(-2z_3\.p)^{-\de'},
\ee
which is completely fixed by Lorentz symmetry. We call (\ref{eq:threeptcelestialblock}) a three-point celestial block.
Although we do not know a compact closed-form expression for it, we can still determine its expansion around the collinear limit, where $z_1,z_2,z_3$ are close to each other. The reason is that the collinear limit is equivalent (up to a Lorentz transformation) to a configuration where $p$ is null. Writing $p$ as $z_0$ in this null limit, the leading term of the three-point celestial block in the collinear limit becomes
\be
&\cC_{12\cP_{\de,j}}(z_1,z_2,\ptl_{z_2},\ptl_{w_2})\cC_{\cP_{\de,j}3\cP_{\de'}}(z_2,z_3,\ptl_{z_3})(-2z_3\.p)^{-\de'} \nn \\
\to\ & \cC_{12\cP_{\de,j}}(z_1,z_2,\ptl_{z_2},\ptl_{w_2})\cC_{\cP_{\de,j}3\cP_{\de'}}(z_2,z_3,\ptl_{z_3})(-2z_3\.z_0)^{-\de'} 
= g^{(\de_1,\de_2,\de_3,\de')}_{\de,j}(z_1,z_2,z_3,z_0),
\ee
where the second line is simply the definition of a four-point conformal block.

\begin{figure}[t]
\centering
\includegraphics[width=10cm]{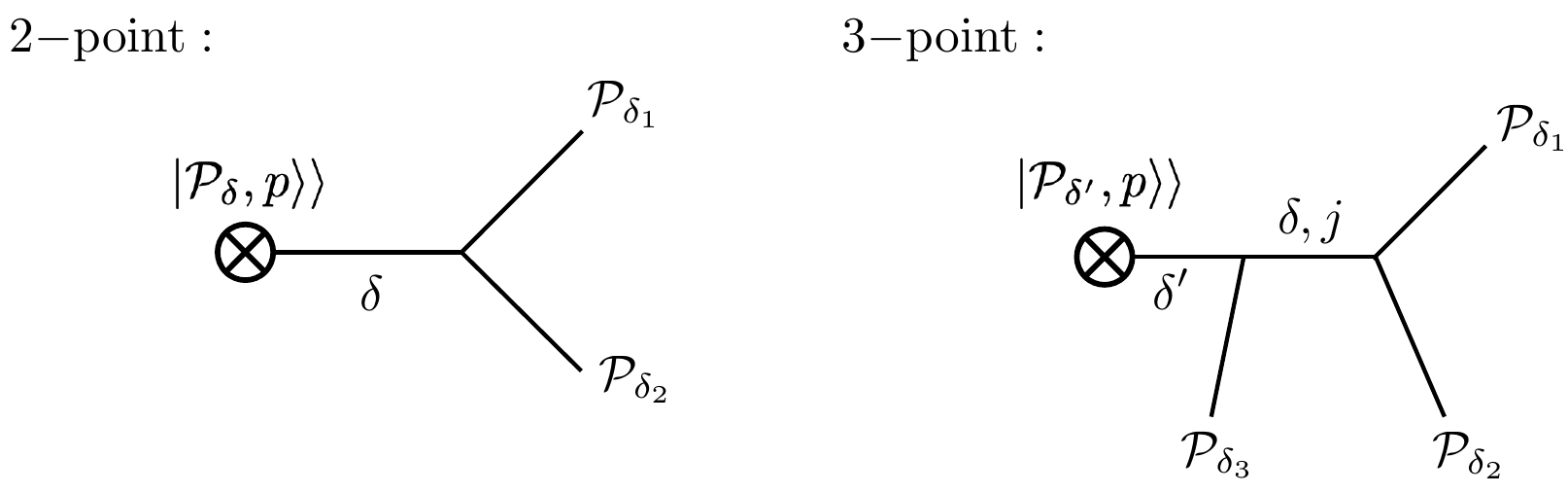}
\caption{Diagrams representing the two-point and three-point celestial blocks. Each vertex should be understood as an OPE differential operator $\cC_{ijk}$ in the $\mathrm{CFT}_{d-2}$. The symbol $\otimes$ represents the factor \eqref{eq:AdSstate_twoptstructure}.}
\label{fig:celestialOPE}
\end{figure}

To characterize subleading terms in the expansion around the collinear limit (or equivalently the null $p$ limit), it is helpful to introduce $|\cP_{\de'},p\>\>$, defined as the state in the conformal multiplet of $|\cP_{\de'}\>$ that is invariant under the little group $\SO(d-1)$ that fixes $p$. From the point of view of conformal symmetry, $p$ is a point in the center of EAdS$_{d-1}$, so the overlap of $|\cP_{\de'},p\>\>$ with $|\cP_{\de'}(z_3)\>$ is a bulk-to-boundary propagator:
\be\label{eq:AdSstate_twoptstructure}
\<\cP_{\de'}(z_3)|\cP_{\de'},p\>\>=(-2z_3\.p)^{-\de'}.
\ee
The three-point celestial block can be written as
\be\label{eq:threeptcelestialblock_AdSstate}
\cC_{12\cP_{\de,j}}(z_1,z_2,\ptl_{z_2},\ptl_{w_2})\cC_{\cP_{\de,j}3\cP_{\de'}}(z_2,z_3,\ptl_{z_3})\<\cP_{\de'}(z_3)|\cP_{\de'},p\>\>.
\ee
This expression can be represented by the diagram in figure \ref{fig:celestialOPE}, where we also include the diagram for the two-point case. Theses diagrams can be thought of as the OPE decomposition of the usual three-point and four-point function in the fictitious $\mathrm{CFT}_{d-2}$, but with one of the external operators replaced by $|\cP_{\de'},p\>\>$.

Let us determine a more explicit expression for $|\cP_{\de'},p\>\>$.
Treating $\SO(d-1,1)$ as the $(d{-}2)$-dimensional conformal group, its generators are $\{D,M_{ab},P_{a},K_{a}\}$, where $a,b=1,\cdots,d-2$, $M_{ab}$ are $\SO(d-2)$ rotation generators, and $P_{a}$ and $K_{a}$ are the translation and special conformal generators in $\R^{d-2}$. For $p=(p^0,\vec p)=(1,\vec 0)$, the little group that fixes $p$ is generated by $M_{ab}$ and $P_{a}+K_{a}$. Therefore, the $|\cP_{\de'},p\>\>$ must satisfy the conditions
\be
M_{ab}|\cP_{\de'},(1,\vec 0)\>\>&=0, \nn \\
(P_a+K_a)|\cP_{\de'},(1,\vec 0)\>\>&=0.
\ee
The solution to these conditions was obtained in \cite{Nakayama:2015mva} with a somewhat different motivation (studying local probes of a dual AdS geometry). The result is
\be\label{eq:AdSstate_expansion}
|\cP_{\de'},(1,\vec 0)\>\>=\G(\de'+2-\tfrac{d}{2})\p{\frac{\sqrt{P^2}}{2}}^{\frac{d-2}{2}-\de'}J_{\de'-\frac{d-2}{2}}\p{\sqrt{P^2}}|\cP_{\de'}\>,
\ee
where $|\cP_{\de'}\>$ is the CFT primary state killed by $K_a$, and $J_{\nu}(x)$ is a Bessel function of the first kind.

The action of momentum generators $P_a$ on $|\cP_{\de'}\>$ can be written as derivatives of $|\cP_{\de'}(z)\>$ with respect to $z$.
Equation~(\ref{eq:AdSstate_expansion}) thus expresses $|\cP_{\de'},(1,\vec 0)\>\>$ as an infinite-order differential operator acting on $|\cP_{\de'}(z)\>$. Plugging this into \eqref{eq:threeptcelestialblock_AdSstate}, we obtain the three-point celestial block as an infinite-order differential operator acting on a four-point conformal block:
\be
&\cC_{12\cP_{\de,j}}(z_1,z_2,\ptl_{z_2},\ptl_{w_2})\cC_{\cP_{\de,j}3\cP_{\de'}}(z_2,z_3,\ptl_{z_3})(-2z_3\.p)^{-\de'}
\nn\\
&=\G(\de'+2-\tfrac{d}{2})\p{\frac{\sqrt{\ptl_{\vec y}^2}}{2}}^{\frac{d-2}{2}-\de'}J_{\de'-\frac{d-2}{2}}\p{\sqrt{\ptl_{\vec y}^2}}g_{\de,j}^{(\de_1,\de_2,\de_3,\de')}(z_1,z_2,z_3,z_0),
\label{eq:firstexpressionforcelestialbock}
\ee
where $z_0=(1,\vec y^2,\vec y)$. Note that only even powers of $\sqrt{\ptl_{\vec y}^2}$ appear in the expansion of the Bessel function, so the above differential operator is well-defined order-by-order in this expansion. Even though intermediate terms depend on the point $z_0$, the final result must be independent of $z_0$. In appendix \ref{app:celestialblock}, we present an alternative derivation of this identity that leads to an expression where Lorentz invariance is more manifest.

Note that the state $|\cP_{\de'},p\>\>$ breaks the Lorentz group $\SO(d-1,1)$ to the little group $\SO(d-1)$. This is the same pattern of symmetry breaking that occurs in the presence of a codimension-1 spherical boundary or defect \cite{Gadde:2016fbj}. Thus, celestial blocks are equivalent to boundary/defect conformal blocks \cite{Liendo:2012hy}. Often, boundaries and defects are studied in Euclidean space, where the symmetry breaking pattern in CFT$_{d-2}$ is $\SO(d)\to\SO(d-1)$. Since the signature of the corresponding orthogonal groups is different, our celestial blocks are related by analytic continuation to those blocks.

Expanding our expression to leading and subleading in the collinear limit $p^2\to 0$, we find
\be\label{eq:threept_celestialblock_crossratios}
&\cC_{12\cP_{\de,j}}(z_1,z_2,\ptl_{z_2},\ptl_{w_2})\cC_{\cP_{\de,j}3\cP_{\de'}}(z_2,z_3,\ptl_{z_3})(-2z_3\.p)^{-\de'} \nn \\
&=T_{123\de'}(z_1,z_2,z_3,p)\bigg( g_{\de,j}^{(123\de')}(u,v) +\frac{\z_{13}}{d-4-2\de'}\cD_{u,v}^{(1)}g_{\de,j}^{(123\de')}(u,v) + O((-p^2)^2) \bigg),
\ee
where $u$ and $v$ are defined as
\be\label{eq:uandv_definition}
u=\frac{(-2z_1\.z_2)(-2z_3\.p)}{(-2z_1\.z_3)(-2z_2\.p)},\quad v=\frac{(-2z_2\.z_3)(-2z_1\.p)}{(-2z_1\.z_3)(-2z_2\.p)},
\ee
and the overall factor $T_{123\de'}(z_1,z_2,z_3,p)$ is
\be\label{eq:homogeneityfactor}
T_{123\de'}(z_1,z_2,z_3,p)=\frac{\p{\frac{-2z_2\.p}{-2z_1\.p}}^{\frac{\de_1-\de_2}{2}}\p{\frac{-2z_1\.p}{-2z_1\.z_3}}^{\frac{\de_3-\de'}{2}}}{(-2z_1\.z_2)^{\frac{\de_1+\de_2}{2}}(-2z_3\.p)^{\frac{\de_3+\de'}{2}}}.
\ee
 The differential operator $\cD_{u,v}^{(1)}$ generating the first subleading term is given by
\be
\label{eq:funnyD}
\cD_{u,v}^{(1)}=&\frac{1}{2}(\de_1-\de_2)(\de_1-\de_2-\de_3+\de')u+\frac{1}{2}(\de_3+\de')((\de_1-\de_2)v-\de_1+\de_2+\de_3-\de') \nn \\
&+((2-2\de_1+2\de_2+\de_3-\de')u-(\de_3+\de')(v-1))v\ptl_v \nn \\
&+((2-\de_1+\de_2-\de_3-\de')v-(\de_1-\de_2-\de_3+\de')(u-1))u\ptl_u \nn \\
&+2uv(u\ptl_u^2+v\ptl_v^2+(u+v-1)\ptl_u\ptl_v).
\ee
As an example, let us specialize to a three-point energy correlator in a 4d CFT. In this case, we have $\de_1=\de_2=\de_3=3$, and (\ref{eq:funnyD}) becomes
\be
\cD_{u,v}^{(1)\cE\cE\cE}=&\frac{(3-\de')(3+\de')}{2}+((5-\de')u-(3+\de')(v-1))v\ptl_v-((-3+\de')(u-1)+(1+\de')v)u\ptl_u \nn \\
&+2uv(u\ptl_u^2+v\ptl_v^2+(u+v-1)\ptl_u\ptl_v).
\ee

Using \eqref{eq:threept_celestialblock_crossratios}, we can finally write down the expansion of the three-point event shape in the collinear limit:
\be
&\<\f_5(p)|\wL[\f_1](\oo, z_1)\wL[\f_2](\oo, z_2)\wL[\f_3](\oo, z_3)|\f_4(p)\> \nn \\
&=\sum_{\de'}\sum_{\de,j}r_{12\cP_{\de,j}}r'_{\cP_{\de,j}3\cP_{\de'}}s'_{45\de'}(-p^2)^{\frac{\de_4+\de_5+\de'-1-d}{2}}T_{123\de'}(z_1,z_2,z_3,p)\nn \\
&\qquad\qquad\qquad \x\bigg( g_{\de,j}^{(123\de')}(u,v) +\frac{\z_{13}}{d-4-2\de'}\cD_{u,v}^{(1)}g_{\de,j}^{(123\de')}(u,v) + O((-p^2)^2) \bigg).
\ee
In sections~\ref{sec:EEEC_LO} and \ref{sec:allorderEEEC} of this paper, we will mostly focus on the leading term, which is simply a four-point conformal block. In section~\ref{sec:strongcoupling}, when we study $\cN=4$ at strong coupling, we will derive results related to the full structure of the three-point celestial block.

We can also take the OPE of \eqref{eq:threepteventshape_def} in a different order. If we first take the 23 OPE, we obtain
\be
&\<\f_5(p)|\wL[\f_1](\oo, z_1)\wL[\f_2](\oo, z_2)\wL[\f_3](\oo, z_3)|\f_4(p)\> \nn \\
&=\sum_{\de'}\sum_{\de,j}r_{23\cP_{\de,j}}r'_{\cP_{\de,j}1\cP_{\de'}}s'_{45\de'}(-p^2)^{\frac{\de_4+\de_5+\de'-1-d}{2}}T_{321\de'}(z_3,z_2,z_1,p)\nn \\
&\qquad\qquad\qquad \x\bigg( g_{\de,j}^{(321\de')}(v,u) +\frac{\z_{13}}{d-4-2\de'}\cD_{v,u}^{(1)}g_{\de,j}^{(321\de')}(v,u) + O((-p^2)^2) \bigg).
\ee
Thus, we obtain a crossing equation
\be\label{eq:collinearEEEC_crossing_subleading}
&\sum_{\de'}\sum_{\de,j}r_{12\cP_{\de,j}}r'_{\cP_{\de,j}3\cP_{\de'}}s'_{45\de'}(-p^2)^{\frac{\de_4+\de_5+\de'-1-d}{2}}T_{123\de'}(z_1,z_2,z_3,p)\nn \\
&\qquad\qquad\qquad \x\bigg( g_{\de,j}^{(123\de')}(u,v) +\frac{\z_{13}}{d-4-2\de'}\cD_{u,v}^{(1)}g_{\de,j}^{(123\de')}(u,v) + O((-p^2)^2) \bigg) \nn \\
&=\sum_{\de'}\sum_{\de,j}r_{23\cP_{\de,j}}r'_{\cP_{\de,j}1\cP_{\de'}}s'_{45\de'}(-p^2)^{\frac{\de_4+\de_5+\de'-1-d}{2}}T_{321\de'}(z_3,z_2,z_1,p)\nn \\
&\qquad\qquad\qquad \x\bigg( g_{\de,j}^{(321\de')}(v,u) +\frac{\z_{13}}{d-4-2\de'}\cD_{v,u}^{(1)}g_{\de,j}^{(321\de')}(v,u) + O((-p^2)^2) \bigg).
\ee
The leading term of this equation looks like a usual four-point crossing equation in a $(d{-}2)$-dimensional CFT. We will study some of its implications in section \ref{sec:analytic_bootstrap} and appendix \ref{app:lightcone_bootstrap_largej}. It would also be interesting to study \eqref{eq:collinearEEEC_crossing_subleading} with subleading terms included.

\subsection{Expansion of the EEEC in the collinear limit}\label{sec:EEEC_collinear_expansion}

We are now ready to study the expansion of the three-point energy correlator in the collinear limit. In what follows, we only keep the leading term (the conformal block). We also specialize to four dimensions, for simplicity. We can essentially follow section \ref{sec:threepteventshape}, replacing the scalar operators $\f_1,\f_2,\f_3$ with the stress-tensor $T$. Note that the ANEC operator $\cE$ is still a scalar on the celestial sphere, so the only difference from our earlier analysis  is the homogeneity in the momentum $p$. For sink/source states created by a scalar operator $\cO$, the result is\footnote{For later convenience, we relabel the points as $1\to 2, 2\to 3,3\to 1$.}
\be\label{eq:EEEC_decomposition_expr0}
&\<\cO(p)|\cE(\vec{n}_1)\cE(\vec{n}_2)\cE(\vec{n}_3)|\cO(p)\> \nn \\
&=\sum_{\de,j}r_{\cE\cE\cP_{\de,j}}r'_{\cP_{\de,j}\cE\cP_{\de'_{*}}}s'_{\cO\cO\cP_{\de'_{*}}}\frac{(-p^2)^{\frac{2\De_{\cO}+\de'_{*}-1}{2}}\p{\frac{-2z_2\.p}{-2z_1\.z_2}}^{\frac{3-\de'_{*}}{2}}}{(-2z_2\.z_3)^3(-2z_1\.p)^{\frac{3+\de'_{*}}{2}}}g_{\de,j}^{(\cE\cE\cE\cP_{\de'_{*}})}(z,\bar{z}) + \dots,
\ee
where ``$\dots$" denotes subleading terms in the collinear limit, and $\de'_{*}$ is the smallest value of $\de'$ that appears in the $\cP_{\de,j}\x \cE$ OPE in the fictitious $\mathrm{CFT}_{2}$. We assume for now that $\de'_*$ is isolated.\footnote{We study a case where two operators have degenerate $\de_*'$ at the lowest order in perturbation theory in section~\ref{sec:degeneracies}.}  We have also changed variables from $u,v$ to $z,\bar z$, defined by
\be\label{eq:zzbar_definition}
u=z \bar z=\frac{\z_{23}}{\z_{12}}, \quad v=(1-z)(1-\bar z)=\frac{\z_{13}}{\z_{12}},
\ee
where $z, \bar z$ are complex conjugates of each other.\footnote{Note that earlier we used $z$ as a null polarization vector $z\in\R^{d-1,1}$, whereas here it is a complex number $z\in \C$. We hope that no confusion will arise from this overloaded notation.}

Recall that the EEEC is defined by \eqref{eq:EEEC_def_CFT}. Note that $\<\cO(p)|\cE(\vec{n}_1)\cE(\vec{n}_2)\cE(\vec{n}_3)|\cO(p)\>$ depends only on angles between $\vec n_1,\vec n_2,\vec n_3$, which are localized by the delta functions in the first line of (\ref{eq:EEEC_def_CFT}). The remaining Jacobian factor is
\be
&\int d\O_{\vec{n}_1}d\O_{\vec{n}_2}d\O_{\vec{n}_3}\de(\z_{12}-\tfrac{1-\vec{n}_1\.\vec{n}_2}{2})\de(\z_{13}-\tfrac{1-\vec{n}_1\.\vec{n}_3}{2})\de(\z_{23}-\tfrac{1-\vec{n}_2\.\vec{n}_3}{2}) \nn \\
&\to \frac{64\pi^2}{|(\vec n_1\x \vec n_2) \. \vec n_3|} \nn \\
&=\frac{32\pi^2}{\sqrt{-\z_{12}^2-\z_{13}^2-\z_{23}^2+2\z_{12}\z_{13}+2\z_{12}\z_{23}+2\z_{13}\z_{23}-4\z_{12}\z_{13}\z_{23}}}.
\ee
Also, the total cross section is given by
\be\label{eq:crosssection_O}
\s^{\cO}_{\mathrm{total}}\equiv \int d^4x~e^{ip\.x}\<0|\cO^\dag(x)\cO(0)|0\>=\frac{2^{5-2\De_{\cO}}\pi^{3}}{\G(\De_{\cO}-1)\G(\De_{\cO})}(-p^2)^{\frac{2\De_{\cO}-4}{2}}\th(p).
\ee

Combining these results with \eqref{eq:EEEC_decomposition_expr0}, the EEEC in the collinear limit is given by
\be\label{eq:EEEC_Gfunction_definition}
\mathrm{EEEC}(\z_{12},\z_{13},\z_{23})
&=\frac{\z_{12}}{\z_{23}^3\sqrt{-\z_{12}^2-\z_{13}^2-\z_{23}^2+2\z_{12}\z_{13}+2\z_{12}\z_{23}+2\z_{13}\z_{23}}}\cG(\z_{12}, z,\bar z) + \dots,
\ee
where
\be\label{eq:Gfunc_decomposition}
\cG(\z_{12}, z,\bar z)&\equiv \z_{12}^{\frac{\de'_{*}-5}{2}}\sum_{\de,j}R_{\de,j;\de'_{*}}g_{\de,j}^{(\cE\cE\cE\cP_{\de'_{*}})}(z,\bar{z}),\nn\\
R_{\de,j;\de'_{*}}&\equiv\frac{2^{2\De_{\cO}-9}\G(\De_{\cO}-1)\G(\De_{\cO})}{\pi}r_{\cE\cE\cP_{\de,j}}r'_{\cP_{\de,j}\cE\cP_{\de'_{*}}}s'_{\cO\cO\cP_{\de'_{*}}}.
\ee
Here, we have set $z_i=(1,\vec n_i)$ and $\z_{ij}=\frac{1-\vec n_i\.\vec n_j}{2}$.

Thus, the function $\cG(\z_{12}, z,\bar z)$ describing the leading behavior of the EEEC in the collinear limit can be written as a sum of conformal blocks, up to a power of $\z_{12}$. The coefficients $R_{\de,j;\de'_{*}}$ appearing in the expansion are products of light-ray OPE coefficients $r_{\cE\cE\cP_{\de,j}}$ and $r'_{\cP_{\de,j}\cE\cP_{\de'_{*}}}$ and 1-point functions $s'_{\cO\cO\cP_{\de'_{*}}}$. Since most of these quantities are unknown a-priori, we will mostly just work with the coefficients $R_{\de,j;\de'_{*}}$ in this paper. The detailed definition of $R_{\de,j;\de'_{*}}$ would become useful if one could derive a three-point light-ray OPE formula that relates $r_{\cE\cE\cP_{\de,j}}r'_{\cP_{\de,j}\cE\cP_{\de'_{*}}}s'_{\cO\cO\cP_{\de'_{*}}}$ to the OPE data of the CFT (similar to \eqref{eq:rscoeff_lightrayOPE}).

Although most of the coefficients in the expansion (\ref{eq:Gfunc_decomposition}) are a-priori unknown, we {\it do know} a lot about the quantum numbers $\de'_*$ and $\de$ that appear. The dimension $\de'_*$ is associated with the lowest dimension light-ray operator in the triple-$\cE$ OPE. It is natural to guess that it is given by the lowest-twist spin-4 operator in the theory: $\de'_*=\De_\mathrm{min}(J=4)$, as we discuss in section~\ref{sec:lightrayOPE_prediction}. Furthermore, the dimensions $\de$ that appear in the conformal block expansion (\ref{eq:Gfunc_decomposition}) are controlled by the two-$\cE$ light-ray OPE, which we understand much better --- they are associated to light-ray operators with spin $J=3,5,\dots$ \cite{Kologlu:2019mfz,Chang:2020qpj}. Below, we will confirm these expectations in examples.

\subsection{Lightcone bootstrap constraints}
\label{sec:analytic_bootstrap}
The leading term of the crossing equation \eqref{eq:collinearEEEC_crossing_subleading} can be written as\footnote{Here, we use two different notations for conformal blocks, and we hope the meaning hereafter will be clear from context. The first notation is $g_{\de,j}^{(\cO_1\cO_2\cO_3\cO_4)}$, where the block is labeled by the conformal multiplets of each individual external operator $\cO_1,\cdots,\cO_4$. The second notation is $g_{\de,j}^{\de_{12},\de_{34}}$, where we use the fact that the block depends only on the differences of scaling dimensions $\de_{12}=\de_1-\de_2$ and $\de_{34}=\de_3-\de_4$.} 
\be\label{eq:collinearEEEC_crossing_leading}
\cG(\z_{12},z,\bar{z})&=\z_{12}^{\frac{\de'_{*}-5}{2}}\sum_{\de,j}R_{\de,j;\de'_*}g_{\de,j}^{(\cE\cE\cE\cP_{\de'_{*}})}(z,\bar{z}) \nn \\
&=\p{\frac{z\bar{z}}{(1-z)(1-\bar{z})}}^{3}\z_{12}^{\frac{\de'_{*}-5}{2}}\sum_{\de,j}R_{\de,j;\de'_*}g_{\de,j}^{0,\frac{3-\de'_{*}}{2}}(1-z,1-\bar{z}).
\ee
This looks like a four-point crossing equation in a $(d{-}2)$-dimensional CFT. It is thus interesting to ask what we can deduce about the original $d$-dimensional CFT from it. Unfortunately, the coefficients $R_{\de,j;\de'_*}$ do not satisfy any simple positivity conditions, so numerical bootstrap methods do not apply in an obvious way. In this section, we will instead study (\ref{eq:collinearEEEC_crossing_leading}) from the point of view of the lightcone bootstrap \cite{Fitzpatrick:2012yx,Komargodski:2012ek,Simmons-Duffin:2016wlq}, which does not require positivity conditions.

The usual lightcone bootstrap analysis begins by analytically continuing the four-point function into Lorentzian signature, and then considering the double lightcone limit $z\ll 1-\bar z \ll1$. In our setting, this would require analytically continuing  celestial cross-ratios away from the Euclidean regime. However, our ``correlator" $\cG(\z_{12},z,\bar z)$ does not come from a local, reflection-positive Euclidean CFT$_{d-2}$, and thus it is not guaranteed that we can analytically continue it to $(d{-}2)$-dimensional Lorentzian signature (which would be $(2,d-2)$ signature from the point of view of the full $d$-dimensional theory).

However, we believe that the main conclusion of the analytic bootstrap, i.e.\ the existence of double-twist families at large spin, can still be obtained by staying in Euclidean signature. The idea is that by plugging the leading $t$-channel singularity into the Euclidean inversion formula, one can still deduce that the OPE coefficient density $C(\De,J)$ should behave as the lightcone bootstrap predicts at large spin $J$. Similarly, it should be possible to compute subleading corrections in $J$ from subleading terms in the $t$-channel singularity. Hence, we expect that analytic continuation in $z,\bar z$ can be thought of as a proxy for a more complicated analysis using Euclidean partial waves.

Thus, let us proceed to studying implications of the lightcone bootstrap for the celestial crossing equation (\ref{eq:collinearEEEC_crossing_leading}).
We will actually consider a more general three-point event shape $\wL[\cO]\wL[\cO]\wL[\cO]$, where $\cO$ is an operator with spin $J$. The crossing equation reads
\be\label{eq:celestialcrossing_generalJ_1}
&\sum_{\de,j}R_{\de,j,\de'_*}g_{\de,j}^{(0,s)}(z,\bar{z})=\p{\frac{z\bar{z}}{(1-z)(1-\bar{z})}}^{\de_{\cO}}\sum_{\de,j}R_{\de,j,\de'_*}g_{\de,j}^{(0,s)}(1-z,1-\bar{z}),
\ee
where $s=\frac{\de_{\cO}-\de'_{*}}{2}$. As we discuss later in section \ref{sec:lightrayOPE_prediction}, we expect that $\de'_{*}=\De'_{*}-1$ is the celestial scaling dimension of the lowest-twist operator with spin $J'=3J-2$ in the $\cO\x\cO\x\cO$ OPE. Suppose the lowest celestial sphere twist $\tau_c=\de-j$ appearing in the sum on the left-hand side is $\tau_c^*=2h^{*}$. Then in the double lightcone limit $z\ll 1-\bar z\ll 1$, we have
\be\label{eq:celestialcrossing_generalJ_2}
R^{*}_{h^*}z^{h^*-2h_{\cO}}k_{2\bar h^{*}}^{0,s}(\bar z)+ \ldots =\sum_{h,\bar h}R_{h,\bar h}(1-\bar z)^{h-2h_{\cO}}k_{2\bar h}^{0,s}(1-z) + \dots,
\ee
where we have introduced
\be
h=\frac{\de-j}{2}=\frac{\tau_c}{2},\qquad \bar{h}=\frac{\de+j}{2}.
\ee
Near the $\bar z \to 1$ limit, the $\SL(2,\R)$ block $k_{2\bar{h}}^{0,s}(\bar{z})$ has the expansion \cite{Simmons-Duffin:2016wlq}
\be
k_{2\bar{h}}^{0,s}(\bar{z})&=K_0^{0,s}(\bar{h})+ \ldots +K_0^{s,0}(\bar{h})(1-\bar{z})^{-s}+\dots, \nn \\
K_0^{r,s}(\bar{h})&\equiv \frac{\G(r-s)\G(2\bar{h})}{\G(\bar{h}+r)\G(\bar{h}-s)},
\ee
where ``$\dots$" represents terms like $(1-\bar{z})^{k}$ or $(1-\bar{z})^{k-s}$ where $k$ is a positive integer. Therefore, \eqref{eq:celestialcrossing_generalJ_2} becomes
\be
&R^{*}_{h^*}z^{h^*-2h_{\cO}}\p{K_0^{0,s}(\bar{h})+\dots+K_0^{s,0}(\bar{h})(1-\bar{z})^{-s}+\dots}+ \dots \nn \\
&=\sum_{h,\bar h}R_{h,\bar h}(1-\bar z)^{h-2h_{\cO}}k_{2\bar h}^{0,s}(1-z) + \dots.
\ee
Note that on the left-hand side, $z^{h^*-2h_{\cO}}$ is Casimir-singular in $z$ (i.e.\ it can be made arbitrarily singular by repeatedly applying the quadratic Casimir), while on the right-hand side $k_{2\bar h}^{0,s}(1-z)$ is Casimir-regular in $z$. In order to reproduce the Casimir-singular term with the correct $\bar z$ behavior on the left hand side, we must have an infinite family of operators with $h\to 2h_{\cO}$ and $h\to 2h_{\cO}-s=h_{\cO}+h'_{*}$ as $\bar h\to \oo$. They can be thought of as the ``celestial double-twist operators" $[\cP_{\cO}\cP_{\cO}]$ and $[\cP_{\cO}\cP_{\cO'_{*}}]$. 

Since $[\cP_{\cO}\cP_{\cO}]$ has $h=2h_{\cO}$ and large-$j$ (which means that they must be higher transverse spin terms), the light-ray OPE \cite{Kologlu:2019mfz,Chang:2020qpj} predicts that they should come from operators with conventional twist $\tau=2\de_{\cO}+2=2\De_{\cO}$. Thus, the existence of $[\cP_{\cO}\cP_{\cO}]$ predicts that for the maximally allowed transverse spin ($j_{max}=2J$ in this case) in the $\cO\x\cO$ OPE, there must be a trajectory with $\tau=2\tau_{\cO}+2J$ at large spin. Similarly, the existence of $[\cP_{\cO}\cP_{\cO'_{*}}]$ predicts that there should be a trajectory with the maximally allowed transverse spin that has twist $\tau=\De_{\cO}+\De'_{*}$ at large spin.

The coefficients $R_{h,\bar h}$ at large $\bar h$ for the two types of celestial double-twist operators can also be determined. Using the formula \cite{Simmons-Duffin:2016wlq}
\be\label{eq:SL2Rblock_infinitesum}
\sum_{\substack{\bar h=j+ \bar h_0 \\ j=0,2,\dots}}S_{a}^{r,s}(\bar h)k^{r,s}_{2h}(1-z)=\frac{1}{2}\p{\frac{z}{1-z}}^{a}+\text{Casimir-regular}, \nn \\
S^{r,s}_a(\bar h)=\frac{1}{\G(-a-r)\G(-a-s)}\frac{\G(\bar h-r)\G(\bar h-s)}{\G(2\bar h-1)}\frac{\G(\bar h-a-1)}{\G(\bar h+a+1)},
\ee
we find
\be\label{eq:lightconeRcoeff_generalJ}
R_{h=2h_{\cO}}(\bar h) \sim& 2R^{*}_{h^*} K_0^{0,s}(\bar{h}^{*})S_{h^{*}-2h_{\cO}}^{0,s}(\bar h), \nn \\
R_{h=h_{\cO}+h'_{*}}(\bar h) \sim& 2R^{*}_{h^*} K_0^{s,0}(\bar{h}^{*})S_{h^{*}-2h_{\cO}}^{0,s}(\bar h),
\ee
where $\sim$ means that both sides have the same leading behavior at large $\bar h$.

In \eqref{eq:celestialcrossing_generalJ_2}, the left-hand side will have a $\log(1-\bar z)$ term when $s=0$. In the usual four-point lightcone bootstrap, the $\log(1-\bar z)$ term determines the behavior of the anomalous dimensions of double-twist operators. However, unlike in the usual lightcone bootstrap, \eqref{eq:celestialcrossing_generalJ_2} does not possess an identity operator. As a result, the interpretation of $\log(1-\bar z)$ as coming from anomalous dimensions does not work in this case. To see this, consider the case where $\cO$ is a spin-1 conserved current $\cJ$. In this case, $\cO'_{*}$ should be the lowest twist operator with spin $J'=1$ in the $\cJ\x\cJ\x\cJ$ OPE, which is just $\cJ$ itself. So, $\de'_{*}=\de_{\cJ}$ and $s=0$. For $s=0$, the leading term of left hand side of \eqref{eq:celestialcrossing_generalJ_2} becomes
\be\label{eq:lightconecrossing_conservedJ}
-R^{*}_{h^*}z^{h^*-2h_{\cO}}\frac{\G(2\bar{h}^*)}{\G(\bar{h}^*)^2}(2\psi(\bar{h}^*)-2\psi(1)+\log(1-\bar{z})).
\ee
To see how the above expression can be produced by the right hand side of \eqref{eq:celestialcrossing_generalJ_2}, note that near $s=0$, the coefficients $R_{h=2h_{\cO}}(\bar h)$ and $R_{h=h_{\cO}+h'_{*}}(\bar h)$ in \eqref{eq:lightconeRcoeff_generalJ} are singular and take the form
\be
R_{h=2h_{\cO}}(\bar h) =&\frac{\tl{R}}{s} + \tl{R}^{(0)} + O(s), \nn \\
R_{h=h_{\cO}+h'_{*}}(\bar h) =&-\frac{\tl{R}}{s} + \tl{R}^{(0)} + O(s),
\ee
where
\be
\tl{R}=-2R^{*}_{h^*} \frac{\G(2\bar h^{*})}{\G(\bar h^{*})^2}S_{h^{*}-2h_{\cO}}^{0,s}(\bar h),\qquad \tl{R}^{(0)}=(-\psi(1)+\psi(\bar h^{*}))\tl{R}.
\ee
Therefore, the contribution from the two celestial double-twist operators becomes (going back to the notation $g_{\de,j}$ for conformal blocks)
\be
&\lim_{s\to 0}R_{h=2h_{\cO}}(\bar h) g^{0,s}_{4h_{\cO}+j,j} + R_{h=h_{\cO}+h'_{*}}(\bar h)g^{0,s}_{2h_{\cO}-2s+j,j} \nn \\
&=\lim_{s\to 0}\p{\frac{\tl{R}}{s} + \tl{R}^{(0)}}\p{g^{0,0}_{4h_{\cJ}+j,j}+s\ptl_sg^{0,s}_{4h_{\cJ}+j,j}} + \p{-\frac{\tl{R}}{s} + \tl{R}^{(0)}}\p{g^{0,0}_{4h_{\cJ}+j,j}+s\ptl_sg^{0,s}_{4h_{\cJ}-2s+j,j}} \nn \\
&=2\tl{R}^{(0)}g^{0,0}_{4h_{\cJ}+j,j} +2\tl{R}\ptl_{\de}g^{0,0}_{\de,j}|_{\de\to 4h_{\cJ}+j},
\ee
which correctly reproduces \eqref{eq:lightconecrossing_conservedJ} after performing the sum over $\bar h$. We see that the $\log(1-\bar z)$ of \eqref{eq:lightconecrossing_conservedJ} comes from a near-cancellation of coefficients between the two celestial double-twist families at the degenerate point $s=0$.

\section{Extracting light-ray OPE data from the leading order collinear EEEC}
\label{sec:EEEC_LO}
In the previous section, we argued that the EEEC in the collinear limit can be decomposed into conformal blocks (up to a Jacobian factor). In this section, we study the decomposition of the leading-order collinear EEEC in $\cN=4$ SYM and QCD recently computed in \cite{Chen:2019bpb}. In the case of $\cN=4$ SYM, the authors of \cite{Chen:2019bpb} consider sink/source states created by the operator $\mathrm{Tr}F^2$ (with $\De_{\cO}=4$). For the QCD case, they consider both the gluon jet, created by $\Tr F^2$, and the quark jet, created by the quark contribution to the electromagnetic current $J^{\mu}$. Specifically, they contract indices between the bra and the ket, so that the quark jet event shape is an expectation value in the density matrix $|J^\mu(p)\>\<J_\mu(p)|$. 

Note that \cite{Chen:2019bpb} worked at low enough loop order that the $\beta$-function doesn't enter, so QCD can be thought of as conformal for the purposes of studying their results. At higher orders in perturbation theory, selection rules in $J$, such as those discussed in \cite{Hofman:2008ar,Kologlu:2019mfz} will be broken. However, the celestial block expansion should still be valid, since it relies on Lorentz invariance alone. We leave an investigation of these effects to the future.

 The results of \cite{Chen:2019bpb} for the leading-order collinear EEEC can be summarized by three functions of cross ratios, $G_{\cN=4}(z,\bar z)$ (equation (5.2) and (5.3) in \cite{Chen:2019bpb}), $G_{\mathrm{QCD}}^{g}(z,\bar z)$ (square bracket in equation (5.14) in \cite{Chen:2019bpb}), and $G^{q}_{\mathrm{QCD}}(z,\bar z)$ (square bracket in equation (5.16) in \cite{Chen:2019bpb}). The relation between $G(z,\bar z)$ and the EEEC is
\be\label{eq:EEEC_LO_definition}
\mathrm{EEEC}(\z_{12},\z_{13},\z_{23})&=\frac{g^4N_c^2}{64\pi^5\z_{12}^2\sqrt{-\z_{12}^2-\z_{13}^2-\z_{23}^2+2\z_{12}\z_{13}+2\z_{12}\z_{23}+2\z_{13}\z_{23}}}\x G_{\cN=4}(z,\bar{z}) + \dots, \nn \\
\mathrm{EEEC}(\z_{12},\z_{13},\z_{23})&=\frac{g^4}{32\pi^5\z_{12}^2\sqrt{-\z_{12}^2-\z_{13}^2-\z_{23}^2+2\z_{12}\z_{13}+2\z_{12}\z_{23}+2\z_{13}\z_{23}}}\x G_{QCD}^{g/q}(z,\bar{z}) + \dots.
\ee

In the weak coupling limit, we can expand \eqref{eq:Gfunc_decomposition} as
\be\label{eq:weakcouplingexpansion_1}
\cG&=a_0\p{\cG^{(0)}+a\cG^{(1)}+a^2\cG^{(2)}+\dots}, \nn \\
\de_i&=\de^{(0)}_i+a\g^{(1)}_i+a^2\g^{(2)}_i+\dots, \nn \\
\de'_{*}&=5+a\g'^{(1)}_*+a^2\g'^{(2)}_*+\dots, \nn \\
R_{\de,j;\de'_{*}}&=a_0\p{R^{(0)}_{\de,j;\de'_{*}}+aR^{(1)}_{\de,j;\de'_{*}}+\dots},
\ee
where for $\cN=4$ SYM we have $a_0=\frac{g^2N_c}{16\pi^3}$, $a=\frac{g^2N_c}{4\pi^2}$, and for QCD $a_0=\frac{g^2}{2\pi^3}$, $a=\frac{g^2}{16\pi^2}$.\footnote{We choose $a_0$ such that there is no prefactor in \eqref{eq:Gfunc_relation}.} As we explain later in section \ref{sec:lightrayOPE_prediction}, we also set $\de_{*}^{'(0)}=5$ since it is the $J'=4$ point on the twist-2 trajectory. Comparing \eqref{eq:EEEC_Gfunction_definition}, \eqref{eq:EEEC_LO_definition} and \eqref{eq:weakcouplingexpansion_1}, we immediately see that $\cG^{(0)}=0$ and\footnote{Note that \eqref{eq:Gfunc_decomposition} assumes that the sink/source states are created by scalar operators. Even though for the QCD quark jet case, the current $J^{\mu}$ is spin-1, we can still treat it as a scalar here since we contract indices between the bra and ket. This will produce an additional factor of 2 due to the fact that for a spin-1 operator $V_{\mu}$, $\<0|V^{\mu}(x)V_{\mu}(0)|0\>=2\<0|\f(x)\f(0)|0\>$ where $\f$ is a scalar with dimension $\De_{V}$, and we use the conventional two-point structures \cite{Costa:2011mg} for operators with spin 0 and 1 in CFT.}
\be\label{eq:Gfunc_relation}
\cG^{(1)}(\z_{12},z,\bar z)=(z\bar z)^3G(z,\bar z)
\ee
for both $\cN=4$ SYM and QCD. The expansion of \eqref{eq:Gfunc_decomposition} in the weak coupling limit is given by
\be
\cG^{(0)}(\z_{12},z,\bar{z})=&\sum_{\de^{(0)},j}\<R^{(0)}_{\de,j;\de'_*}\>g_{\de^{(0)},j}^{(\cE\cE\cE\cP_{\de'_{*}=5})}(z,\bar{z}) \nn \\
\cG^{(1)}(\z_{12},z,\bar{z})=&\sum_{\de^{(0)},j}\p{\<R^{(1)}_{\de,j;\de'_*}\>g_{\de^{(0)},j}^{(\cE\cE\cE\cP_{\de'_{*}=5})}(z,\bar{z})+\<R^{(0)}_{\de,j;\de'_*}\g^{(1)}_{\de,j}\>\ptl_{\de}g_{\de^{(0)},j}^{(\cE\cE\cE\cP_{\de'_{*}=5})}(z,\bar{z})} \nn \\
&+\g'^{(1)}_*\sum_{\de^{(0)},j}\p{\<R^{(0)}_{\de,j;\de'_*}\>\ptl_{\de'_*}g_{\de^{(0)},j}^{(\cE\cE\cE\cP_{\de'_{*}})}(z,\bar{z})+\frac{1}{2}\log(\z_{12})\cG^{(0)}(z,\bar{z})},
\ee
where the notation $\<\dots\>$ represents a sum of the contributions from possibly degenerate operators, following e.g.\ \cite{Alday:2017vkk}.\footnote{The degeneracy can come from either operators with the same $\de,j$ or operators with the same $\de'_{*}$.} Since $\cG^{(0)}=0$, we must have $\<R^{(0)}_{\de,j;\de'_*}\>=0$. The function $\cG^{(1)}$ then becomes independent of $\z_{12}$ (which is consistent with \eqref{eq:Gfunc_relation}), and it can be written as
\be\label{eq:cG1_conformalblockdecomposition}
\cG^{(1)}(z,\bar{z})=&\sum_{\de^{(0)},j}\p{\<R^{(1)}_{\de,j;\de'_*}\>g_{\de^{(0)},j}^{(\cE\cE\cE\cP_{\de'_{*}=5})}(z,\bar{z})+\<R^{(0)}_{\de,j;\de'_*}\g^{(1)}_{\de,j}\>\ptl_{\de}g_{\de^{(0)},j}^{(\cE\cE\cE\cP_{\de'_{*}=5})}(z,\bar{z})}.
\ee

\begin{figure}[t]
\centering
\includegraphics[width=8cm]{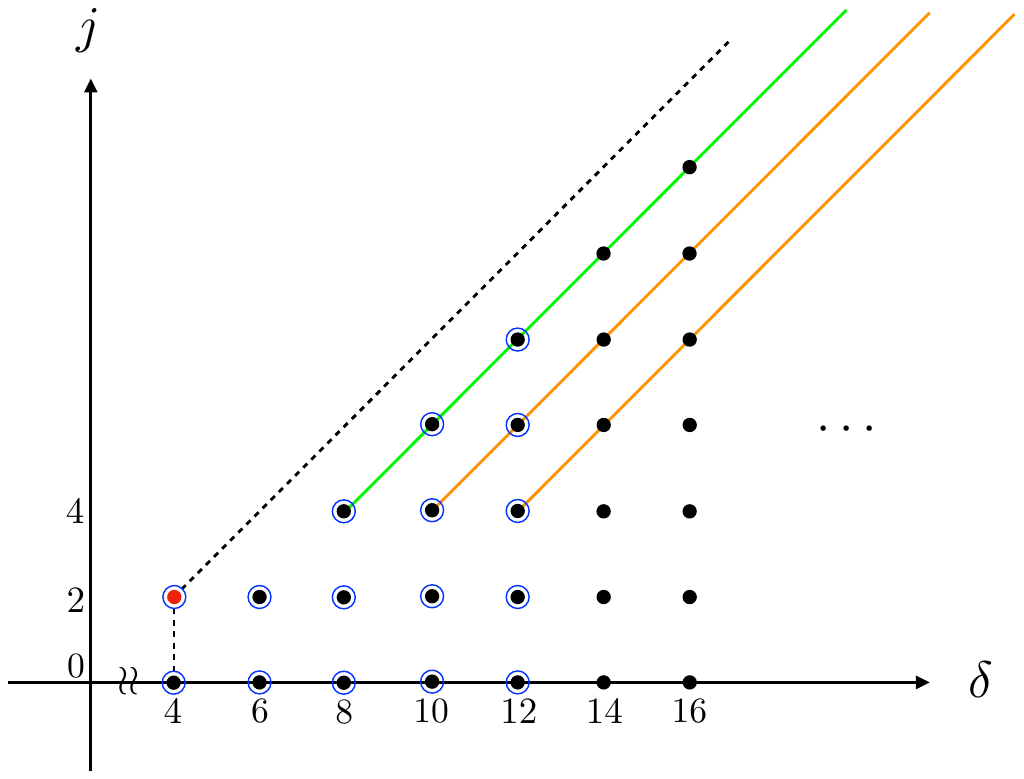}
\caption{A plot showing the values of $\de,j$ appearing in the conformal block decomposition \eqref{eq:cG1_conformalblockdecomposition}. Black dashed line is the improved unitarity bound from \eqref{eq:4dunitaritybound}. The red dot appears only in QCD, and the black dots appear in both QCD and $\cN=4$ SYM. The coefficients $\<R^{(1)}_{\de,j;\de'_*}\>$ and $\<R^{(0)}_{\de,j;\de'_*}\g^{(1)}_{\de,j}\>$ of dots with a blue circle can be obtained using the direct decomposition method. The $\<R^{(1)}_{\de,j;\de'_*}\>$ coefficient of the green line and the $\<R^{(0)}_{\de,j;\de'_*}\g^{(1)}_{\de,j}\>$ coefficient of the orange line can be obtained using the Lorentzian inversion formula.}
\label{fig:lightrayOPEdata}
\end{figure}

The values of $\de^{(0)}$ and $j$ appearing in the decomposition \eqref{eq:cG1_conformalblockdecomposition} can be related to the spectrum of the theory using the light-ray OPE formula. We describe this relation in section \ref{sec:lightrayOPE_prediction}. We then obtain coefficients $\<R^{(1)}_{\de,j;\de'_*}\>$ and $\<R^{(0)}_{\de,j;\de'_*}\g^{(1)}_{\de,j}\>$  using both a direct series expansion around the OPE limit (section \ref{sec:Rcoeff_directdecomposition}) and the Lorentzian inversion formula (section \ref{sec:Rcoeff_Lorentzianinversionformula}).  Our results are summarized in figure \ref{fig:lightrayOPEdata}, where we plot the allowed values of $\de,j$ in \eqref{eq:cG1_conformalblockdecomposition} and indicate points for which we obtained the ceofficients $\<R^{(1)}_{\de,j;\de'_*}\>$ and $\<R^{(0)}_{\de,j;\de'_*}\g^{(1)}_{\de,j}\>$ using either of the two methods.

\subsection{Predictions from the light-ray OPE}\label{sec:lightrayOPE_prediction}
Schematically, the light-ray OPE for two stress tensors is given by \cite{Kologlu:2019mfz,Chang:2020qpj}
\be\label{eq:TT_lightrayOPE}
\wL[T]\x \wL[T] \sim \sum_{i} \mathbb{O}^{+}_{i,J=3,j=0} + \mathbb{O}^{+}_{i,J=3,j=2} +  \mathbb{O}^{+}_{i,J=3,j=4} +\sum_{n,i}\cD_{2n}\mathbb{O}^{+}_{i,J=3+2n,j=4}.
\ee
This formula allows us to predict which quantum numbers $\de^{(0)}, j$ appear in the decomposition \eqref{eq:cG1_conformalblockdecomposition}. First, we immediately see that only even values of $j$ can appear. This is consistent with the fact that the OPE of two identical scalars on the celestial sphere only includes operators with even transverse spin.

To see the allowed values of $\de$, let us denote the conventional twist $\tau=\De-J$ of a trajectory with transverse spin $j$ by $\tau_j$.\footnote{We are working in perturbation theory, so the twist of each Regge trajectory is fixed and $\tau_j$ doesn't depend on $J$. In fact, we will simply use $\tau_j$ to label each Regge trajectory.} Note that the $T\x T$ OPE only contains trajectories with transverse spin $j=0,2,4$. We also define a ``celestial twist" $\tau_c\equiv\de-j$, where $\de,j$ are the quantum numbers appearing in \eqref{eq:cG1_conformalblockdecomposition}.

Using \eqref{eq:TT_lightrayOPE}, we can see that for $j=0,2$, the relation between $\tau_j$ and $\tau_c$ is given by
\be\label{eq:tauc_tau_relation_lowj}
\tau_c(j=0,2)=\de-j=\De(J=3)-1-j=\tau_j+2-j,
\ee
while for $j\geq 4$ we have
\be
\tau_c(j)=\de-j=\De(J=-1+j)-1-j=\tau_{j=4}-2.
\ee
In 4d, the conventional twist $\tau_j$ should satisfy the improved unitarity bound \cite{Cordova:2017dhq}
\be\label{eq:4dunitaritybound}
\tau_j\geq \mathrm{max}\{2,j\}.
\ee
Consequently, one might expect that the values of $\tau_j$ and $\tau_c(j)$ that can appear are
\be
\tau_{j=0}&= 2,4,6, \dots \quad \Rightarrow \quad \tau_c(j=0)=4,6,8,\dots,  \nn \\
\tau_{j=2}&= 2,4,6, \dots \quad \Rightarrow \quad \tau_c(j=2)=2,4,6, \dots,  \nn \\
\tau_{j=4}&=4,6,8, \dots \quad \Rightarrow \quad \tau_c(j\geq 4)=2,4,6, \dots.
\ee
However, as discussed in \cite{Chen:2020adz,Chen:2021gdk}, the contribution of the $j=4$, $\tau_{j=4}=4$ operator vanishes at the leading order for both $\cN=4$ SYM and QCD, and the contribution of the $j=2$, $\tau_{j=2}=2$ operator vanishes in $\cN=4$ SYM due to supersymmetry. So the actual values of $\tau_j$ and $\tau_c(j)$ appearing at the leading order should be
\be
\tau_{j=0}&= 2,4,6, \dots \quad \Rightarrow \quad \tau_c(j=0)=4,6,8, \dots,  \nn \\
\tau_{j=2}&= \mathbf{2},4,6, \dots \quad \Rightarrow \quad \tau_c(j=2)=\mathbf{2},4,6,\dots,  \nn \\
\tau_{j=4}&= 6,8,10, \dots \quad \Rightarrow \quad \tau_c(j\geq 4)=4,6,8, \dots,
\ee
where the bold $\tau_{j=2}=\mathbf{2}$ appears only in QCD.

What operators realize these quantum numbers? We will focus on operators with leading $\tau_c$. We will identify light-ray operators by writing local operators on their Regge trajectories, with $J$ as a free parameter. The actual light-ray operators are obtained by light-transforming and analytically continuing $J$ to the appropriate value according to (\ref{eq:TT_lightrayOPE}).

For $\cN=4$ SYM, the leading $\tau_c$ is $4$, coming from operators with $j=0,2,4$. For $j=0$, $\tau_{j=0}=2$, the operators can be schematically written as\footnote{We use the $\mathfrak{su}(2)\oplus \mathfrak{su}(2)$ spinor indices for the 4d Lorentz indices. The derivative is $\ptl_{\dot \a \b}$, and the field contents are $\bar \f, \f, \bar \psi_{\dot \a}, \psi_{\b}, \bar F_{\dot \a_1 \dot \a_2}, F_{\b_1 \b_2}$. In \eqref{eq:j0operator_N4SYM}-\eqref{eq:tauc2j2_nondegenerateop}, the Lorentz indices are implicitly symmetrized and the gauge indices are implicitly contracted.}
\be\label{eq:j0operator_N4SYM}
\bar \f\ptl^{J}\f, \quad \bar \psi\ptl^{J-1}\psi, \quad \bar F\ptl^{J-2}F.
\ee
For $j=2$, $\tau_{j=2}=4$, we have
\be\label{eq:j2operator_N4SYM}
\psi \ptl^{J_1} \psi \ptl^{J_2} \psi \ptl^{J-J_1-J_2-2} \psi, \quad F\ptl^{J-2}\Box F, \quad \bar F \ptl^{J_1} F \ptl^{J_2} F \ptl^{J-J_1-J_2-4} F,
\ee
and for $j=4$, $\tau_{j=4}=6$,
\be\label{eq:j4operator_N4SYM}
F\ptl^{J_1}F\ptl^{J_2} F \ptl^{J-J_1-J_2-4} \Box F, \quad \bar F \ptl^{J_1} F \ptl^{J_2} F \ptl^{J_3} F \ptl^{J_4} F \ptl^{J-J_1-J_2-J_3-J_4-6} F.
\ee
Note that there are degeneracies for all three values of $j$. (Supersymmetry relates some of these trajectories, but we do not study the consequences of supersymmetry here.) On the other hand, in QCD the leading $\tau_c$ is $2$, and it is carried by one operator with $j=2,\tau_{j=2}=2$:
\be\label{eq:tauc2j2_nondegenerateop}
F \ptl^{J-2} F.
\ee
Thus, the leading $\tau_c$ operator is non-degenerate in QCD.

Finally, let us discuss the value of $\de'_{*}$ in \eqref{eq:cG1_conformalblockdecomposition}. The value of $\de'_{*}$ should be determined by a generalized light-ray OPE formula
\be
\wL[T]\x\mathbb{O}_{i,J=3,j} \sim \mathbb{O}'_{i,J=4,\l} ,
\ee
where $\mathbb{O}'_{i,J=4,\l}$ is some unknown object that transforms like a primary operator with scaling dimension $\De_{\mathbb{O}'}=1-J=-3$. Though we do not have a rigorous definition of the object $\mathbb{O}'_{i,J=4,\l}$ for a general nonperturbative CFT, we expect that it should be related to light transforms of operators in the $T\x T\x T$ OPE with $J=4$. Indeed, it has been shown in \cite{Chen:2020adz,Chen:2021gdk} that for perturbative QCD, this object is just the light transformed operator $\wL[\cO_{i,J=4,\l}]$ itself (at least at the leading order). Recently, there is also evidence from LHC data showing that the scaling behavior of the three-point energy correlator in the perturbative regime is governed by the twist-2 spin-4 anomalous dimension \cite{Komiske:2022enw}. Therefore, in this paper we will assume that $\de'_{*}$ is given by $\de'_{*}=\De'_{*}-1=5$ (at leading order in perturbation theory), since $\De'_{*}=6$ is the scaling dimension of the leading twist-$2$, spin-$4$ operator in the $T\x T\x T$ OPE.\footnote{See also \cite{Holguin:2022epo}, where they show that in QCD, the leading correction to the scaling of collinear EEEC determined by $\de'_{*}=5$ can be used for top quark mass measurements.}

\subsection{Celestial block coefficients from direct decomposition}\label{sec:Rcoeff_directdecomposition}
We now explain how to obtain the coefficients $\<R^{(1)}_{\de,j;\de'_{*}}\>$ and $\<R^{(0)}_{\de,j;\de'_{*}}\g^{(1)}_{\de,j}\>$ in \eqref{eq:cG1_conformalblockdecomposition} using the known result for $\cG^{(1)}(z,\bar z)$ computed in \cite{Chen:2019bpb}. Firstly, we can simply expand \eqref{eq:cG1_conformalblockdecomposition} in the OPE limit $z=re^{i\th},\bar z=re^{-i\th}$ with $r\to 0$ and compare both sides order-by-order in $r$. Recall that the 2d block $g_{\de,j}^{(\cE\cE\cE\cP_{\de'_{*}=5})}(z,\bar z)$ in \eqref{eq:cG1_conformalblockdecomposition} is given by
\be
g_{\de,j}^{(\cE\cE\cE\cP_{\de'_{*}=5})}(z,\bar z)=&\frac{1}{1+\de_{0,j}}(k^{0,-1}_{\de+j}(z)k^{0,-1}_{\de-j}(\bar z) + k^{0,-1}_{\de-j}(z)k^{0,-1}_{\de+j}(\bar z)), \nn \\
k^{r,s}_{\b}(x)=&x^{\frac{\b}{2}}{}_2F_1\p{\tfrac{\b}{2}-r,\tfrac{\b}{2}+s,\b,x}.
\ee

The OPE limit of $\cG^{(1)}$ corresponds to a ``squeezed limit" on the celestial sphere, where two of the detectors are taken to be even closer after the collinear limit. The expansion of \cite{Chen:2019bpb} in the squeezed limit has been studied in \cite{Chen:2020adz,Chen:2021gdk} up to $O(r^{10})$.\footnote{We thank Hao Chen, Ian Moult, and Hua Xing Zhu for sending us a mathematica notebook containing the expansion.} Using these results, we find that the first few coefficients $\<R^{(1)}_{\de,j}\>$ and $\<R^{(0)}_{\de,j}\g^{(1)}_{\de,j}\>$ in $\cN=4$ SYM are given by
\be\label{eq:Rcoeff_N4SYM_directdecomposition1}
&\<R^{(1)}_{4,0}\>=1, \quad \<R^{(1)}_{6,0}\>=\frac{41}{12}-\frac{\pi^2}{4}, \quad \<R^{(1)}_{6,2}\>=\frac{107}{60}-\frac{\pi^2}{6}, \nn \\
&\<R^{(1)}_{8,0}\>=\frac{471}{600}-\frac{\pi^2}{15},\quad \<R^{(1)}_{8,2}\>=\frac{4883}{2800}-\frac{9\pi^2}{56}, \quad \<R^{(1)}_{8,4}\>=\frac{2843}{5040}-\frac{\pi^2}{18}, \nn \\
&\<R^{(0)}_{8,0}\g^{(1)}_{8,0}\>=\frac{2}{5}, \quad \<R^{(0)}_{8,2}\g^{(1)}_{8,2}\>=-\frac{1}{20}.
\ee

In QCD, we use $R^{(1)g/q}_{\de,j}$ to denote coefficients for the gluon/quark jet. For the QCD gluon jet, we find
\be\label{eq:Rcoeff_QCD_gluon_directdecomposition1}
&\<R^{(1)g}_{4,0}\>=\frac{98 C_A^2+14 C_A n_f T_F+15 C_F n_f T_F}{1600}, \nn \\
&R^{(1)g}_{4,2}=\frac{C_A (C_A-2 n_f T_F)}{2880}, \nn \\
&\<R^{(1)g}_{6,0}\>= \frac{\left(636386-63000 \pi ^2\right) C_A^2+2 \left(12600 \pi ^2-120899\right) C_A n_f T_F-819 C_F n_f T_F}{403200}, \nn \\
&\<R^{(1)g}_{6,2}\>= \frac{\left(834469-84000 \pi ^2\right) C_A^2+4 \left(13125 \pi ^2-129587\right) C_A n_f T_F+1260 C_F n_f T_F}{504000}. 
\ee
Note that we do not use the bracket notation $\<\cdots\>$ for the coefficient $R^{(1)}_{4,2}$, because from the discussion in section \ref{sec:lightrayOPE_prediction} it is non-degenerate. For the QCD quark jet we find
\be\label{eq:Rcoeff_QCD_quark_directdecomposition1}
&\<R^{(1)q}_{4,0}\>= \frac{C_F (91 C_A+240 C_F+13 n_f T_F)}{4800}, \nn \\
&R^{(1)q}_{4,2}= \frac{C_F (C_A-2 n_f T_F)}{2880}, \nn \\
&\<R^{(1)q}_{6,0}\>= \frac{C_F \left(\left(109200 \pi ^2-1077733\right) C_A-28 \left(5100 \pi ^2-50929\right) C_F+3 \left(111199-11200 \pi ^2\right) n_f T_F\right)}{403200}, \nn \\
&\<R^{(1)q}_{6,2}\>= \frac{C_F \left(\left(157500 \pi ^2-1548703\right) C_A-210 \left(1300 \pi ^2-12859\right) C_F+326 n_f T_F\right)}{1008000}. 
\ee
We have further expanded the results of \cite{Chen:2019bpb} up to $O(r^{12})$, which will be helpful when comparing to the Lorentzian inversion formula result in section \ref{sec:Rcoeff_Lorentzianinversionformula}. We record the coefficients up to $\de=12$ in appendix \ref{app:Rcoeff_directdecomposition}.

\subsection{The Lorentzian inversion formula on the celestial sphere}
\label{sec:Rcoeff_Lorentzianinversionformula}
Direct decomposition yields OPE data at low dimensions $\de$, but becomes cumbersome as $\de$ gets larger. Alternatively, we can use the Lorentzian inversion formula (LIF) \cite{Caron-Huot:2017vep} to extract OPE data from a four-point correlator. Like the lightcone bootstrap, the LIF requires us to analytically continue the correlator to Lorentzian signature --- which in our case means complexifying the celestial sphere. Again, it is not clear whether this analytic continuation is admissible nonperturbatively. However, nothing prevents us from using the LIF as a tool in perturbation theory, as long as perturbative correlators are well-behaved. Indeed, we do not observe any pathologies when analytically continuing the results of \cite{Chen:2019bpb} in $z,\bar z$. It would be interesting to study the analytic structure of the collinear EEEC as a function of $z,\bar z$ at higher orders in perturbation theory.

The LIF is only valid for $j>j_0$, where the ``Regge intercept" $j_0$ controls the behavior of the correlator in the Regge limit. To reach the Regge limit, one should first take $\bar z$ around the branch point at $1$, and then take both $z$ and $\bar z$ to zero. From the point of view of celestial CFT, this is a strange kinematic regime, and we do not have rigorous bounds (much less physical intuition) for how the correlator should behave there. Note that the Regge limit on the celestial sphere has nothing to do with the Regge limit in $d$-dimensional Minkowski space (as far as we know), and that the celestial Regge intercept $j_0$ is not obviously related to the usual Regge intercept $J_0$. However, we can study this limit in perturbation theory. 

In $\cN=4$ SYM, we find that the leading term of $\cG^{(1)}(z,\bar z)$ in the celestial Regge limit is given by
\be
&\cG^{(1)}_{\mathrm{Regge}}(z,\bar z) \nn \\
&=\frac{i \pi  r^3 (w^2+1) \left(2 (w^2-1) \left((w^2-1)^2 \log (w+\frac{1}{w})-2 w^2\right)-2 (w^6-3 w^4-3 w^2+1) \log w\right)}{w (w^2-1)^3},
\ee
where we set $z=rw$ and $\bar z =r/w$. The scaling $r^3$ implies that the celestial Regge intercept is given by $j_0=-2$ at this order in perturbation theory. The expressions for $\<R^{(1)}_{\de,j;\de'_*}\>$ and $\<R^{(0)}_{\de,j;\de'_*}\g^{(1)}_{\de,j}\>$ for $\cN=4$ SYM obtained from the LIF should then be valid for all $j=0,2,4,\dots$.

For the QCD gluon jet, $\cG^{(1)g}$ can be written as
\be
\cG^{(1)g}=C_F n_f T_F \cG^{(1)g}_1 + C_A n_f T_F \cG^{(1)g}_2 + C_A^2 \cG^{(1)g}_3.
\ee
We find that the Regge intercept for $\cG^{(1)g}_1$ is $j_0=0$, while for $\cG^{(1)g}_2$ and $\cG^{(1)g}_3$ the intercept is $j_0=2$. Therefore, for the coefficient proportional to $C_F n_f T_F$, the LIF result will agree with the result from direct decomposition only for $j>0$. For the other two flavor structures ($C_A n_f T_F$ and $C_A^2$), we expect the result to agree only for $j>2$. Similarly, for QCD quark jet, $\cG^{(1)q}$ can be written as
\be
\cG^{(1)q}=C_F n_f T_F \cG^{(1)q}_1 + (C_A-2C_F)C_F \cG^{(1)q}_2 + \cG^{(1)q}_3,
\ee
where $\cG^{(1)q}_3$ contains both $C_F^2$ and $C_F C_A$ factors. We find that the Regge intercept for $\cG^{(1)q}_2$ is $j_0=0$, and the intercept for $\cG^{(1)q}_1$ and $\cG^{(1)q}_3$ is $j_0=2$.

Let us now briefly review the LIF. Consider a four-point function $g(z,\bar z)$ in 2d with conformal block expansion
\be
g(z,\bar z)=\sum_{\de,j}p_{\de,j}g^{\de_i}_{\de,j}(z,\bar z),
\ee
where only even $j$ appear in the sum. The LIF in this case can be written as
\be\label{eq:Lorentzianinversionformula_1}
C^{+}(\de,j)=\frac{\kappa_{\de+j}}{2}\int_{0}^{1}\int_0^{1}\frac{dz d\bar z}{(z \bar z)^2} g^{\tl{\de_i}}_{j+1,\de-1}(z,\bar z)\mathrm{dDisc}_t[g(z,\bar z)],
\ee
where $\tl{\de}_i=2-\de_i$, and
\be\label{eq:kappaandmudef}
\kappa_{\b}&=\frac{\G(\frac{\b+\de_1-\de_2}{2})\G(\frac{\b-\de_1+\de_2}{2})\G(\frac{\b+\de_3-\de_4}{2})\G(\frac{\b-\de_3+\de_4}{2})}{2\pi^2\G(\b-1)\G(\b)}.
\ee
The double discontinuity $\mathrm{dDisc}_t[g(z,\bar z)]$ is
\be
\mathrm{dDisc}_t[g(z,\bar z)]&=\cos(\pi \f)g(z, \bar z)-\frac{1}{2}e^{i\pi\f}g^{\circlearrowleft}(z,\bar z)-\frac{1}{2}e^{-i\pi\f}g^{\circlearrowright}(z,\bar z), \nn \\
\f&=\frac{\de_2-\de_1+\de_3-\de_4}{2},
\ee
where $g^{\circlearrowleft}$ and $g^{\circlearrowright}$ indicate that we should take $\bar z$ around $1$ in the direction shown, with $z$ held fixed. Finally, the OPE coefficients $p_{\de,j}$ are given by
\be\label{eq:CdeltajvsOPEcoefficient}
p_{\de^*,j}=-\mathrm{Res}_{\de=\de^*}C^{+}(\de,j).
\ee

To compute $p_{\de,j}$, it is convenient to define a generating functional
\be
C(z,\de,j)&=\kappa_{\de+j}z^{\frac{\de-j}{2}-1}\int_0^1\frac{d\bar{z}}{\bar{z}^2}g^{\tl{\de}_i}_{j+1,\de-1}(z,\bar{z})\mathrm{dDisc}_t[g(z,\bar{z})],  \nn \\
\label{eq:Lorentzianinversionformula_2}
C^{+}(\de,j)&=\int_0^1\frac{dz}{2z}z^{-\frac{\de-j}{2}}C(z,\de,j).
\ee
In the small $z$ limit, $C(z,\de,j)$ should have the expansion
\be
C(z,\de,j)=\sum_{\tau_c}C(\tau_c,\de,j)z^{\frac{\tau_c}{2}}.
\ee
Using \eqref{eq:Lorentzianinversionformula_1}, we see that $C^{+}(\de,j)$ has a pole when $\de=j+\tau_c$, and therefore the OPE coefficient $p_{\de,j}$ can be written as\footnote{In general, there will be an additional Jacobian factor coming from the dependence of $\tau_c$ on $j$. At the order we are working in, $\tau_c$ on each trajectory is a constant and this Jacobian factor is just $1$.}
\be
p_{j+\tau_c,j}=2C(\tau_c,j+\tau_c,j),
\ee
where the additional factor of $2$ is due to $z \leftrightarrow \bar z$ symmetry.

Now, suppose we have a weak-coupling expansion
\be
g(z,\bar z)&=a\p{g^{(0)}(z,\bar z)+a g^{(1)}(z,\bar z)+\cdots}, \nn \\
\de&=\de^{(0)}+a\g^{(1)}+\cdots, \nn \\
p_{\de,j}&=a\p{p^{(0)}_{\de,j}+ap^{(1)}_{\de,j} + \cdots},
\ee
with coupling constant $a$, and we are interested in finding $p^{(1)}_{\de,j}$ and $p^{(0)}_{\de,j}\g^{(1)}_{\de,j}$. Expanding \eqref{eq:CdeltajvsOPEcoefficient} near $\de^{(0)}$, we have
\be\label{eq:Cdeltaj_weakcoupling}
C^{+}(\de,j) \sim a\p{-\frac{\<p^{(0)}_{\de,j}\>}{\de-\de^{(0)}}-a\frac{\<p^{(1)}_{\de,j}\>}{\de-\de^{(0)}} -a\frac{\<p^{(0)}_{\de,j}\g^{(1)}_{\de,j}\>}{(\de-\de^{(0)})^2} + \cdots},
\ee
where $\<\cdots\>$ indicates a sum over possible degenerate operators. For the generating functional, the expansion near $\de^{(0)}$ is
\be\label{eq:generatingfunctional_weakcoupling}
C(z,\de,j)\sim& a(C^{(0)}(\tau_c,\de,j)+aC^{(1)}(\tau_c,\de,j))z^{\frac{\de^{(0)}+a\g^{(1)}-j}{2}} +\cdots \nn \\
\sim& a\p{C^{(0)}(\tau_c,\de,j)+aC^{(1)}(\tau_c,\de,j)+a\frac{\g^{(1)}}{2}C^{(0)}(\tau_c,\de,j)\log z}z^{\frac{\tau_c^{(0)}}{2}},
\ee
where $\tau_c^{(0)}=\de^{(0)}-j$. Let us plug \eqref{eq:generatingfunctional_weakcoupling} into \eqref{eq:Lorentzianinversionformula_2} and compare it with \eqref{eq:Cdeltaj_weakcoupling}. We see that after integrating over $z$, the $C^{(1)}(\tau_c,\de,j)$ term becomes a simple pole corresponding to $\<p^{(1)}_{\de,j}\>$. On the other hand, the $\g^{(1)}C^{(0)}(\tau_c,\de,j)$ term has an additional $\log z$ and becomes a double pole, corresponding to $\<p^{(0)}_{\de,j}\g^{(1)}_{\de,j}\>$. The precise formula is
\be\label{eq:Lorentzianinversionformula_3}
\<p^{(1)}_{j+\tau_c,j}\>&=2C(z,\de,j)\bigg|_{z^{\frac{\tau_c}{2}}}, \nn \\
\<p^{(0)}_{j+\tau_c,j}\g^{(1)}_{j+\tau_c,j}\>&=4C(z,\de,j)\bigg|_{z^{\frac{\tau_c}{2}}\log z}.
\ee

\subsubsection{$\cN=4$ SYM}
Let us now apply the Lorentzian inversion formula to $\cG^{(1)}(z,\bar z)$ in $\cN=4$ SYM. After plugging in $g(z,\bar z) = \cG^{(1)}_{\cN=4}(z,\bar z)$ and $\de_1,\de_2,\de_3 = 3, \de_4 = 5$ in \eqref{eq:Lorentzianinversionformula_1}, we find that $\<R^{(1)}_{\de,j;\de'_*}\>$ is nonzero for even $j$ and ``celestial twists" $\tau_c=\de-j=4,6,8,\dots$. Furthermore, $\<R^{(0)}_{\de,j;\de'_*}\g^{(1)}_{\de,j}\>$ is nonzero for even $j$ and $\tau_c=6,8,10,\dots$ (except that $\<R^{(0)}_{\de=6,j=0;\de'_*}\g^{(1)}_{\de=6,j=0}\>=0$). We can find analytical expressions for $\<R^{(0)}_{j+\tau_c,j;\de'_*}\g^{(1)}_{j+\tau_c,j}\>$ for general $\tau_c$, and also $\<R^{(1)}_{j+4,j;\de'_*}\>$.\footnote{It would be interesting to calculate $\<R^{(1)}_{j+\tau_c,j;\de'_*}\>$ for general $\tau_c$ as well, but we leave that for future work.} The results agree with those obtained from direct decomposition in section \ref{sec:Rcoeff_directdecomposition} and appendix \ref{app:Rcoeff_directdecomposition}. 

As an example, let us describe the detailed calculation for $\<R^{(0)}_{\de,j;\de'_*}\g^{(1)}_{\de,j}\>$ at celestial twist $\tau_c=6$. For higher twists and $\<R^{(1)}_{j+4,j;\de'_*}\>$, we simply present the final result and leave details to appendix \ref{app:inversionformula_details}. Using \eqref{eq:Lorentzianinversionformula_3}, we have
\be
\<R^{(0)}_{j+6,j;\de'_*}\g^{(1)}_{j+6,j}\>=&4\p{\kappa_{2j+6}z^{2}\int_0^1\frac{d\bar{z}}{\bar{z}^2}g^{0,1}_{j+1,j+5}(z,\bar{z})\mathrm{dDisc}_t[\cG(z,\bar{z})]}\bigg|_{z^3\log z} \nn \\
=&4\kappa_{2j+6}\int_0^1\frac{d\bar{z}}{\bar{z}^2} k_{2j+6}^{0,1}(\bar z)\mathrm{dDisc}_t[\cG(z,\bar{z})]|_{z^3\log z},
\ee
where we have used the fact that the $\log z$ term of $\cG(z,\bar z)$ starts at $z^3$. We have also used the $\SL(2,\R)$ expansion of the conformal block
\be\label{eq:conformalblock_SL2Rexpansion}
g^{r,s}_{\de,j}(z,\bar z)=&z^{\frac{\de-j}{2}}k_{\de+j}^{r,s}(\bar z) + O(z^{\frac{\de-j}{2}+1}), \nn \\
k_{\b}^{r,s}(x)=&x^{\frac{\b}{2}}{}_2F_1(\tfrac{\b}{2}-r,\tfrac{\b}{2}+s,\b,x), \nn \\
r=\frac{\de_1-\de_2}{2},&\qquad s=\frac{\de_3-\de_4}{2}.
\ee

Next we should compute the double discontinuity $\mathrm{dDisc}_t[\cG(z,\bar z)]$. Note that the singularities of $\cG(z, \bar z)|_{z^3\log z}$ at $\bar z =1$ only have integer powers or single logarithms of $1-\bar z$. For example,
\be
\cG(z,\bar z)|_{z^3\log z}=-\frac{1}{4(1-\bar{z})}-\frac{3}{2}\log(1-\bar{z})+O((1-\bar{z})^{0}).
\ee
 Naively, such terms have vanishing dDisc. However the correct interpretation is that their dDiscs are distributions localized at $\bar z = 1$. See \cite{Henn:2019gkr} for examples of dealing with such distributions. To compute them, we insert a regulator $\e$ so that the powers become non-integer, removing the regulator after taking the $\mathrm{dDisc}$. For example, inserting a regulator $1/(1-\bar z)^\e$, we have
\be\label{eq:oneloop_anomalous_expr1}
&\<R^{(0)}_{j+6,j;\de'_*}\g^{(1)}_{j+6,j}\> \nn \\
&=\lim_{\e\to 0}4\kappa_{\b}\int_0^1\frac{d\bar{z}}{\bar{z}^2}k^{0,1}_{\b}(\bar{z})\mathrm{dDisc}_t\left[-\frac{1}{4(1-\bar{z})^{1+\e}}-\frac{3}{2(1-\bar{z})^{\e}}\log(1-\bar{z})+O((1-\bar{z})^{-\e})\right] \nn \\
&=\lim_{\e\to 0} \bigg(8\sin^2(\pi\e)\kappa_{\b}\int_0^1\frac{d\bar{z}}{\bar{z}^2}k^{0,1}_{\b}(\bar{z})\p{\frac{1}{4(1-\bar{z})^{1+\e}}+\frac{3}{2(1-\bar{z})^{\e}}\log(1-\bar{z})+O((1-\bar{z})^{-\e})} \nn \\
&\qquad\qquad-8\pi\sin(2\pi\e)\kappa_{\b}\int_0^1\frac{d\bar{z}}{\bar{z}^2}k^{0,1}_{\b}(\bar{z})\frac{3}{2(1-\bar{z})^{\e}}\bigg),
\ee
where $\b=2j+6$ and we have used \eqref{eq:dDisc_identities} for the double discontinuities. The $\bar z$ integrals will localize to $\bar z =1$ when taking the $\e\to 0$ limit, so we can expand the SL$_2$ block $k_\b^{0,1}(\bar z)$ in this limit. For example, in the last line of (\ref{eq:oneloop_anomalous_expr1}), we have
\be
&-8\pi\kappa_{\b}\lim_{\e\to 0}\sin(2\pi\e)\int_0^1\frac{d\bar{z}}{\bar{z}^2}k^{0,1}_{\b}(\bar{z})\frac{3}{2(1-\bar{z})^{\e}} \nn \\
=&-8\pi\kappa_{\b}\lim_{\e\to 0}\sin(2\pi\e)\int_0^1\frac{d\bar{z}}{\bar{z}^2}\p{\frac{\G(\b)}{\G(\tfrac{\b}{2})\G(\tfrac{\b}{2}+1)}\frac{1}{1-\bar z}+\cdots}\frac{3}{2(1-\bar{z})^{\e}} \nn \\
=&-8\pi\kappa_{\b}\frac{3\G(\b)}{2\G(\tfrac{\b}{2})\G(\tfrac{\b}{2}+1)}\lim_{\e\to 0}\sin(2\pi\e)\int_0^1\frac{d\bar{z}}{\bar{z}^2}\p{-\frac{1}{\e}\de(1-\bar z)+O(\e^0)} \nn \\
=&\frac{12\G(\frac{\b}{2}-1)\G(\frac{\b}{2})}{\G(\b-1)},
\ee
where we have used the expansion of $k_{\b}^{0,1}(\bar z)$ around $\bar z=1$ and the distributional identity $\frac{1}{x^{n+\e}}\th(x)=\frac{1}{\e}\frac{(-1)^n}{(n-1)!}\de^{(n-1)}(x)+O(\e^0)$. Performing similar calculations for the other terms, we  obtain
\be\label{eq:oneloop_anomalous_expr2}
\<R^{(0)}_{j+6,j;\de'_*}\g^{(1)}_{j+6,j}\>=-\frac{j(j+5)\G(j+2)^2}{2\G(2j+4)}.
\ee

Similarly, we can use \eqref{eq:Lorentzianinversionformula_3} to calculate $\<R^{(0)}_{j+\tau_c,j;\de'_*}\g^{(1)}_{j+\tau_c,j}\>$ with higher $\tau_c$ and $\<R^{(1)}_{j+4,j}\>$. For example, for $\tau_c=8$ we have (see \eqref{eq:Rgamma_generaltauc} for the expression for general $\tau_c$)
\be
\<R^{(0)}_{j+8,j}\g^{(1)}_{j+8,j}\>=\frac{\p{12(j+3)(j+4)S_1(j+3)-(19j^2+133j+192)}\G(j+3)\G(j+4)}{3\G(2j+7)},
\ee
where $S_1(n)=\sum_{k=1}^n \frac{1}{k}$ is the harmonic number. For $\<R^{(1)}_{j+4,j}\>$, the result is
\be\label{eq:oneloop_coeff_analytical}
\<R^{(1)}_{j+4,j}\>=&\frac{\G(j+2)^2}{6(j+1)\G(2j+3)}\nn \\
&\x\p{-9+2\pi^2+j(j+3)(-6+\pi^2)+\frac{6(j+1)}{(j+3)^2(j+4)}{}_3F_2\left(\begin{matrix} &2,\quad 3,\quad j+3 & \\ &j+4,\quad j+5&\end{matrix};1\right)}.
\ee

\subsubsection{QCD}
We can perform a similar calculation for the QCD case by replacing $g(z,\bar z)$ with $\cG^{(1)g/q}$ in \eqref{eq:Lorentzianinversionformula_1}. For the gluon jet case, we again find that $\<R^{(0)g}_{j+\tau_c,j}\g^{(1)}_{j+\tau_c,j}\>$ is nonzero for $\tau_c=6,8,\cdots$ and even $j$. The expressions for $\tau_c=6$ and $\tau_c=8$ are given by
\be\label{eq:R0gamma1tau6_QCD_gluon}
\<R^{(0)g}_{j+6,j}\g^{(1)}_{j+6,j}\>=-\frac{\G(j+3)\G(j+4)}{80\G(2j+5)}C_A n_F T_F -\frac{(7j^2+35j-18)\G(j+3)\G(j+2)}{80\G(2j+5)}C_A^2,
\ee
and
\be\label{eq:R0gamma1tau8_QCD_gluon}
&\<R^{(0)g}_{j+8,j}\g^{(1)}_{j+8,j}\>\nn \\
&=-\frac{3\G(j+4)\G(j+5)}{40\G(2j+7)}C_F n_F T_F -\frac{\left(28 j^4+392 j^3+1697 j^2+2275 j+204\right) \G (j+3) \G (j+4)}{336 \G (2 j+7)} C_A n_F T_F \nn \\
&\quad + \frac{\left(420 \left(j^2+7 j+12\right) S_1(j+3)+35 j^4+490 j^3+1052 j^2-4641 j-10056\right) \G (j+3) \G (j+4)}{840 \G (2 j+7)} C_A^2.
\ee
For the $\<R^{(1)g}_{j+\tau_c,j}\>$ coefficient, although one can see from \eqref{eq:Rcoeff_QCD_gluon_directdecomposition1} that it has leading twist $\tau_c=2$, the Lorentzian inversion formula will only give nonzero $\<R^{(1)g}_{j+\tau_c,j}\>$ for $\tau_c=4,6,\dots$. This is because the only nonzero $\<R^{(1)g}_{j+\tau_c,j}\>$ with $\tau_c=2$ has $j=2$, and only contains flavor structures with Regge intercept $j_0=2$ ($C_A^2$ and $C_A n_f T_F$). Therefore we don't expect the Lorentzian inversion formula to give the correct result for $\tau_c=2$. For $\tau_c=4$, we obtain
\be\label{eq:R1tau4_QCD_gluon}
&\<R^{(1)g}_{j+4,j}\>\nn \\
&=\frac{\G(j+2) \G (j+3)}{80 \G(2 j+3)}C_F n_F T_F \nn \\
&+ \left(-\frac{\left(31 j^2+93 j+17\right) \G (j+1) \G (j+2)}{7200 \G (2 j+3)} \right. \nn \\
&\qquad + \frac{\G(j+1)\G(j+2)^2\G(j+3)}{4\G(2j+3)\G(j+4)\G(j+5)}\left( 2 {}_3F_2\left(\begin{matrix} & 2,\quad 3 ,\quad j-2 & \\ & j+4 ,\quad j+5 &\end{matrix};1\right) \right. \nn \\
&\qquad\qquad \left.\left. -2 {}_3F_2\left(\begin{matrix} & 2 ,\quad 3,\quad j-1 & \\ & j+4 ,\quad j+5 &\end{matrix};1\right) + {}_3F_2\left(\begin{matrix} & 2 ,\quad 3,\quad j & \\ & j+4 ,\quad j+5 &\end{matrix};1\right)\right)\right)C_A n_F T_F \nn \\
&+ \left(-\frac{\left(315+(j+1)(j+2)(-1117+150\pi^2)\right) \G (j+1) \G (j+2)}{7200 \G (2 j+3)} \right. \nn \\ 
&\qquad \left. +  \frac{\G(j+1)\G(j+2)^2}{8\G(2j+3)\G(j+4)}\sum_{k=0}^4 c_k {}_3F_2\left(\begin{matrix} & 1,\quad 2,\quad j-3+k & \\ & j+3 ,\quad j+4 &\end{matrix};1\right) \right)C_A^2,
\ee
where
\be
c_0=1,\quad c_1=-2,\quad c_2=3,\quad c_3=-2,\quad c_4=1.
\ee
As expected from the value of the Regge intercept, the results \eqref{eq:R0gamma1tau6_QCD_gluon}, \eqref{eq:R0gamma1tau8_QCD_gluon} and \eqref{eq:R1tau4_QCD_gluon} agree with \eqref{eq:Rcoeff_QCD_gluon_directdecomposition1}, \eqref{eq:Rcoeff_QCD_gluon_directdecomposition2} and \eqref{eq:Rcoeff_QCD_gluon_directdecomposition3} for $j>2$, but for the flavor structure $C_F n_f T_F$ they also agree at $j=2$.

For the quark jet, the calculation is almost identical. We find that $\<R^{(0)q}_{j+\tau_c,j}\g^{(1)}_{j+\tau_c,j}\>$ is nonzero for $\tau_c=6,8,\cdots$, and for $\tau_c=6$ and $\tau_c=8$ the expressions are
\be\label{eq:R0gamma1tau6_QCD_quark}
\<R^{(0)q}_{j+6,j}\g^{(1)}_{j+6,j}\>=\frac{C_F(C_A(19j^2+95j-6)-32C_F(2j^2+10j-3))\G(j+2)\G(j+3)}{480\G(2j+5)},
\ee
and
\be\label{eq:R0gamma1tau8_QCD_quark}
&\<R^{(0)q}_{j+8,j}\g^{(1)}_{j+8,j}\>\nn \\
&=-\frac{(10j^4+140j^3+637j^2+1029j+204)\G(j+3)\G(j+4)}{120\G(2j+7)}C_F n_F T_F \nn \\
&\quad -\frac{3(3j^2+21j+16)\G(j+3)\G(j+4)}{160\G(2j+7)}C_F(C_A-2C_F)-\frac{(23j^2+161j-24)\G(j+3)\G(j+4)}{60\G(2j+6)}C_F^2  \nn \\
&\quad +\frac{(10 j^4+140 j^3+401 j^2-623 j-3018 + 120(j+3)(j+4)S_1(j+3))\G(j+3)\G(j+4)}{240\G(2j+7)}C_F C_A.
\ee
For the $\<R^{(1)q}_{j+\tau_c,j}\>$ coefficient, the leading twist $R^{(1)q}_{4,2}$ given in \eqref{eq:Rcoeff_QCD_quark_directdecomposition1} also only contains flavor structure with Regge intercept $j_0=2$. So from the Lorentzian inversion formula the leading nonzero coefficient starts at $\tau_c=4$, and it is given by
\be\label{eq:R1tau4_QCD_quark}
&\<R^{(1)q}_{j+4,j}\>\nn \\
&= \left(\frac{\left(1193-120\pi^2\right) \G (j+2) \G (j+3)}{5760 \G (2 j+3)} \right. \nn \\
&\qquad - \frac{\G(j+1)\G(j+2)^2\G(j+3)}{160\G(2j+3)\G(j+4)\G(j+5)}\left( -(j^2+3j+26) {}_3F_2\left(\begin{matrix} & 2,\quad 3,\quad j & \\ & j+4 ,\quad j+5 &\end{matrix};1\right) \right. \nn \\
&\qquad\qquad\qquad\qquad\qquad\qquad \left.\left. +(j^2+3j+40) {}_3F_2\left(\begin{matrix} & 2, \quad 3,\quad j+1 & \\ & j+4 ,\quad j+5 &\end{matrix};1\right) \right)\right)(C_A-2C_F)C_F \nn \\
&+\frac{C_F \left(C_A \left(167 j^2+501 j+220\right)+12 C_F \left(13 j^2+39 j+58\right)\right)\G(j+1) \G (j+2)}{5760 \G(2 j+3)} \nn \\
&+ \frac{\G(j+1)\G(j+2)^2}{16\G(2j+3)\G(j+4)}\left({}_3F_2\left(\begin{matrix} & 1,\quad 2,\quad j-2 & \\ & j+3 ,\quad j+4 &\end{matrix};1\right)+ {}_3F_2\left(\begin{matrix} & 1,\quad 2,\quad j & \\ & j+3 ,\quad j+4 &\end{matrix};1\right)\right)C_AC_F.
\ee
Similar to the gluon jet case, \eqref{eq:R0gamma1tau6_QCD_quark}, \eqref{eq:R0gamma1tau8_QCD_quark} and \eqref{eq:R1tau4_QCD_quark} agree with \eqref{eq:Rcoeff_QCD_quark_directdecomposition1}, \eqref{eq:Rcoeff_QCD_quark_directdecomposition2} and \eqref{eq:Rcoeff_QCD_quark_directdecomposition3} for $j>2$, but for the flavor structure $(C_A-2C_F)C_F$ they also agree at $j=2$.

\section{Higher-order collinear EEEC}
\label{sec:allorderEEEC}

In this section, we use the celestial block decomposition \eqref{eq:Gfunc_decomposition} for the collinear EEEC and the leading order coefficients obtained in section \ref{sec:EEEC_LO} to make predictions for higher-order terms in the expansion of the EEEC in the coupling constant. Specifically, we study the $(n+1)$-st order expansion of $\cG(\z_{12},z ,\bar z)$ in $a$, which we denote by $\cG^{(n+1)}(\z_{12},z,\bar z)$. Our key physical input is that contributions to \eqref{eq:Gfunc_decomposition} come from individual light-ray operators, whose contributions are fixed by symmetry in terms of their quantum numbers. In particular, this implies that anomalous dimensions should ``exponentiate" to create the power laws predicted by symmetry.

Exponentiation is most powerful when the operators of interest are non-degenerate in perturbation theory. Thus, the $\tau_c=2,j=2$ non-degenerate operator \eqref{eq:tauc2j2_nondegenerateop} in QCD will play a crucial role in our arguments. Expanding \eqref{eq:Gfunc_decomposition} in the weak coupling limit, we find that $\cG^{(n+1)}$ contains a term
\be
\cG^{(n+1)} \supset R^{(1)}_{4,2}(\g^{(1)}_{4,2})^n \ptl_{\de}^ng^{(\cE\cE\cE\cP_{\de'_{*}}=5)}_{\de,2}(z,\bar z)|_{\de \to 4}.
\ee
Since $R^{(1)}_{4,2}$ is non-degenerate, and the anomalous dimension $\g^{(1)}_{4,2}$ is known \cite{Chen:2021gdk}, we can predict the coefficient of this term in $\cG^{(n+1)}$ using available perturbative data. Moreover, it turns out that $\ptl_{\de}^ng^{(\cE\cE\cE\cP_{\de'_{*}}=5)}_{\de,j}(z,\bar z)$ dominates in certain kinematics limits, and thus we have a prediction for the behavior of $\cG^{(n+1)}$ in those limits. It is harder to apply the same argument for $\cG^{(n+1)}$ in $\cN=4$ SYM due to the fact that all the operators contributing to $\cG^{(1)}$ in $\cN=4$ SYM have tree-level degeneracies (see \eqref{eq:j0operator_N4SYM}, \eqref{eq:j2operator_N4SYM} and \eqref{eq:j4operator_N4SYM} for the leading $\tau_c$ operators). This problem could be circumvented by using higher-order perturbative data to disentangle the degeneracies, or perhaps by organizing the EEEC in $\cN=4$ into an appropriate super-celestial-block expansion. Regardless, we will focus on QCD in this section.

Before proceeding, let us comment on the nonzero $\b$-function of QCD. Note that even in the presence of a nonzero $\b$-function, the celestial block decomposition \eqref{eq:Gfunc_decomposition}, which follows from Lorentz symmetry, should exist. However, some features will be different. Firstly, the spin selection rule $J=3$ for operators in the $\cE\x\cE$ OPE will be violated in the absence of conformal symmetry. We now expect light ray operators with spins $J=3+\de J$ to appear, where contributions proportional to $\de J$ come with additional factors of the $\b$-function. In addition, without conformal symmetry, the quantum numbers of light-ray operators are no longer simply related to quantum numbers of local operators via the rule $(\De,J)\to (1-J,1-\De)$. Instead, light-ray operators at null infinity carry so-called ``timelike" anomalous dimensions \cite{Basso:2006nk,Dixon:2019uzg}. 

While these issues are interesting to explore, here we will sidestep them by making predictions for ``conformal QCD." Specifically, we work in dimensional regularization $d=4-\e$, and tune the coupling constant $a$ to a conformal fixed-point $\beta(\e,a)=0$. The resulting predictions constrain the perturbative expansion of QCD away from the fixed point, up to terms proportional to the $\beta$-function. In fact, we expect that terms proportional to $\beta$ do not affect the central predictions of this section. The reason is that the $\b$-function corrections to the spin selection rule should only affect $R^{(n\geq 2)}_{\de,j}$ and $\g^{(n\geq 2)}_{\de,j}$, and the difference between spacelike and timelike anomalous dimensions should affect $\g^{(n\geq 2)}_{\de,j}$. When deriving the results in this section, we only use known values of $R^{(1)}_{4,2}$ and $\g^{(1)}_{4,2}$. Therefore, they should still be true in the usual 4d QCD.

Let us now give concrete predictions for $\cG^{(n+1)}(\z_{12},z,\bar z)$ in conformal QCD.  We define
\be\label{eq:tildeG_definition}
\tl{\cG}(z,\bar z)=\sum_{\de,j}R_{\de,j;\de'_{*}}g^{(\cE\cE\cE\cP_{\de'_{*}}=5)}_{\de,j}(z,\bar z),
\ee
so the collinear EEEC is 
\be\label{eq:cG_cGtilde_relation}
\cG(\z_{12},z,\bar z)=\z_{12}^{\frac{\de'_{*}-5}{2}}\tl{\cG}(z,\bar z).
\ee
We will first derive the leading behavior of $\tl{\cG}^{(n+1)}(z,\bar z)$ in three different kinematic limits, and then study the behavior of $\cG^{(n+1)}(\z_{12},z,\bar z)$. Our main predictions for $\cG^{(n+1)}(\z_{12},z,\bar z)$ are given by \eqref{eq:cG_allorderleadinglog_gluon}, \eqref{eq:cG_allorderleadinglog_quark}, \eqref{eq:cG_allorderOPE_gluon},  and \eqref{eq:cG_allordersmallz_gluon}.

\subsection{Predictions for $\tl{\cG}^{(n+1)}(z,\bar z)$}\label{sec:Gtilde_prediction}
To make predictions for $\tl{\cG}^{(n+1)}(z,\bar z)$, we must take a kinematic limit where the $\tau_c=2,j=2$ non-degenerate operator gives the leading contribution. The first limit we consider is the OPE limit/squeezed limit, where $z,\bar z\ll1$ with $z/\bar z$ fixed. We parametrize $z$ and $\bar z$ as $z=re^{i\th}, \bar z=re^{-i\th}$. Using \eqref{eq:tildeG_definition}, we can fix the leading log term of the $(n+1)$-th order expansion of $\tl{\cG}$. This is because $\log(r)$ can only come from a derivative of $g_{\de,j}$ with respect to $\de$, and each derivative introduces an anomalous dimension factor $\g^{(1)}$. Hence, the leading log term, which has the most powers of $\log(r)$, should take the form $\<R^{(1)}(\g^{(1)})^n\> \log^n(r) \ptl_{\de}^n g_{\de,j}$. The quantum numbers with the lowest value of $\de$ and nonzero $R^{(1)}$ are $(\de,j)=(4,0)$ and $(\de,j)=(4,2)$. Plugging these in, we obtain
\be
&\tl{\cG}^{(n+1)}(r \to 0,\th) \nn \\
&=\frac{r^4 \log^{n}(r)}{n!}\p{\<R_{4,0}^{(1)}(\g^{(1)}_{4,0})^n\> + 2 R^{(1)}_{4,2}(\g^{(1)}_{4,2})^n \cos(2\th)} + O(r^4 \log^{n-1}(r)).
\ee
As discussed in \ref{sec:lightrayOPE_prediction}, the light-ray OPE formula implies that $\g^{(1)}_{4,2}$ should be the anomalous dimension of the operator \eqref{eq:tauc2j2_nondegenerateop} evaluated at $J=3$. This has been calculated in e.g. \cite{Bukhvostov:1985rn,Chen:2021gdk}, and it is given by
\be
\g^{(1)}_{F\ptl^{J-2}F}(J)=4C_A S_1(J)-\b_0,
\ee
where $\b_0=\frac{11}{3}C_A-\frac{4}{3}n_F T_F$. At $J=3$, we then have
\be\label{eq:gamma_tauc2j2}
\g^{(1)}_{4,2}=\g^{(1)}_{F\ptl^{J-2}F}(3)=\frac{22}{3}C_A-\p{\frac{11}{3}C_A-\frac{4}{3}n_F T_F} = \frac{11}{3}C_A+\frac{4}{3}n_F T_F.
\ee
Thus, the leading log term of $\tl{\cG}^{(n+1)}$ in the OPE limit should be
\be\label{eq:Gtilde_OPE_leadingprediction}
\tl{\cG}^{(n+1)g/q}(r\to 0,\th)&= \<R_{4,0}^{(1)g/q}(\g^{(1)}_{4,0})^n\> \frac{r^4\log^{n}(r)}{n!}  \nn \\
& +2R^{(1)g/q}_{4,2}\p{\frac{11}{3}C_A+\frac{4}{3}n_F T_F}^n \frac{r^4\log^{n}(r)}{n!}\cos(2\th) \nn \\
& + O(r^4\log^{n-1}(r)),
\ee
where the superscript $g/q$ denotes gluon jet or quark jet, and $R^{(1)g/q}_{4,2}$ are given in \eqref{eq:Rcoeff_QCD_gluon_directdecomposition1} and \eqref{eq:Rcoeff_QCD_quark_directdecomposition1}.

We see that the coefficient of the leading term $r^{4}\log^{n}(r)\cos(2\th)$ in the OPE limit is completely fixed. In fact, since the next quantum number with nonzero $R^{(1)}$ starts at $\de=6$, we can also predict the term proportional to $r^{5}\log^{n}(r)\cos(3\th)$ since it should come from the descendant of the $g_{\de=4,j=2}$ block. The result is
\be
\tl{\cG}^{(n+1)g/q}(r\to 0,\th)&=\<R_{4,0}^{(1)g/q}(\g^{(1)}_{4,0})^n\> \frac{r^4\log^{n}(r)}{n!}  \nn \\
& +2R^{(1)g/q}_{4,2}\p{\frac{11}{3}C_A+\frac{4}{3}n_F T_F}^n \frac{r^4\log^{n}(r)}{n!}\cos(2\th) \nn \\
& + r^4(\cdots)\nn \\
& + \<R_{4,0}^{(1)g/q}(\g^{(1)}_{4,0})^n\> \frac{r^5\log^{n}(r)}{n!}\cos(\th) \nn \\
& +2R^{(1)g/q}_{4,2}\p{\frac{11}{3}C_A+\frac{4}{3}n_F T_F}^n \frac{r^5\log^{n}(r)}{n!}\cos(3\th) \nn \\
&+O(r^5\log^{n-1}(r)).
\ee

To study other interesting limits where the non-degenerate operator gives the leading contribution, we can go to Lorentzian signature on the celestial sphere (where $z$ and $\bar z$ are independent real variables) and consider the $z\ll 1$, fixed $\bar z$ limit. From the collider physics point of view, this limit might not be so useful since the kinematic region one can explore using collider experiments is intrinsically Euclidean (where $z$ and $\bar z$ are complex conjugates of each other). Nevertheless, we still find a nontrivial constraint on the analytic expression of $\tl{\cG}^{(n+1)}(z,\bar z)$. In the limit $z \ll 1$, we can use the $\SL(2,\R)$ expansion for the conformal block \eqref{eq:conformalblock_SL2Rexpansion}, and the leading term is from the operator with the leading celestial twist $\tau_c$, which is exactly the non-degenerate operator with $\tau_c=2,j=2$. Using this, we find that the leading log of $\tl{\cG}^{(n+1)}$ in the limit $z \ll 1$ is given by
\be\label{eq:Gtilde_smallz_expr1}
&\tl{\cG}^{(n+1)g/q}(z \ll 1, \bar z) \nn \\
&=R^{(1)g/q}_{4,2}(\g^{(1)}_{4,2})^n\frac{z\log^n(z)}{2^n n!} k^{0,-1}_{6}(\bar z) + O(z\log^{n-1}z) \nn \\
&=10R^{(1)g/q}_{4,2}\p{\frac{11}{3}C_A+\frac{4}{3}n_F T_F}^n\frac{z\log^n(z)}{2^n n!}\frac{(\bar z(12-12\bar z+\bar z^2)+6(2-3\bar z+ \bar z^2)\log(1-\bar z))}{\bar z^2} \nn \\
&\qquad + O(z\log^{n-1}z).
\ee
If we also write down the subleading $O(z\log^{n-1}z)$ order, we find
\be
&\tl{\cG}^{(n+1)g/q}(z \ll 1, \bar z) \nn \\
&=R^{(1)g/q}_{4,2}(\g^{(1)}_{4,2})^n\frac{z\log^n(z)}{2^n n!} k^{0,-1}_{6}(\bar z) \nn \\
&\quad +  \left(\sum_{j=2,4,6,\cdots}\<R^{(2)g/q}_{j+2,j}(\g^{(1)}_{j+2,j})^{n-1}\> k^{0,-1}_{2j+2}(\bar z) + R^{(1)g/q}_{4,2}(\g^{(1)}_{4,2})^n \ptl_{\de}k^{0,-1}_{\de+2}(\bar z)|_{\de\to 4} \right. \nn \\
&\qquad \left. + \<R^{(1)g/q}_{j+2,2}\g'_{*}\>(\g^{(1)}_{j+2,2})^{n-1} \ptl_{\de'_{*}}k_{6}^{0,\frac{3-\de'_{*}}{2}}(\bar z)|_{\de'_{*}\to 5} \right)\frac{z\log^{n-1}(z)}{2^{n-1} (n-1)!} + O(z\log^{n-2}z).
\ee
Note that for the subleading order coefficient $\<R^{(2)}\>$, we should expect to get contribution from operators with $\tau_c=2$ and all even $j$. The only term we know is $R^{(1)g/q}_{4,2}(\g^{(1)}_{4,2})^n \ptl_{\de}k^{0,-1}_{\de+2}(\bar z)|_{\de\to 4}$. We can actually isolate this term by further taking the small $\bar z$ limit, in which we find $k^{0,-1}_{2j+2}(\bar z) = \bar z^{j+1} + O(\bar z^{j+2})$, $\ptl_{\de}k^{0,-1}_{\de+2}(\bar z)|_{\de\to 4} = \frac{1}{2}\bar z^3 \log \bar z + O(\bar z^4)$, and $\ptl_{\de'_{*}}k_{6}^{0,\frac{3-\de'_{*}}{2}}(\bar z)|_{\de'_{*}\to 5} = O(\bar z^4)$. Therefore, in the $z \ll \bar z \ll 1$ limit, we can also predict the coefficient of the  $z\log^{n-1}z \bar z^3\log\bar z$ term. It is given by
\be\label{eq:Gtilde_smallz_smallzbar}
&\tl{\cG}^{(n+1)g/q}(z \ll \bar z \ll 1) \nn \\
&=10R^{(1)g/q}_{4,2}\p{\frac{11}{3}C_A+\frac{4}{3}n_F T_F}^n\frac{z\log^n(z)}{2^n n!}\frac{(\bar z(12-12\bar z+\bar z^2)+6(2-3\bar z+ \bar z^2)\log(1-\bar z))}{\bar z^2}   \nn \\
&\quad + \frac{1}{2}R^{(1)g/q}_{4,2}\p{\frac{11}{3}C_A+\frac{4}{3}n_F T_F}^n\frac{z\log^{n-1}(z)}{2^{n-1} (n-1)!}\p{\bar z^3\log \bar z + O(\bar z^3)} \nn \\
&\quad + O(z\log^{n-2}z).
\ee

It is also interesting to study the leading behavior of $\tl{\cG}^{(n+1)}(z,\bar z)$ in the double lightcone limit $z\ll 1-\bar z\ll 1$. From \eqref{eq:Gtilde_smallz_expr1}, we obtain
\be\label{eq:Gtilde_doublelightcone}
&\tl{\cG}^{(n+1)g/q}(z \ll 1-\bar z \ll 1) \nn \\
&=R^{(1)g/q}_{4,2}(\g^{(1)}_{4,2})^n\frac{z\log^n(z)}{2^n n!}(10 + O(1-\bar z))+ O(z\log^{n-1}z),
\ee
and one can try to study how this term can be created using the crossing equation \eqref{eq:collinearEEEC_crossing_leading} and the lightcone bootstrap \cite{Simmons-Duffin:2016wlq}. One will find that they come from the $R^{(n+1)}$ coefficients with $\tau_c=6$ and large-$j$ (which corresponds to double-trace operators with conventional twist $\tau=8$ and $j=4$). From the crossing equation, we can also predict the large-$j$ behavior of $\<R^{(n+1)}_{j+6,j}\>$. For example, we find that at large-$j$,
\be\label{eq:Rtwistc6_largej}
\<R^{(1)g/q}_{j+6,j}\> \sim \frac{5\sqrt{\pi}}{8}R^{(1)g/q}_{4,2} 4^{-j}j^{\frac{7}{2}}.
\ee
This result can be generalized to $\<R^{(n+1)g/q}_{j+6,j}\>$ at any order. We describe the result and the details of the calculation in appendix \ref{app:lightcone_bootstrap_largej}. 

\subsection{Degeneracies of $\cO'_{*}$}
\label{sec:degeneracies}

So far in our analysis, we have assumed there is a unique isolated spin-4 operator $\cO'_*$ that controls the collinear limit. However, it can happen that the leading-twist spin-4 operator is degenerate at tree-level, and thus we must take into account the contribution of multiple $\cO'_*$'s in perturbation theory.

As discussed in section \ref{sec:lightrayOPE_prediction}, $\De'_{*}=\de'_{*}+1$ should be the scaling dimension of the leading twist, spin-$4$ operator $\cO'_{*}$ in the $T\x T\x T$ OPE. Also, since the sink/source states we consider are rotationally-invariant, $\cO'_{*}$ should have zero transverse spin. There are only two such Regge trajectories in QCD:\footnote{We are using mostly positive metric, and our convention for indices symmetrization is $T^{\{\mu_1\cdots\mu_n\}}=\frac{1}{n!}\sum_{P\in S_n}T^{\mu_{P(1)}\cdots\mu_{P(n)}}$}
\be
\cO_{q}(J)=&\frac{(-1)^{J}}{2^J}\sum_{k=1}^{J}(-1)^{k-1}\a^{q}_k(J)\bar \psi \g^{\{\mu_1}(i\overleftarrow{D}^{\mu_2})\cdots(i\overleftarrow{D}^{\mu_k})(iD^{\mu_{k+1}})\cdots(iD^{\mu_J\}})\psi - \mathrm{traces}, \nn \\ 
\cO_{g}(J)=&\frac{(-1)^{J}}{2^J}\sum_{k=1}^{J-1}(-1)^{k-1}\a^{g}_k(J)F_{a\nu}{}^{\{\mu_1}(i\overleftarrow{D}^{\mu_2})\cdots(i\overleftarrow{D}^{\mu_k})(iD^{\mu_{k+1}})\cdots(iD^{\mu_{J-1}})F_a^{\nu \mu_J\}} - \mathrm{traces}.
\ee
where $\overleftarrow{D}^{\mu}$ (acting on the left) and $D^{\mu}$ (acting on the right) are covariant derivatives. The coefficients $\a^{q/g}_{k}(J)$ are chosen such that the entire expression of $\cO_{q/g}(J)$ is a conformal primary, and they satisfy the normalization condition $\sum_{k=1}^{J}\a^{q}_k(J)=\sum_{k=1}^{J-1}\a^{g}_k(J)=1$. For example, for $J=2$ we have
\be
\cO_q(2)=&\frac{1}{8}\p{\bar \psi\g^{\{\mu_1}iD^{\mu_2\}}\psi-\bar \psi\g^{\mu_1\}}i\overleftarrow{D}^{\{\mu_2}\psi} - \mathrm{traces}, \nn \\
\cO_g(2)=&\frac{1}{4}F_{a\nu}{}^{\mu_1}F^{a\nu\mu_2} - \mathrm{traces}.
\ee
One can see that the sum $\cO_q(2)+\cO_g(2)$ is proportional to the stress-energy tensor.

 For general $J$, there are many different methods to determine the coefficients $\a_{k}^{q/g}(J)$ that make $\cO_{q/g}(J)$ a conformal primary \cite{Korchemsky:2021htm, Braun:2003rp, Makeenko:1980bh, Ohrndorf:1981qv}. Here, we are only interested in the $J=4$ case. So we take a simple approach: apply the generator $K_{\mu}$ of the special conformal transformation on $\cO_{q/g}(4)$ and demand that its action vanishes. Using the basic commutation relation $[K_{\mu},P_{\nu}]=2\eta_{\mu\nu}D-2M_{\mu\nu}$, we find
\be\label{eq:OqJ4_definition}
\cO_q(4)=\frac{1}{224}&\left( \bar \psi\g^{\{\mu_1}(iD^{\mu_2})(iD^{\mu_3})(iD^{\mu_4\}})\psi-6(iD^{\{\mu_2})\bar \psi\g^{\mu_1}(iD^{\mu_3})(iD^{\mu_4\}})\psi\right. \nn \\
& \left.+6(iD^{\{\mu_2})(iD^{\mu_3})\bar \psi\g^{\mu_1}(iD^{\mu_4\}})\psi - (iD^{\{\mu_2})(iD^{\mu_3})(iD^{\mu_4})\bar \psi\g^{\mu_1\}}\psi\right) - \mathrm{traces},
\ee
and
\be\label{eq:OgJ4_definition}
\cO_g(4)=\frac{3}{112}\left(F_{a\nu}{}^{\{\mu_1}(iD^{\mu_2})(iD^{\mu_{3}})F_a^{\nu \mu_4\}}-\frac{4}{3}(iD^{\{\mu_2})F_{a\nu}{}^{\mu_1}(iD^{\mu_{3}})F_a^{\nu \mu_4\}}\right) - \mathrm{traces}.
\ee

The one-loop dilatation matrix for $\cO_q(4)$ and $\cO_g(4)$ is given by \cite{Gross:1973id,Gross:1973ju,Gross:1974cs}
\be
\hat \g=\frac{\a_s}{4\pi}\begin{pmatrix}\frac{157 C_F}{30} && -\frac{11 n_F T_F}{15} \\ -\frac{11 C_F}{15} && \frac{21 C_A}{5}+\frac{4 n_F T_F}{3}\end{pmatrix}.
\ee 
Its eigenvalues are
\be
\frac{21}{10}C_A + \frac{157}{60}C_F +\frac{2}{3}n_F T_F \mp \frac{1}{60}\sqrt{(126C_A-157C_F)^2+32(315C_A-332 C_F)n_F T_F+1600n_F^2 T_F^2},
\ee
and its left eigenvectors are
\be
\begin{pmatrix} \a_{\pm} && 1 \end{pmatrix},
\ee
where
\be\label{eq:alphapm_def}
\a_{\pm}=\frac{126C_A -157C_F + 40n_F T_F \pm \sqrt{(126C_A-157C_F)^2 + 32(315C_A-332 C_F)n_F T_F +1600n_F^2 T_F^2 }}{44 n_F T_F}.
\ee

To resolve the degeneracy at 1-loop, we can define the following two operators:
\be\label{eq:degenerateoperator_diagonalbasis}
\cO'_{*1}&=\a_{+}\cO_{q}(4)+ \cO_{g}(4), \nn \\
\cO'_{*2}&=\cO_{q}(4)+\frac{1}{\a_{-}} \cO_{g}(4).
\ee
In the $\cO'_{*1}, \cO'_{*2}$ basis, the anomalous dimension matrix is then diagonal. The first operator has anomalous dimension (suppressing the $\frac{\a_s}{4\pi}$ factor)
\be\label{eq:gammaprimestar_1}
\g'_{*1}\equiv \frac{21}{10}C_A + \frac{157}{60}C_F +\frac{2}{3}n_F T_F - \frac{1}{60}\sqrt{(126C_A-157C_F)^2+32(315C_A-332 C_F)n_F T_F+1600n_F^2 T_F^2}.
\ee
and $\g_{*2}'$ is given by replacing $-\sqrt{\cdot}\to \sqrt{\cdot}$.

Taking into account the degeneracies of $\cO'_{*}$, the decomposition \eqref{eq:Gfunc_decomposition} for $\cG(\z_{12},z,\bar z)$ should become
\be\label{eq:Gfunc_decomposition_degeneracy}
\cG(\z_{12}, z,\bar z)=&\frac{\pi^2}{16}\sum_{i=1,2}\frac{(-p^2)^{\frac{2\De_{\cO}-4}{2}}}{\s^{\cO}_{\mathrm{total}}}s^{\prime(i)}_{\cO\cO\cP_{\de'_{*}}}\z_{12}^{\frac{\de'_{*}-5}{2}}\sum_{\de,j}r_{\cE\cE\cP_{\de,j}}r^{\prime(i)}_{\cP_{\de,j}\cE\cP_{\de'_{*}}}g_{\de,j}^{(\cE\cE\cE\cP_{\de'_{*}})}(z,\bar{z}) \nn \\
=&\sum_{i=1,2}c^{\cO}_{i}\z_{12}^{\frac{\de'_{*}-5}{2}}\sum_{\de,j}\tl{R}_{\de,j;\de'_{*};i}g_{\de,j}^{(\cE\cE\cE\cP_{\de'_{*}})}(z,\bar{z}),
\ee
where in the first line we have used \eqref{eq:crosssection_O} to restore the total cross section in order to emphasize the dependence on the sink/source states, and in the second line we have defined
\be
c^{\cO}_{i}\equiv \frac{(-p^2)^{\frac{2\De_{\cO}-4}{2}}}{\s^{\cO}_{\mathrm{total}}}s^{\prime(i)}_{\cO\cO\cP_{\de'_{*}}} ,\qquad \tl{R}_{\de,j;\de'_{*};i} \equiv \frac{\pi^2}{16}r_{\cE\cE\cP_{\de,j}}r^{\prime(i)}_{\cP_{\de,j}\cE\cP_{\de'_{*}}}.
\ee
Note that unlike $R_{\de,j;\de'_{*}}$, the newly defined coefficient $\tl{R}_{\de,j;\de'_{*};i}$ is independent of the operator $\cO$, which creates the sink/source states. The only dependence on $\cO$ is in the coefficient $c^{\cO}_{i}$. Using \eqref{eq:scoeff_defiition}, one can show that $c^{\cO}_{i}$ can be determined using
\be
\frac{\<\cO(p)|\wL[\cO'_{*;i}](\oo,z)|\cO(p)\>}{\<\cO(p)|\cO(p)\>}=c^{\cO}_{i}(-2p\.z)^{-\de'_{*}}(-p^2)^{\frac{\de'_{*}+3}{2}},
\ee
where we have replaced the object $\mathbb{W}_{\de'}$ with $\wL[\cO'_{*;i}]$\footnote{The coefficient relating $\mathbb{W}_{\de'}$ and $\wL[\cO'_{*;i}]$ can be absorbed into $r^{\prime(i)}_{\cP_{\de,j}\cE\cP_{\de'_{*}}}$.} based on the assumptions made in section \ref{sec:lightrayOPE_prediction}. For us, the two degenerate operators can be chosen to be $\cO'_{*1}$ and $\cO'_{*2}$ defined in \eqref{eq:degenerateoperator_diagonalbasis}. The two different $\cO$'s are $\Tr F^2$ corresponding to the gluon jet and $J^{\mu}$ corresponding to the quark jet (averaged over polarization), so we will simply use $c^{g/q}_{i}$ for two cases.

 We are interested in $c^{(0)g/q}_{i}$ at the leading order in the coupling constant. To find $c^{(0)g/q}_{i}$, we first determine $\<\cO(x_1)\cO_{g/q}(4)(x_2,z_2)\cO(x_3)\>$ in position space using Wick contractions, where $\cO$ is either $\Tr F^2$ or $J^{\mu}$. After doing the light transform and Fourier transform, we then obtain that at the leading order
\be
\frac{\<\Tr F^2(p)|\wL[\cO_g(4)](\oo,z)|\Tr F^2(p)\>}{\<\Tr F^2(p)|\Tr F^2(p)\>}=\frac{1}{64\pi}(-2p\.z)^{-5} (-p^2)^4 + O(\a_s), \nn \\
\frac{\<J_{\mu}(p)|\wL[\cO_q(4)](\oo,z)|J^{\mu}(p)\>}{\<J_{\mu}(p)|J^{\mu}(p)\>}=\frac{1}{64\pi}(-2p\.z)^{-5} (-p^2)^4 + O(\a_s),
\ee
which implies that
\be\label{eq:c0coeff_solution}
c^{(0)g}_1=&\frac{1}{64\pi}, \qquad c^{(0)g}_2=\frac{1}{\a_{-}}\frac{1}{64\pi},\qquad c^{(0)q}_1=\a_{+}\frac{1}{64\pi}, \qquad c^{(0)q}_2=\frac{1}{64\pi}.
\ee
Therefore, the coefficients appearing in \eqref{eq:cG1_conformalblockdecomposition} can be rewritten as (we focus on $\<R^{(1)}_{\de,j;\de'_*}\>$)
\be\label{eq:R1gq_degeneracy}
\<R^{(1)g}_{\de,j;\de'_*}\>&=\frac{1}{64\pi}\p{\<\tl{R}^{(1)}_{\de,j;\de'_*;1}\>+\frac{1}{\a_{-}}\<\tl{R}^{(1)}_{\de,j;\de'_*;2}\>}, \nn \\
\<R^{(1)q}_{\de,j;\de'_*}\>&=\frac{1}{64\pi}\p{\a_{+}\<\tl{R}^{(1)}_{\de,j;\de'_*;1}\>+\<\tl{R}^{(1)}_{\de,j;\de'_*;2}\>}.
\ee
Solving \eqref{eq:R1gq_degeneracy} for $\<\tl{R}^{(1)}_{\de,j;\de'_*;1/2}\>$, we find
\be\label{eq:R1tilde_solution}
\<\tl{R}^{(1)}_{\de,j;\de'_*;1}\>=&\frac{64\pi}{\a_{+}-\a_{-}}\p{\<R^{(1)q}_{\de,j;\de'_*}\>-\a_{-}\<R^{(1)g}_{\de,j;\de'_*}\>}, \nn \\
\<\tl{R}^{(1)}_{\de,j;\de'_*;2}\>=&\frac{64\pi \a_{-}}{\a_{+}-\a_{-}}\p{\a_{+}\<R^{(1)g}_{\de,j;\de'_*}\>-\<R^{(1)q}_{\de,j;\de'_*}\>}.
\ee

\subsection{Predictions for $\cG^{(n+1)}(\z_{12},z,\bar z)$}\label{eq:cG_higheroder_prediction}
We now explain how to use \eqref{eq:Gfunc_decomposition_degeneracy} and \eqref{eq:R1tilde_solution} to deal with the degeneracy of $\cO'_{*}$ and make predictions for $\cG^{(n+1)}(\z_{12},z,\bar z)$. If we expand \eqref{eq:cG_cGtilde_relation} in small coupling assuming there are no degeneracies, we will get
\be\label{eq:Ghigherorderexpansion_nodegeneracy}
&\cG^{(n+1)}(\z_{12},z, \bar z)\nn \\
&=\sum_{k=0}^{n} \frac{\log^{n-k}\z_{12} }{2^{n-k}(n-k)!}\p{\sum_{p=0}^{k}\sum_{\substack{i_1+i_2+\cdots i_{p+1}=n-k \\ i_1+2i_2+\cdots + (p+1)i_{p+1}=n-k+p}}\frac{(n-k)!}{i_1! \cdots i_{p+1}!}\p{\g^{\prime(1)}_{*}}^{i_1}\cdots\p{\g^{\prime(p+1)}_{*}}^{i_{p+1}}\tl{\cG}^{(k+1-p)}}.
\ee
More explicitly, the first few terms are given by
\be
&\cG^{(n+1)}(\z_{12},z, \bar z)\nn \\
&=\frac{\log^{n}\z_{12}}{2^{n}n!}\p{\g^{\prime(1)}_{*}}^{n}\tl{\cG}^{(1)}(z,\bar z) \nn \\
&+ \frac{\log^{n-1}\z_{12}}{2^{n-1}(n-1)!}\p{\p{\g^{\prime(1)}_{*}}^{n-1}\tl{\cG}^{(2)}(z,\bar z) +(n-1) \p{\g^{\prime(1)}_{*}}^{n-2}\g^{\prime(2)}_{*}\tl{\cG}^{(1)}(z,\bar z)} \nn \\
&+ \frac{\log^{n-2}\z_{12}}{2^{n-2}(n-2)!}\left(\p{\g^{\prime(1)}_{*}}^{n-2}\tl{\cG}^{(3)}(z,\bar z) +(n-2) \p{\g^{\prime(1)}_{*}}^{n-3}\g^{\prime(2)}_{*}\tl{\cG}^{(2)}(z,\bar z) \right. \nn \\
&\qquad \qquad \qquad \quad\left. +\frac{(n-2)(n-3)}{2} \p{\g^{\prime(1)}_{*}}^{n-4}\p{\g^{\prime(2)}_{*}}^2\tl{\cG}^{(1)}(z,\bar z) + (n-2)\p{\g^{\prime(1)}_{*}}^{n-3}\g^{\prime(3)}_{*}\tl{\cG}^{(1)}(z,\bar z)  \right) \nn \\
&+ \cdots.
\ee
The above expression will be modified in the presence of degeneracies. In general, if we know the anomalous dimension $\g^{\prime(k)}_{*}$ to the $k$-th order, we will be able to rewrite the above expression up to the $\log^{n-k+1}\z_{12}$ term. In the previous section, we have only diagonalized the $\g^{\prime(1)}_{*}$ matrix, which then allows us to rewrite all the $(\g^{\prime(1)}_{*})^{k}$ terms. 

First, let us focus on the leading logarithmic divergence $\log^n(\z_{12})$. Using \eqref{eq:Gfunc_decomposition_degeneracy}, we find that in the presence of degeneracies, the leading log term of \eqref{eq:Ghigherorderexpansion_nodegeneracy} should become
\be
&\cG^{(n+1)}(\z_{12},z, \bar z) \nn \\
&=\frac{\log^{n}\z_{12}}{2^{n}n!}\p{c_1^{(0)\cO}\p{\g^{\prime}_{*1}}^{n}\sum_{\de,j}\<\tl{R}^{(1)}_{\de,j;\de'_*;1}\>g^{(\cE\cE\cE\cP_{\de'_{*}=5})}_{\de,j}(z,\bar z)+c_2^{(0)\cO}\p{\g^{\prime}_{*2}}^{n}\sum_{\de,j}\<\tl{R}^{(1)}_{\de,j;\de'_*;2}\>g^{(\cE\cE\cE\cP_{\de'_{*}=5})}_{\de,j}(z,\bar z)} \nn \\
&\quad+O(\log^{n-1}\z_{12})\,,
\ee
where $\g'_{*1}$ and $\g'_{*2}$ are given by \eqref{eq:gammaprimestar_1}. Plugging in \eqref{eq:R1tilde_solution}, we obtain for the gluon and quark jets
\be\label{eq:cG_allorderleadinglog_gluon}
\cG^{(n+1)g}(\z_{12},z, \bar z) 
&=\frac{\log^{n}\z_{12}}{2^{n}n!}\p{\frac{\a_{+}\p{\g^{\prime}_{*2}}^{n}-\a_{-}\p{\g^{\prime}_{*1}}^{n}}{\a_{+}-\a_{-}}\cG^{(1)g} + \frac{\p{\g^{\prime}_{*1}}^{n}-\p{\g^{\prime}_{*2}}^{n}}{\a_{+}-\a_{-}}\cG^{(1)q}}\nn \\
&\quad+O(\log^{n-1}\z_{12})\,,
\\
\label{eq:cG_allorderleadinglog_quark}
\cG^{(n+1)q}(\z_{12},z, \bar z) 
&=\frac{\log^{n}\z_{12}}{2^{n}n!}\p{\frac{\a_{+}\p{\g^{\prime}_{*1}}^{n}-\a_{-}\p{\g^{\prime}_{*2}}^{n}}{\a_{+}-\a_{-}}\cG^{(1)q} + \frac{\a_{+}\a_{-}\p{\p{\g^{\prime}_{*2}}^{n}-\p{\g^{\prime}_{*1}}^{n}}}{\a_{+}-\a_{-}}\cG^{(1)g}}\nn \\
&\quad+O(\log^{n-1}\z_{12})\,,
\ee
where $\cG^{(1)g/q}$ are simply the leading order results for the gluon/quark jet, related to the known result of \cite{Chen:2019bpb} by \eqref{eq:Gfunc_relation}. Equations \eqref{eq:cG_allorderleadinglog_gluon} and \eqref{eq:cG_allorderleadinglog_quark} show that the leading logarithmic divergence of the $(n+1)$-th order EEEC $\cG^{(n+1)}(\z_{12},z, \bar z)$ is $\log^{n}(\z_{12})$, and it is completely determined by the leading order result $\cG^{(1)g}$ and $\cG^{(1)q}$.

In fact, since we can rewrite all the $(\g^{\prime(1)}_{*})^{k}$ terms in \eqref{eq:Ghigherorderexpansion_nodegeneracy}, we can make further predictions for $\cG^{(n+1)}(\z_{12},z, \bar z)$ if we know how to separate out the contributions from $(\g^{\prime(1)}_{*})^{k}$. This can be done by taking the limits considered in section \ref{sec:Gtilde_prediction}. For example, in the OPE limit we find
\be
&\cG^{(n+1)g/q}(\z_{12}, r\to 0,\th)\nn \\
&=\sum_{k=0}^{n}\frac{\log^{n-k}\z_{12}}{2^{n-k} (n-k)!}\Big(c_{1}^{(0)g/q}(\g'_{*1})^{n-k} \tl{\cG}_{1}^{(k+1)}(r \to 0,\th) + c_{2}^{(0)g/q}(\g'_{*2})^{n-k} \tl{\cG}_{2}^{(k+1)}(r \to 0,\th)\nn\\
&\qquad \qquad \qquad \qquad \qquad + O(r^4\log^{k-1}r)\Big).
\ee
From \eqref{eq:Gtilde_OPE_leadingprediction} we know that $\tl{\cG}^{(k+1)}\sim r^4\log^kr$ in the OPE limit. Thus, at each $\log^{n-k}\z_{12}$ power, all the terms involving higher-loop anomalous dimensions ($\g^{\prime(p)}_{*}$ with $p>1$) in \eqref{eq:Ghigherorderexpansion_nodegeneracy} go as at most $r^4\log^{k-1}r$, while the term involving $(\g^{\prime(1)}_{*})^{k}$ has the most divergent piece $r^4\log^{k}r$. Therefore, we can determine the leading behavior in small $r$ for each $\log^{n-k}\z_{12}$ power. The functions $\tl{\cG}^{(k+1)}_1$ and $\tl{\cG}^{(k+1)}_2$ only include the contribution from $\cO'_{*1}$ and $\cO'_{*2}$ respectively. They are defined as
\be
\tl{\cG}^{(k+1)}_1=&\frac{64\pi}{\a_{+}-\a_{-}}\p{\tl{\cG}^{(k+1)q}-\a_{-}\tl{\cG}^{(k+1)g}}, \nn \\
\tl{\cG}^{(k+1)}_2=&\frac{64\pi \a_{-}}{\a_{+}-\a_{-}}\p{\a_{+}\tl{\cG}^{(k+1)g}-\tl{\cG}^{(k+1)q}}.
\ee
Using \eqref{eq:Gtilde_OPE_leadingprediction}, we then obtain, for example,
\be\label{eq:cG_allorderOPE_gluon}
&\cG^{(n+1)g}(\z_{12},r \to 0, \th) \nn \\
&=\sum_{k=0}^{n}\frac{\log^{n-k}\z_{12}}{2^{n-k-1} k!(n-k)!}\x \nn \\
&\Bigg(\p{\g^{(1)}_{4,2}}^k\p{\frac{\a_{+}\p{\g^{\prime}_{*2}}^{n-k}-\a_{-}\p{\g^{\prime}_{*1}}^{n-k}}{\a_{+}-\a_{-}}R^{(1)g}_{4,2} + \frac{\p{\g^{\prime}_{*1}}^{n-k}-\p{\g^{\prime}_{*2}}^{n-k}}{\a_{+}-\a_{-}}R^{(1)q}_{4,2}}r^4\log^kr\cos(2\th) \nn \\
& + (\cdots) r^4\log^kr + O(r^4\log^{k-1}r)\Bigg),
\ee
where $\g^{(1)}_{4,2}$ and $R^{(1)g/q}_{4,2}$ are given in \eqref{eq:gamma_tauc2j2}, \eqref{eq:Rcoeff_QCD_gluon_directdecomposition1} and \eqref{eq:Rcoeff_QCD_quark_directdecomposition1} respectively. Thus, we have a prediction for the spin-$2$ part of the leading term of $\cG^{(n+1)}$ in the OPE limit, at each logarithmic order in $\log\z_{12}$.

Similarly, $(\g^{\prime(1)}_{*})^{k}$ terms in \eqref{eq:Ghigherorderexpansion_nodegeneracy} also give the dominant contribution in the $z\ll 1$ limit. Following the same calculation as that of the OPE limit, we obtain (using \eqref{eq:Gtilde_smallz_expr1})
\be\label{eq:cG_allordersmallz_gluon}
&\cG^{(n+1)g}(\z_{12},z\ll 1,\bar z) \nn \\
&=\sum_{k=0}^{n}\frac{\log^{n-k}\z_{12}}{2^{n}k!(n-k)!}\x \nn \\
&\Bigg(\p{\g^{(1)}_{4,2}}^k\p{\frac{\a_{+}\p{\g^{\prime}_{*2}}^{n-k}-\a_{-}\p{\g^{\prime}_{*1}}^{n-k}}{\a_{+}-\a_{-}}R^{(1)g}_{4,2} + \frac{\p{\g^{\prime}_{*1}}^{n-k}-\p{\g^{\prime}_{*2}}^{n-k}}{\a_{+}-\a_{-}}R^{(1)q}_{4,2}}z\log^{k}(z) k^{0,-1}_{6}(\bar z) \nn \\
& + O(z\log^{k-1}(z))\Bigg),
\ee
and similarly for $\cG^{(n+1)q}$.
It is also straightforward to repeat the analysis in the $z \ll \bar z\ll1$ limit and predict the $z\log^{k-1}(z)\bar z^3\log(\bar z)$ term at each $\log\z_{12}$ order using \eqref{eq:Gtilde_smallz_smallzbar}.

\section{Contact terms and Ward identities}\label{sec:Wardidentities_contactterm}

In this section, we study Ward identities satisfied by the EEEC, and also compute the EEEC at $O(g^0)$ and $O(g^2)$ order. Note that conventionally $O(g^4)$ is called the ``leading order" (LO) since it is the lowest order at which the EEEC is nonzero for generic detector positions. However, contributions at $O(g^0)$ and $O(g^2)$ also exist: they are proportional to delta functions, which we call ``contact terms", and so only become nonzero in special configurations. It was shown in \cite{Dixon:2019uzg,Kologlu:2019mfz,Korchemsky:2019nzm} that Ward identities can be used to determine the contact terms in the two-point EEC. Here, we perform a similar analysis for the EEEC, and use perturbation theory and Ward identities to obtain the EEEC at $O(g^0)$ and $O(g^2)$ order.

\subsection{$\EEEC'$ and Ward identities}

To study Ward identities, it is more convenient to write the EEEC as a function of the explicit positions on the celestial sphere $\vec n_i$, instead of the cross-ratios $\zeta_{ij}$. Thus, we define
\be
\label{eq:eeecnparam}
\EEEC'(\vec n_1,\vec n_2,\vec n_3)&=\frac{\int d^dx~e^{ip\.x}\<0|\cO^\dag(x)\cE(\vec{n}_1)\cE(\vec{n}_2)\cE(\vec{n}_3)\cO(0)|0\>}{(-p^2)^{\frac{3}{2}}\int d^dx~e^{ip\.x}\<0|\cO^\dag(x)\cO(0)|0\>} 
\nn\\
&=\sum_{i,j,k}\int d\s \frac{E_i E_j E_k}{Q^3} \de\p{\vec n_1, \frac{\vec p_i}{E_i}}\de\p{\vec n_2, \frac{\vec p_j}{E_j}}\de\p{\vec n_3, \frac{\vec p_k}{E_k}},
\ee
where the spherical delta function $\de(\vec n_1,\vec n_2)$ is defined by
\be
\int d\O_{\vec n_2}\de\p{\vec n_1,\vec n_2}f(\vec n_2)=f(\vec n_1).
\ee
The first line in (\ref{eq:eeecnparam}) is a nonperturbative definition, while the second line is suitable for perturbation theory. As before, $d\s$ represents an integration over phase space weighted by the scattering cross section, and $(E_i,\vec p_i)$ are energy and momentum of outgoing particles. The parametrization (\ref{eq:eeecnparam}) of the EEEC is related to (\ref{eq:EEEC_def}) by
\be
\EEEC'(\vec n_1,\vec n_2,\vec n_3)=\frac{\sin\th_1\sin\th_2|\sin \f|}{64\pi^2}\EEEC(\z_{12},\z_{13},\z_{23}),
\ee\label{eq:EEECprime_EEECcomparison}
where
\be\label{eq:nvectors_frame}
\vec n_1=(\sin\th_1,0,\cos\th_1),\quad \vec n_2=(\sin\th_2\cos\f,\sin\th_2\sin\f,\cos\th_2),\quad \vec n_3=(0,0,1).
\ee

To derive Ward identities, we simply integrate \eqref{eq:eeecnparam} over $\vec n_1,\vec n_2,\vec n_3$ and use energy and momentum conservation. For example,
\be
\int d\O_{\vec n_1}d\O_{\vec n_2}d\O_{\vec n_3}\EEEC'(\vec n_1,\vec n_2,\vec n_3) =\sum_{i,j,k}\int d\s \frac{E_i E_j E_k}{Q^3}  =1,
\ee
where we have used energy conservation $\sum_i E_i=Q$. Similarly, we also have
\be
\int d\O_{\vec n_1}d\O_{\vec n_2}d\O_{\vec n_3} (1-\vec n_1\.\vec n_2)\EEEC'(\vec n_1,\vec n_2,\vec n_3) =\sum_{i,j,k}\int d\s \frac{(E_i E_j -\vec p_i\.\vec p_j)E_k}{Q^3}=1,
\ee
where we have used momentum conservation $\sum_{i} \vec p_i=0$. In this way, we can derive the following Ward identities:
\be\label{eq:EEECp_Wardidentities}
&\int d\O_{\vec n_1}d\O_{\vec n_2}d\O_{\vec n_3}\EEEC'(\vec n_1,\vec n_2,\vec n_3)=1, \nn \\
&\int d\O_{\vec n_1}d\O_{\vec n_2}d\O_{\vec n_3}(1-\vec n_1\.\vec n_3)\EEEC'(\vec n_1,\vec n_2,\vec n_3)=1, \nn \\
&\int d\O_{\vec n_1}d\O_{\vec n_2}d\O_{\vec n_3}(1-\vec n_2\.\vec n_3)\EEEC'(\vec n_1,\vec n_2,\vec n_3)=1, \nn \\
&\int d\O_{\vec n_1}d\O_{\vec n_2}d\O_{\vec n_3}(1-\vec n_1\.\vec n_2)\EEEC'(\vec n_1,\vec n_2,\vec n_3)=1, \nn \\
&\int d\O_{\vec n_1}d\O_{\vec n_2}d\O_{\vec n_3}(1-\vec n_1\.\vec n_3)(1-\vec n_2\.\vec n_3)\EEEC'(\vec n_1,\vec n_2,\vec n_3)=1, \nn \\
&\int d\O_{\vec n_1}d\O_{\vec n_2}d\O_{\vec n_3}(1-\vec n_1\.\vec n_3)(1-\vec n_1\.\vec n_2)\EEEC'(\vec n_1,\vec n_2,\vec n_3)=1, \nn \\
&\int d\O_{\vec n_1}d\O_{\vec n_2}d\O_{\vec n_3}(1-\vec n_2\.\vec n_3)(1-\vec n_1\.\vec n_2)\EEEC'(\vec n_1,\vec n_2,\vec n_3)=1.
\ee

Using the Ward identities, we immediately see that the $O(g^0)$ $\EEEC'$ must be given by
\be\label{eq:EEECprime_g0order}
\EEEC'(\vec n_1,\vec n_2,\vec n_3)=\frac{1}{16\pi}\bigg(&\de(\vec n_1,\vec n_2)\de(\vec n_1,\vec n_3)+\de(\vec n_1,\vec n_2)\de(\vec n_1,-\vec n_3) \nn \\
&+\de(\vec n_1,\vec n_3)\de(\vec n_3,-\vec n_2)+\de(\vec n_2,\vec n_3)\de(\vec n_2,-\vec n_1)\bigg).
\ee
The structure of (\ref{eq:EEECprime_g0order}) is easy to understand. The first term is supported when all three detectors are coincident. The other three terms appear when two of the detectors are coincident and the other is diametrically opposite on the celestial sphere. Physically, the $O(g^0)$ EEEC gets contributions only from two particle states. By energy and momentum conservation, these particles must fly in opposite directions, and thus can only be observed by detectors that are either coincident (observing the same particle) or diametrically opposite (observing the two different particles).

\subsection{Tree-level ($O(g^2)$) $\EEEC'$ in $\cN=4$ SYM}
Let us now consider the EEEC at $O(g^2)$. We work in $\cN=4$ SYM and consider the case where the sink/source states are created by the operator $\Tr F^2$. At this order, there are at most three particles in the outgoing state. Therefore, if the detectors are at different positions, the three vectors $\vec n_1, \vec n_2, \vec n_3$ must be coplanar in order for the total momentum to be zero. It follows that the $O(g^2)$ EEEC must take the form:
\be\label{eq:EEECprime_g2_exprgeneral}
&\EEEC'(\vec n_1,\vec n_2,\vec n_3)|_{O(g^2)}\nn\\
&=\cF_0(\vec n_1,\vec n_2,\vec n_3)\de\p{(\vec n_1\x \vec n_2)\.\vec n_3} \nn \\ 
& \quad+ \cF_1(\vec n_1,\vec n_2)\de\p{\vec n_1, \vec n_3} + \cF_1(\vec n_1,\vec n_3)\de\p{\vec n_2, \vec n_3} + \cF_1(\vec n_2,\vec n_3)\de\p{\vec n_1, \vec n_2} \nn \\
& \quad+ 2\pi c_1 \de(\vec n_1, \vec n_2)\de(\vec n_1,\vec n_3) \nn \\
& \quad+ 2\pi c_2[\de(\vec n_1,\vec n_2)\de(\vec n_1,-\vec n_3) +\de(\vec n_1,\vec n_3)\de(\vec n_3,-\vec n_2)+\de(\vec n_2,\vec n_3)\de(\vec n_2,-\vec n_1)].
\ee
On the right-hand side, the first line describes the configuration where the detectors are coplanar. The second line are the contact terms that appear when two of the detectors are at the same position and the third detector is at a generic position. The third and the fourth lines appear in the same configurations as \eqref{eq:EEECprime_g0order}, and can be thought of as the higher-order corrections.

Using perturbation theory, we can obtain the functions $\cF_1(\vec n_1,\vec n_2)$ and $\cF_0(\vec n_1,\vec n_2,\vec n_3)$. We leave the details of the calculation in appendix \ref{app:treeEEEC_F0F1}. The result for $\cF_1$ is\footnote{Note that $\cF_1(\vec n_1,\vec n_2)$ is invariant under an overall rotation of $\vec n_1,\vec n_2$, so it can be written as a function of the cross ratio $\z=\tfrac{1-\vec n_1\.\vec n_2}{2}$.}
\be\label{eq:F1_fullexpr}
&\cF_1(\z)=-\frac{\z(-60+102\z-44\z^2+3\z^3)+(-60+132\z-90\z^2+19\z^3)\log(1-\z)}{256\pi^4(1-\z)\z^6}.
\ee
This expression is only valid for $0<\z<1$. The final expression should be a distribution and include contact terms at $\z=0$ and $\z=1$. To see the contact terms, we can further separate $\cF_1$ into a singular part and a regular part:
\be
\label{eq:f1split}
\cF_1(\z)=&\frac{1}{512\pi^4\z}-\frac{1+\log(1-\z)}{256\pi^4(1-\z)}+\cF_1^{\mathrm{reg}}(\z).
\ee
The singular terms can be interpreted as
\be\label{eq:F1singular_def}
\cF_1^{\mathrm{sing}}(\z)=\frac{1}{512\pi^4}\left[\frac{1}{\z}\right]_{0}-\frac{1}{256\pi^4}\left[\frac{1}{1-\z}\right]_1-\frac{1}{256\pi^4}\left[\frac{\log(1-\z)}{1-\z}\right]_1 + a_1\de\p{\z}+b_1\de\p{1-\z},
\ee
where the distribution $\left[\frac{1}{\z}\right]_{0}$ is defined as the unique distribution that agrees with $\frac{1}{\z}$ for $\z>0$ and satisfies
\be
\int_0^1 d\z\left[\frac{1}{\z}\right]_{0}=0,
\ee
and $[\cdots]_{1}$ is defined in a similar way with $\z \to 1-\z$.  The expressions (\ref{eq:f1split}) and (\ref{eq:F1singular_def}) now specify $\cF_1(\zeta)$ as a distribution. However, this distribution depends on unknown coefficients $a_1,b_1$.  We will determine them using Ward identities.

Now consider the function $\cF_0$. We find that it can be written as\footnote{Again due to rotational invariance, we can write $\cF_0$ as a function of two cross ratios $\z_1,\z_2$, where $\z_1=\tfrac{1-\vec n_1\.\vec n_3}{2}$, $\z_2=\tfrac{1-\vec n_2\.\vec n_3}{2}$.}
\be\label{eq:F0_intildeF0}
\cF_0(\z_1,\z_2)&=\sqrt{\z_1\z_2(1-\z_1)(1-\z_2)}\tl{\cF}_0(\z_1,\z_2)\theta(\z_1+\z_2-1),
\ee
where $\th(\cdots)$ is a step function (we explain its appearance in appendix \ref{app:treeEEEC_F0F1}). The function $\tl{\cF}_0$ is given by
\be\label{eq:F0tilde_fullexpr}
\tl{\cF}_0(\z_1,\z_2)&=\frac{1}{256\pi^4 \z_1^2 \z_2^2 \sqrt{\z_1\z_2(1-\z_1)(1-\z_2)} (\sqrt{\z_1(1-\z_2)}+\sqrt{\z_2(1-\z_1)})^4} \nn \\
&\quad\x\bigg(-14 \z_1^2 \z_2^2-18 \z_1 \z_2+19 \z_1 \z_2 (\z_1+\z_2)-6 (\z_1+\z_2)^2+12 (\z_1+\z_2)-6 \nn \\
&\quad\quad\quad + 2\sqrt{\z_1\z_2(1-\z_1)(1-\z_2)} \left(6-6(\z_1+\z_2)+7\z_1 \z_2 \right)\bigg).
\ee
It is convenient to study $\tl{\cF}_0$ instead of $\cF_0$. To interpret $\tl{\cF}_0$ as a distribution, we again separate it into a singular part and regular part as $\tl{\cF}_0=\tl{\cF}^{\mathrm{sing}}_0+\tl{\cF}^{\mathrm{reg}}_0$. The singular part is given by
\be\label{eq:F0tildesingular_def}
\tl{\cF}^{\mathrm{sing}}_0(\z_1,\z_2)=\frac{f(\th_1)}{r_1^2} + \frac{f(\th_2)}{r_2^2} + \frac{g(\th_3)}{r_3^2},
\ee
where
\be\label{eq:rtheta_def}
r_1&=\sqrt{\z_1^2+ (1-\z_2)^2},\qquad\qquad \th_1=\mathrm{tan}^{-1}\p{\frac{1-\z_2}{\z_1}}, \nn \\
r_2&=\sqrt{\z_2^2+ (1-\z_1)^2},\qquad\qquad \th_2=\mathrm{tan}^{-1}\p{\frac{1-\z_1}{\z_2}}, \nn \\
r_3&=\sqrt{(1-\z_1)^2 + (1-\z_2)^2}, \quad \th_3=\mathrm{tan}^{-1}\p{\frac{1-\z_2}{1-\z_1}}.
\ee
\begin{figure}[ht]
\centering
\includegraphics[width=8cm]{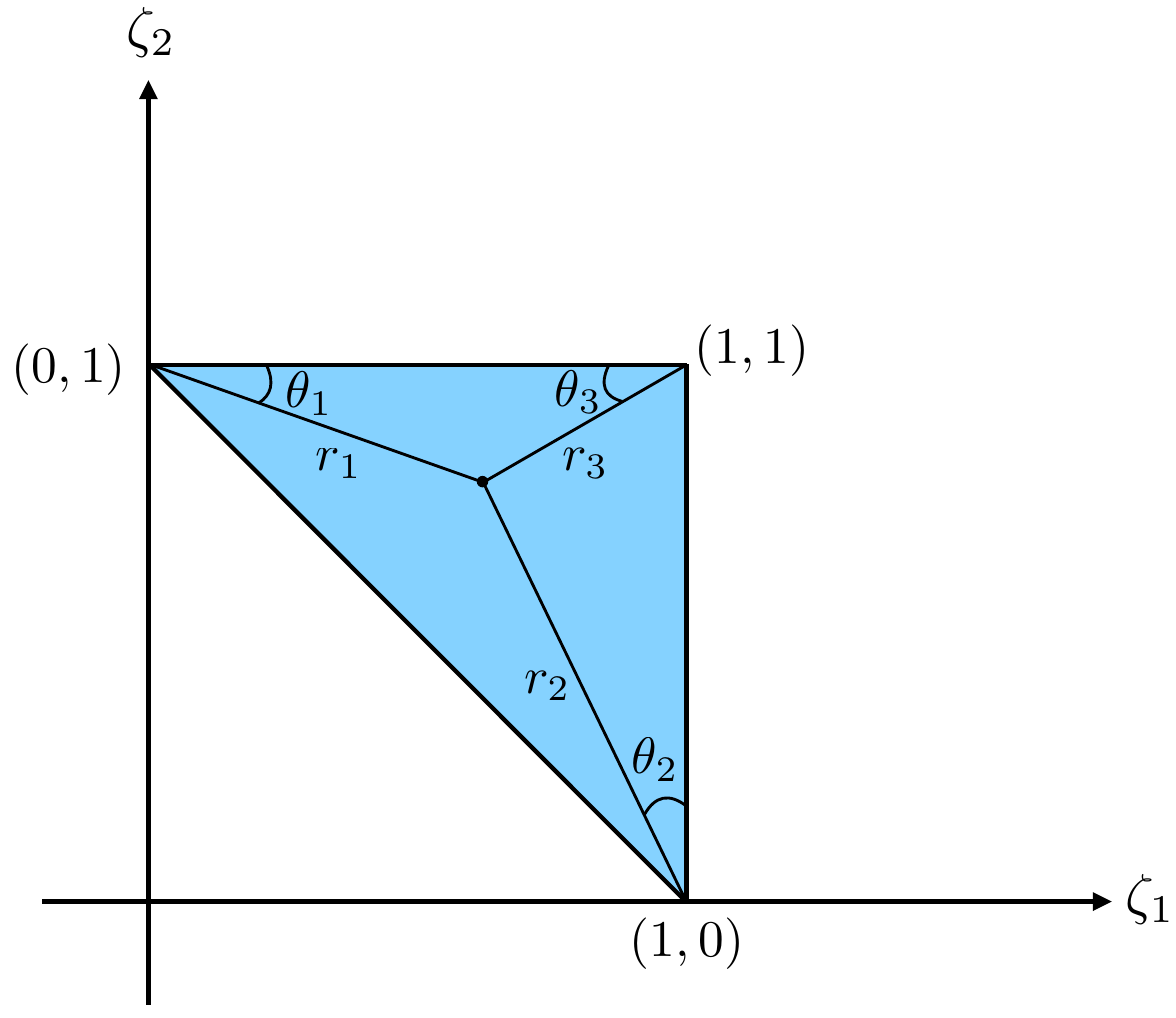}
\caption{Definition of $(r_i,\th_i)$. Due to the step function in \eqref{eq:F0_intildeF0}, $\cF_0(\z_1,\z_2)$ is nonzero only in the blue region. The three points $(1,0),(0,1),(1,1)$ are where $\cF_0(\z_1,\z_2)$ becomes singular.}
\label{fig:crossratio}
\end{figure}
$(r_i,\th_i)$ are the polar coordinates with respect to the points where $\tl{\cF}_0$ becomes singular. See figure \ref{fig:crossratio}. The two functions in \eqref{eq:F0tildesingular_def} are given by
\be
f(\th)&= \frac{1}{256 \pi ^4 \sqrt{\sin (\th)} \cos ^{\frac{3}{2}}(\th)},\nn \\
g(\th)&=\frac{1}{256 \pi ^4 \sqrt{\sin (\th)} \sqrt{\cos (\th)} \left(\sqrt{\sin (\th)}+\sqrt{\cos (\th)}\right)^2}.
\ee
One can then show that $\tl{\cF}^{\mathrm{sing}}_0$ as a distribution should be given by
\be\label{eq:F0tilde_singular_expr}
\tl{\cF}^{\mathrm{sing}}_0(\z_1,\z_2)=a_0 \p{\frac{\de(r_1)}{r_1}+\frac{\de(r_2)}{r_2}} + b_0 \frac{\de(r_3)}{r_3} + \frac{f_0(\th_1)}{r_1}\left[\frac{1}{r_1}\right]_{0} + \frac{f_0(\th_2)}{r_2}\left[\frac{1}{r_2}\right]_{0} + \frac{g_0(\th_3)}{r_3}\left[\frac{1}{r_3}\right]_{0}.
\ee
Moreover, from \eqref{eq:EEECprime_g2_exprgeneral} we see that $\cF_0$ must be crossing symmetric. This condition implies\footnote{See appendix \ref{app:EEEC_Wardidentities} for a derivation.}
\be\label{eq:crossing_a0b0condition}
b_0-\frac{1}{2}a_0=& \frac{\pi-\sqrt{2}\pi+2\log 2+\sqrt{2}\log(3-2\sqrt{2})}{256\pi^5} \approx 0.00003291.
\ee

Note that although we have six unknown coefficients ($c_1,c_2,a_1,b_1,a_0,b_0$), there are actually only two types of contact terms in \eqref{eq:EEECprime_g2_exprgeneral}. By integrating against test functions, we find
\be
\frac{\de(r_1)}{r_1}\sqrt{\z_1\z_2(1-\z_1)(1-\z_2)}\de\p{(\vec n_1\x \vec n_2)\.\vec n_3}&=\frac{\pi^2}{2}\de\p{\vec n_1,\vec n_3}\de\p{\vec n_2, -\vec n_3}, \nn \\
\frac{\de(r_2)}{r_2}\sqrt{\z_1\z_2(1-\z_1)(1-\z_2)}\de\p{(\vec n_1\x \vec n_2)\.\vec n_3}&=\frac{\pi^2}{2}\de\p{\vec n_1,-\vec n_3}\de\p{\vec n_2, \vec n_3}, \nn \\
\frac{\de(r_3)}{r_3}\sqrt{\z_1\z_2(1-\z_1)(1-\z_2)}\de\p{(\vec n_1\x \vec n_2)\.\vec n_3}&=\pi^2\de\p{\vec n_1,-\vec n_3}\de\p{\vec n_2, -\vec n_3},
\ee
and in addition
\be
\de(\tfrac{1-\vec n_1\.\vec n_2}{2})\de\p{\vec n_2,\vec n_3}&=4\pi\de(\vec n_1,\vec n_2)\de(\vec n_2,\vec n_3), \nn \\
\de(1-\tfrac{1-\vec n_1\.\vec n_2}{2})\de\p{\vec n_2,\vec n_3}&=4\pi\de(\vec n_1,-\vec n_2)\de(\vec n_2,\vec n_3).
\ee
Collecting all the contact terms together, we find
\be
&2\pi \p{c_1+6a_1}\de(\vec n_1,\vec n_2)\de(\vec n_2,\vec n_3) \nn\\
&+2\pi\p{c_2+2b_1+\frac{\pi}{4}a_0}\p{\de(\vec n_1,\vec n_3)\de(\vec n_2,-\vec n_3)+\de(\vec n_1,-\vec n_3)\de(\vec n_2,\vec n_3)}\nn \\
&+2\pi\p{c_2+2b_1+\frac{\pi}{2}b_0}\de\p{\vec n_1,-\vec n_3}\de\p{\vec n_2, -\vec n_3}.
\ee
Only the above linear combinations of coefficients are physically meaningful.
Note that the coefficients of $\de(\vec n_1,\vec n_3)\de(\vec n_2,-\vec n_3)$ and $\de\p{\vec n_1,-\vec n_3}\de\p{\vec n_2, -\vec n_3}$ are different because of our choice of coordinates \eqref{eq:rtheta_def} to write the distributions $[\cdots]_0$. These coordinates are convenient for analyzing the singularities of $\cF_0$, but they do not manifest crossing symmetry. When we perform a crossing transformation, we rescale the arguments of the distributions $[\cdots]_0$, which can produce $\de$-functions via $[1/(\l x)]_0=\l^{-1}[1/x]_0 - (\log \l/\l) \de(x)$. Taking these $\de$-functions into account, one can show that as long as \eqref{eq:crossing_a0b0condition} is satisfied, the full expression is crossing symmetric. See appendix \ref{app:treeEEEC_F0_crossing} for details.

We can now use the Ward identities \eqref{eq:EEECp_Wardidentities} to determine the coefficients of the contact terms. Plugging \eqref{eq:EEECprime_g2_exprgeneral} into \eqref{eq:EEECp_Wardidentities}, we obtain
\be\label{eq:Wardidentity_final}
&\int_0^1 d\z_1\int_0^1 d\z_2\ \tl{\cF}_0(\z_1,\z_2)\theta(\z_1+\z_2-1)+ 6 \int_0^{1}d\z\ \cF_1(\z)+ c_1 + 3 c_2 = 0, \nn \\
&2\int_0^1 d\z_1\int_0^1 d\z_2\ \z_1\tl{\cF}_0(\z_1,\z_2)\theta(\z_1+\z_2-1) + 8 \int_0^1d\z\ \z \cF_1(\z) + 4 c_2=0, \nn \\
&4\int_0^1 d\z_1\int_0^1 d\z_2\ \z_1\z_2\tl{\cF}_0(\z_1,\z_2)\theta(\z_1+\z_2-1) + 8 \int_0^1d\z\ \z^2 \cF_1(\z) + 4 c_2=0,
\ee
where $\tl{\cF}_0(\z_1,\z_2)$ is defined by \eqref{eq:F0_intildeF0}. The integrals in \eqref{eq:Wardidentity_final} can be computed using \eqref{eq:F1_fullexpr}, \eqref{eq:F1singular_def}, \eqref{eq:F0tilde_fullexpr}, and \eqref{eq:F0tildesingular_def}. The integrals of the $\tl{\cF}_0$ function are given by
\be
\int_0^1 d\z_1\int_0^1 d\z_2\ \tl{\cF}_0(\z_1,\z_2)\theta(\z_1+\z_2-1)&=\frac{\pi}{2}(a_0+b_0)+ 0.00011205, \nn \\ 
\int_0^1 d\z_1\int_0^1 d\z_2\ \z_1\tl{\cF}_0(\z_1,\z_2)\theta(\z_1+\z_2-1)&=\frac{\pi}{2}\p{\frac{a_0}{2} + b_0} + 0.0000815314, \nn \\
\int_0^1 d\z_1\int_0^1 d\z_2\ \z_1\z_2\tl{\cF}_0(\z_1,\z_2)\theta(\z_1+\z_2-1)&=\frac{\pi}{2}b_0 + 0.0000884406,
\ee
and for the $\cF_1$ integrals we get
\be
\int_0^1 d\z \cF_1(\z)&=\frac{51+10\pi^2}{15360\pi^4} +a_1+b_1, \nn \\
\int_0^1 d\z\ \z \cF_1(\z)&=\frac{3+2\pi^2}{3072\pi^4} +b_1,  \nn \\
\int_0^1 d\z\ \z^2 \cF_1(\z)&=\frac{-4+\pi^2}{1536\pi^4} +b_1.
\ee
Using these results, we find the solution to \eqref{eq:Wardidentity_final}:
\be
b_0- \frac{a_0}{2} =& 0.0000329062, \nn \\
c_2+2b_1+\frac{\pi}{4}a_0=& -0.00021859, \nn \\
c_1+6a_1=& -0.000108274.
\ee
We see that the first line is consistent with \eqref{eq:crossing_a0b0condition}, which comes from crossing symmetry of $\cF_0(\z_1,\z_2)$. The second and the third line give the coefficients of the two types of contact terms that can appear in $\mathrm{EEEC}'|_{O(g^2)}$.

\subsection{Contact terms from the conformal block decomposition}
It is interesting to ask how the $\de$-functions in $\mathrm{EEEC}'|_{O(g^2)}$ are reproduced from the conformal block decomposition described in section \ref{sec:blockdecomposition}. Since the conformal block decomposition describes the leading term in the collinear limit, we should study those $\de$-functions that survive when all $\vec n_i$'s are close to each other. More precisely, we have an expansion around the squeezed limit, where we first take $\vec n_2$ and $\vec n_3$ close together, and then $\vec n_1$.

Using \eqref{eq:EEEC_decomposition_expr0}, the collinear limit is
\be
\EEEC'(\vec n_1,\vec n_2,\vec n_3)|_{O(g^2)}&=\frac{1}{32\pi^2}\frac{\z_{12}}{\z_{23}^3}\sum_{\de,j}R_{\de,j}|_{O(g^2)} g_{\de,j}^{\cE\cE\cE\cP_{\de'_{*}}}(z, \bar z) + \ldots\,,
\ee
where $z, \bar z$ are defined by
\be
\frac{\z_{23}}{\z_{12}}= z \bar z,\quad \frac{\z_{13}}{\z_{12}}= (1-z) (1-\bar z).
\ee
The small $z, \bar z$ limit corresponds to the limit where $\vec n_2$ and $\vec n_3$ become close. The contact term $\cF_1(\vec n_1,\vec n_2)\de\p{\vec n_2, \vec n_3}$ can then appear from the exchanged quantum numbers $\de=4, j=0$. More precisely, we have
\be
\EEEC'(\vec n_1,\vec n_2,\vec n_3)|_{O(g^2)}& \sim \frac{g^2 N_c}{512\pi^5}\frac{1}{\z_{12}^2}\lim_{\de \to 4}\<R^{(0)}_{\de,0}\>(z \bar z)^{\frac{\de}{2}-3} 
= \frac{g^2 N_c}{64\pi^4}\p{\lim_{\de \to 4}\frac{\<R^{(0)}_{\de,0}\>}{\de-4}}\frac{1}{\z_{12}}\de(\vec n_2,\vec n_3).
\ee
This result should agree with the $\frac{1}{\z}$ term of $\cF_1(\z)$ in \eqref{eq:F1singular_def}. Matching the two expressions, we obtain
\be\label{eq:Rcoeff0order_contact}
\lim_{\de \to 4}\frac{\<R^{(0)}_{\de,0}\>}{\de-4}=\frac{1}{8}.
\ee
Thus, even though $\<R^{(0)}_{4,0}\>$ vanishes, the zero of $\<R^{(0)}_{\de,0}\>$ at $\de=4$ is related to the EEEC at $O(g^2)$ order. It would be interesting to verify \eqref{eq:Rcoeff0order_contact} using other methods.

\section{EEEC at strong coupling}\label{sec:strongcoupling}
We now consider the EEEC at strong coupling. In \cite{Hofman:2008ar}, Hofman and Maldacena computed the EEEC at strong coupling in $\cN=4$ SYM up to $O(\l^{-3/2})$ order using AdS/CFT for a sink/source state created by a massless closed string. In this paper, we focus on the leading order and $O(1/\l)$ correction. In terms of \eqref{eq:eeecnparam}, the result of \cite{Hofman:2008ar} is
\be
\label{eq:hofmalresult}
&\mathrm{EEEC}_{\mathrm{strong}}'(\vec n_1,\vec n_2,\vec n_3)\nn \\
&=\p{\frac{1}{4\pi}}^3\p{1+\frac{6\pi^2}{\l}\p{(\vec n_1\. \vec n_2)^2+(\vec n_1\. \vec n_3)^2+(\vec n_2\. \vec n_3)^2-1} + O(\l^{-3/2})}.
\ee
Let us lift this expression into a more covariant form, using embedding space coordinates for the celestial sphere $z_i\in \R^{d-1,1}$ with $z_i^2=0$. We define
\be\label{eq:EEECprime_def_embedding}
\cF(z_1,z_2,z_3,p)\equiv \frac{8\int d^dx\ e^{ip\.x}\<0|\cO^{\dag}(x)\wL[T](\oo,z_1)\wL[T](\oo,z_2)\wL[T](\oo,z_3)\cO(0)|0\>}{(-p^2)^{\frac{3}{2}}\int d^dx\ e^{ip\.x}\<0|\cO^{\dag}(x)\cO(0)|0\>},
\ee
which is a homogeneous function of the $z_i$. The original EEEC can be recovered by specializing: $\cF(z_i=(1,\vec n_i),p=(1,\vec 0))=\mathrm{EEEC}'(\vec n_i)$. In embedding space coordinates, the expression (\ref{eq:hofmalresult}) becomes
\be\label{eq:EEEC_strong1}
&\cF_{\mathrm{strong}}(z_1,z_2,z_3,p)\nn \\
&=\frac{8}{\pi^3}\frac{1}{(-2z_1\.p)^3(-2z_2\.p)^3(-2z_3\.p)^3}\p{1+\frac{6\pi^2}{\l}\p{(1-2\z_{12})^2+(1-2\z_{23})^2+(1-2\z_{13})^2-1}} \nn \\
&\quad + O(\l^{-3/2}),
\ee
where the cross ratios $\z_{ij}$ are defined in \eqref{eq:crossratio_definition}. For this section, we also assume that $-p^2=1$, since the $-p^2$ factors can be easily restored using dimensional analysis of $p$.

In the weak coupling limit, the EEEC at tree-level is only nonzero when the three detectors are coplanar, and the EEEC at 1-loop it is only known in the collinear limit. On the other hand, the strong-coupling EEEC given in \eqref{eq:EEEC_strong1} is valid for any detector positions on the celestial sphere.

Just as Mean Field Theory provides a simple example of a crossing-symmetric, conformally-invariant four-point function, the strong-coupling EEEC given in \eqref{eq:EEEC_strong1} gives a simple example of a crossing-symmetric, Lorentz invariant EEEC. It is therefore a perfect target for us to study its celestial block expansion and test the results of section \ref{sec:blockdecomposition}. Because the strong-coupling EEEC is so simple, we will be able to compute its {\it complete\/} 3-point celestial block expansion --- i.e.\ not just its conformal block expansion in the collinear limit. 

The celestial block expansion can be written as\footnote{The relation between $P_{\de,j;\de'}$ and the coefficient $R_{\de,j;\de'}$ defined in section \ref{sec:blockdecomposition} is $P_{\de,j;\de'}=\frac{16}{\pi^2}R_{\de,j;\de'}$.}
\be\label{eq:strongEEEC_celestialblockexpansion1}
&\cF_{\mathrm{strong}}(z_1,z_2,z_3,p)=\sum_{\de'}\sum_{\de,j}P_{\de,j;\de'}G^c_{\de,j;\de'}(z_1,z_2,z_3,p),
\ee
where the celestial block is defined as
\be
G^{c}_{\de,j;\de'}(z_1,z_2,z_3,p) = \cC_{12\cP_{\de,j}}(z_1,z_2,\ptl_{z_2},\ptl_{w_2})\cC_{\cP_{\de,j}3\cP_{\de'}}(z_2,z_3,\ptl_{z_3})(-2z_3\.p)^{-\de'}.
\ee
Let us first focus on the $j=0$ sector of \eqref{eq:strongEEEC_celestialblockexpansion1}. The light-ray OPE formula gives a relation between the OPE data and the value of $\de$ in the sum. In the strong coupling limit, the $T\x T$ OPE should contain double-trace operators with twists $\tau=2\tau_T+2n=4+2n$. Therefore, the values of $\de$ appearing in \eqref{eq:strongEEEC_celestialblockexpansion1} should be
\be\label{eq:strongcoupling_deltaprediction}
\de=\De(J=3)-1=6+2n.
\ee
Similarly, the $T\x T\x T$ OPE should contain triple-trace operators with twists $\tau=3\tau_T+2k=6+2k$. As argued in section \ref{sec:lightrayOPE_prediction}, we expect that the values of $\de'$ appearing in \eqref{eq:strongEEEC_celestialblockexpansion1} are given by
\be\label{eq:strongcoupling_deltaprimeprediction}
\de'=\De(J=4)-1=9+2k.
\ee

To study the celestial block expansion of the strong-coupling EEEC, we will use harmonic analysis for the Euclidean conformal group $\SO(d-1,1)$ \cite{Dobrev:1977qv}. A modern review of harmonic analysis for Euclidean CFTs is given in \cite{Karateev:2018oml}, where they derive a Euclidean inversion formula that expresses OPE data as an integral of CFT four-point functions over Euclidean space. In this section, we use techniques from \cite{Karateev:2018oml} to derive a ``celestial inversion formula" that expresses the celestial block expansion data as an integral of the EEEC over the celestial sphere. We then consider the strong coupling limit and use this celestial inversion formula to obtain the celestial block expansion \eqref{eq:strongEEEC_celestialblockexpansion1} for the strong-coupling EEEC \eqref{eq:EEEC_strong1}.

\subsection{Celestial inversion formula}

We first derive the celestial inversion formula, following the derivation of the Euclidean inversion formula in section 2 of \cite{Karateev:2018oml}. By harmonic analysis for $\SO(d-1,1)$, the EEEC $\cF(z_1,z_2,z_3,p)$ defined in \eqref{eq:EEECprime_def_embedding} can be written as an integral of the form
\be\label{eq:EEEC_celestialpartialwaveexpansion}
\cF(z_1,z_2,z_3,p)=\sum_{j}\int_{\frac{d-2}{2}}^{\frac{d-2}{2}+i\oo}\frac{d\de}{2\pi i}\int_{\frac{d-2}{2}}^{\frac{d-2}{2}+i\oo}\frac{d\de'}{2\pi i} I(\de,j;\de')\Psi^c_{\de,j;\de'}(z_1,z_2,z_3,p),
\ee
where $\Psi^{c}_{\de,j;\de'}$ is a ``celestial partial wave" defined by
\be
\label{eq:celestialpartialwavedefinition}
&\Psi^{c}_{\de,j;\de'}(z_1,z_2,z_3,p)\nn \\
&\equiv \int D^{d-2}zD^{d-2}z' \<\cP_{\de_1}(z_1)\cP_{\de_2}(z_2)\cP_{\de,j}(z)\>\<\tl{\cP}_{\de,j}(z)\cP_{\de_3}(z_3)\cP_{\de'}(z')\>\frac{1}{(-2z'\.p)^{\tl{\de'}}},
\ee
where $\tl{\cP}_{\de,j}$ is the shadow representation of $\cP_{\de,j}$, with scaling dimension $\tl{\de}=d-2-\de$. The measure $D^{d-2}z$ is defined by
\be\label{eq:Dz_definition}
D^{d-2}z=\frac{2d^dz \de(z^2)\th(z^0)}{\vol \R_{+}},
\ee
where $\R_{+}$ acts by rescaling $z$. Note that the operators $\cP_{\de,j}(z)$ and $\tl \cP_{\de,j}(z)$ each carry $j$ tangent-space indices on the celestial sphere. These indices are implicitly contracted in (\ref{eq:celestialpartialwavedefinition}) and below.

By construction, $\Psi^c$ is an eigenfunction of the Casimirs of $\SO(d-1,1)$, acting simultaneously on $z_1,z_2$ and simultaneously on $z_1,z_2,z_3$. So, we can study the behavior of $\Psi^c$ in the OPE limit to determine its relation to the celestial block $G^c$. Following the logic in \cite{Karateev:2018oml}, the relation is
\be\label{eq:celestialpartialwavevsblock}
\Psi^{c}_{\de,j;\de'}=&S(\cP_{\de_3}\cP_{\de'}[\tl{\cP}_{\de,j}])I_{\de'}(p)G^{c}_{\de,j;\de'}  + S(\cP_{\de_3}\cP_{\de'}[\tl{\cP}_{\de,j}])S(\cP_{\de,j}\cP_{\de_3}[\cP_{\de'}])G^{c}_{\de,j;\tl{\de'}} \nn \\
+& S(\cP_{\de_1}\cP_{\de_2}[\cP_{\de,j}])I_{\de'}(p)G^{c}_{\tl{\de},j;\de'} + S(\cP_{\de_1}\cP_{\de_2}[\cP_{\de,j}])S(\tl{\cP}_{\de,j}\cP_{\de_3}[\cP_{\de'}])G^{c}_{\tl{\de},j;\tl{\de'}},
\ee
where the coefficients $S(\cdots), I_{\de'}(p)$ are defined by
\be
\int D^{d-2}z \<\tl{\cP}(z')\tl{\cP}(z)\>\<\cP_1\cP_2\cP(z)\> &= S(\cP_1\cP_2[\cP])\<\cP_1\cP_2\tl{\cP}(z')\> \nn \\
\int D^{d-2}z' \<\cP_{\de'}(z_3)\cP_{\de'}(z')\>\frac{1}{(-2z'\.p)^{\tl{\de'}}}&=I_{\de'}(p) (-2z_3\.p)^{-\de'}.
\ee
More explicitly, their expressions are
\be
S(\cP_1\cP_2[\cP_{\de,j}])=&\frac{\pi^{\frac{d-2}{2}}\G(\de-\tfrac{d-2}{2})\G(\de+j-1)\G(\tfrac{\tl{\de}+\de_1-\de_2+j}{2})\G(\tfrac{\tl{\de}-\de_1+\de_2+j}{2})}{\G(\de-1)\G(d-2-\de+j)\G(\tfrac{\de+\de_1-\de_2+j}{2})\G(\tfrac{\de-\de_1+\de_2+j}{2})}, \nn \\
S(\cP_{\de,j}\cP_3[\cP_{\de'}])=&\frac{\pi^{\frac{d-2}{2}}\G(\de'-\tfrac{d-2}{2})\G(\tfrac{\tl{\de'}+\de_3-\de+j}{2})\G(\tfrac{\tl{\de'}-\de_3+\de+j}{2})}{\G(d-2-\de')\G(\tfrac{\de'+\de_3-\de+j}{2})\G(\tfrac{\de'-\de_3+\de+j}{2})}, \nn \\
I_{\de'}(p)=&\frac{\pi^{\frac{d-2}{2}}\G(\tfrac{d-2}{2}-\de')}{\G(d-2-\de')}(-p^2)^{\de'-\frac{d-2}{2}}.
\ee

Just like the four-point conformal partial wave, the celestial partial wave $\Psi^c$ satisfies an orthogonality relation that can be derived using a ``bubble" formula. Consider two celestial partial waves $\Psi^{c}_{\de_5,j_5;\de'_5}$ and $\Psi^{c(\tl{\de}_i)}_{\tl{\de}_6,j_6;\tl{\de'}_6}$, where $\de_{5,6}=\frac{d-2}{2}+is_{5,6},\de'_{5,6}=\frac{d-2}{2}+is'_{5,6}$ are on the principal series and the external dimensions of $\Psi^{c(\tl{\de}_i)}$ are the shadows of $\Psi^{c}$. The orthogonality relation is (see Appendix \ref{app:strongcoupling} for a derivation)
\be\label{eq:celestialpartialwave_orthogonality}
&\int\frac{D^{d-2}z_1 D^{d-2}z_2 D^{d-2}z_3 d^{d-1}_{\mathrm{AdS}}p}{\vol(\SO(d-1,1))}\ \Psi^{c}_{\de_5,j_5;\de'_5}(z_1,z_2,z_3,p)\Psi^{c(\tl{\de}_i)}_{\tl{\de}_6,j_6;\tl{\de'}_6}(z_1,z_2,z_3,p) \nn \\
&=\frac{1}{2^{d-2}\vol(\SO(d-2))}B_{12\cP_{\de_5,j_5}}B_{\tl{\cP}_{\de_5,j_5}3\cP_{\de'_{5}}}\de_{\cP_5\cP_6}\de_{\cP'_5 \cP'_{6}},
\ee
where the integral measure for $p$ and the delta function $\de_{\cP_5\cP_6}$ are defined by
\be\label{eq:dp_delta56_definition}
d_{AdS}^{d-1}p&=2d^dp\de(p^2+1)\theta(p^0), \nn \\
\de_{\cP_5\cP_6}&=2\pi\de(s_5-s_6)\de_{j_5,j_6}.
\ee
The $B$-coefficients are ``bubble" coefficients defined by
\be\label{eq:bubblecoeff_definition}
B_{12\cP_{\de,j}}=\frac{1}{\mu(\de,j)}\p{\<\cP_1\cP_2\cP_{\de,j}\>,\<\tl{\cP}_1\tl{\cP}_2\tl{\cP}_{\de,j}\>},
\ee
where $\mu(\de,j)$ is the Plancherel measure of $\SO(d-1,1)$, and $\p{\<\cP_1\cP_2\cP_{\de,j}\>,\<\tl{\cP}_1\tl{\cP}_2\tl{\cP}_{\de,j}\>}$ is a conformally-invariant three-point pairing. Their explicit expressions are given in \cite{Karateev:2018oml}.

Integrating both sides of celestial partial wave expansion \eqref{eq:EEEC_celestialpartialwaveexpansion} against a celestial partial wave $\Psi^{c(\tl{\de}_i)}_{\tl{\de},j;\tl{\de}'}$, the orthogonality relation \eqref{eq:celestialpartialwave_orthogonality} gives
\be\label{eq:inversionformula_1}
&I(\de,j;\de') \nn \\
&= \frac{2^{d-2}\vol(\SO(d-2))}{B_{12\cP_{\de,j}}B_{\tl{\cP}_{\de,j}3\cP_{\de'}}}\int \frac{D^{d-2}z_1 D^{d-2}z_2 D^{d-2}z_3 d^{d-1}_{\mathrm{AdS}}p}{\vol(\SO(d-1,1))}\ \cF(z_1,z_2,z_3,p)\Psi^{c(\tl{\de}_i)}_{\tl{\de},j;\tl{\de}'}(z_1,z_2,z_3,p).
\ee
This is the celestial inversion formula that expresses the celestial partial wave expansion data $I(\de,j;\de')$ as an integral of the EEEC $\cF(z_1,z_2,z_3,p)$ over the celestial sphere. Finally, we must find the relation between the celestial partial wave expansion \eqref{eq:EEEC_celestialpartialwaveexpansion} and the celestial block expansion \eqref{eq:strongEEEC_celestialblockexpansion1}. Plugging \eqref{eq:celestialpartialwavevsblock} into \eqref{eq:EEEC_celestialpartialwaveexpansion}, we obtain
\be\label{eq:celestialpartialwave_decomposition_1}
\cF(z_1,z_2,z_3,p)=&\sum_{j}\int_{\frac{d-2}{2}}^{\frac{d-2}{2}+i\oo}\frac{d\de}{2\pi i}\int_{\frac{d-2}{2}}^{\frac{d-2}{2}+i\oo}\frac{d\de'}{2\pi i} I(\de,j;\de')\x \nn \\
&\qquad\left(S(\cP_{\de_3}\cP_{\de'}[\tl{\cP}_{\de,j}])I_{\de'}(p)G^{c}_{\de,j;\de'}  + S(\cP_{\de_3}\cP_{\de'}[\tl{\cP}_{\de,j}])S(\cP_{\de,j}\cP_{\de_3}[\cP_{\de'}])G^{c}_{\de,j;\tl{\de'}} \right.\nn \\
&\qquad+ \left.S(\cP_{\de_1}\cP_{\de_2}[\cP_{\de,j}])I_{\de'}(p)G^{c}_{\tl{\de},j;\de'} + S(\cP_{\de_1}\cP_{\de_2}[\cP_{\de,j}])S(\tl{\cP}_{\de,j}\cP_{\de_3}[\cP_{\de'}])G^{c}_{\tl{\de},j;\tl{\de'}} \right).
\ee
Using the inversion formula \eqref{eq:inversionformula_1}, one can show that above expression remains the same when we keep only the first term $S(\cP_{\de_3}\cP_{\de'}[\tl{\cP}_{\de,j}])I_{\de'}(p)G^{c}_{\de,j;\de'}$ in the parentheses and extend the integration ranges to $\int_{\tfrac{d-2}{2}-i\oo}^{\tfrac{d-2}{2}+i\oo}$. Therefore, we have
\be\label{eq:EEEC_celestialblockexpansion0}
\cF(z_1,z_2,z_3,p)=&\sum_{j}\int_{\frac{d-2}{2}-i\oo}^{\frac{d-2}{2}+i\oo}\frac{d\de}{2\pi i}\int_{\frac{d-2}{2}-i\oo}^{\frac{d-2}{2}+i\oo}\frac{d\de'}{2\pi i} C(\de,j;\de') G^{c}_{\de,j;\de'}(z_1,z_2,z_3,p),
\ee
where
\be\label{eq:Cfunc_definition}
C(\de,j;\de')\equiv I(\de,j;\de')S(\cP_{\de_3}\cP_{\de'}[\tl{\cP}_{\de,j}])I_{\de'}(p).
\ee
Finally, we can close the contour of the $\de$ and $\de'$ integrals in \eqref{eq:EEEC_celestialblockexpansion0} to the right and obtain the celestial block expansion.\footnote{As we will see later, closing the $\de$ contour or the $\de'$ contour first gives the same celestial block expansion, at least for the strong-coupling EEEC we consider in this paper. Moreover, in appendix \ref{app:strongcoupling} we study the celestial block $G^c_{\de,j;\de'}$ at large $\de$ and $\de'$, and show that contributions at infinity of both contours vanish.} In particular, for the strong-coupling EEEC \eqref{eq:strongEEEC_celestialblockexpansion1}, we have
\be
\Res_{\de=\de_i}\Res_{\de'=\de'_i} C_{\mathrm{strong}}(\de,j;\de')= P_{\de_i,j;\de'_i}.
\ee

\subsection{Leading order}
We first consider the leading order term of \eqref{eq:EEEC_strong1},
\be
\cF^{(0)}_{\mathrm{strong}}(z_1,z_2,z_3,p)=\frac{8}{\pi^3(-2z_1\.p)^3(-2z_2\.p)^3(-2z_3\.p)^3}.
\ee
Plugging this into the celestial inversion formula \eqref{eq:inversionformula_1}, we find
\be
&I^{(0)}_{\mathrm{strong}}(\de,j;\de') \nn \\
&=\frac{8}{\pi^3} \frac{2^{d-2}\vol(\SO(d-2))}{B_{\cP_{\cE}\cP_{\cE}\cP_{\de,j}}B_{\tl{\cP}_{\de,j}\cP_{\cE}\cP_{\de'}}}\x \int \frac{D^{d-2}zD^{d-2}z' D^{d-2}z_1 D^{d-2}z_2 D^{d-2}z_3 d^{d-1}_{\mathrm{AdS}}p}{\vol(\SO(d-1,1))}  \nn \\
&\quad\frac{1}{(-2z_1\.p)^3(-2z_2\.p)^3(-2z_3\.p)^3}\<\cP_{\tl{\cE}}(z_1)\cP_{\tl{\cE}}(z_2)\tl{\cP}_{\de,j}(z)\>\<\cP_{\de,j}(z)\cP_{\tl{\cE}}(z_3)\cP_{\tl{\de'}}(z')\>\frac{1}{(-2z'\.p)^{\de'}}.
\ee

Let us study the $z_1,z_2$ integral,
\be\label{eq:z1z2integral_leading}
\int D^{d-2}z_1 D^{d-2}z_2\ \frac{1}{(-2z_1\.p)^3(-2z_2\.p)^3}\<\cP_{\tl{\cE}}(z_1)\cP_{\tl{\cE}}(z_2)\tl{\cP}_{\de,j}(z,w)\>.
\ee
After integration, the polarization vector $w$ must appear in the combination $[z,w]^{\mu\nu}\equiv z^{\mu}w^{\nu}-z^{\nu}w^{\mu}$ due to Lorentz invariance. However, the only remaining vector that is left unintegrated is $p$, and $[z,w]$ cannot be contracted with anything. Therefore, this integral must vanish except when $j=0$. To compute the integral for $j=0$, we can use
\be\label{eq:j0_beforepair}
&\int D^{d-2}z_1D^{d-2}z_2\<\tl{\cP}_{\de_1}(z_1)\tl{\cP}_{\de_2}(z_2)\cP_{\de,j=0}(z)\>(-2p\. z_1)^{-\de_1}(-2p\. z_2)^{-\de_2}\nn \\
&=C_{\de_1,\de_2;\de}(-p^2)^{\frac{\de-\de_1-\de_2}{2}}(-2p\. z)^{-\de},
\ee
where
\be\label{eq:C0_m1}
C_{\de_1,\de_2;\de}=\frac{\pi^{d-2}\G(\frac{\de_1+\de_2+\de-d+2}{2})\G(\frac{\de_1+\de_2-\de}{2})\G(\frac{d-2-\de_2-\de+\de_1}{2})\G(\frac{d-2-\de_1-\de+\de_2}{2})}{\G(\de_1)\G(\de_2)\G(d-2-\de)\G(\frac{d-2}{2})}.
\ee
We give a derivation of \eqref{eq:j0_beforepair} in appendix \ref{app:strongcoupling}. After integrating over $z_1$ and $z_2$, we obtain
\be
I^{(0)}_{\mathrm{strong}}(\de,0;\de')
&=\frac{8}{\pi^3} \frac{2^{d-2}\vol(\SO(d-2))}{B_{\cP_{\cE}\cP_{\cE}\cP_{\de,j}}B_{\tl{\cP}_{\de,j}\cP_{\cE}\cP_{\de'}}}C_{3,3;\tl{\de}}\x \int \frac{D^{d-2}zD^{d-2}z' D^{d-2}z_3 d^{d-1}_{\mathrm{AdS}}p}{\vol(\SO(d-1,1))}  \nn \\
&\quad\quad\frac{1}{(-2z_3\.p)^3(-2z\.p)^{\tl{\de}}}\<\cP_{\de,0}(z)\cP_{\tl{\cE}}(z_3)\cP_{\tl{\de'}}(z')\>\frac{1}{(-2z'\.p)^{\de'}}.
\ee

The remaining $p$-integral is just a three-point Witten diagram, and it is given by \cite{Freedman:1998tz}
\be\label{eq:Witten_3pt}
&\int d_{AdS}^{d-1}p~(-2p\. z_1)^{-\de_1}(-2p\. z_2)^{-\de_2}(-2p\. z_3)^{-\de_3}\, \nn \\
=&D_{\de_1,\de_2,\de_3}\frac{1}{(-2z_1\. z_2)^{\frac{\de_1+\de_2-\de_3}{2}}(-2z_1\. z_3)^{\frac{\de_1+\de_3-\de_2}{2}}(-2z_2\. z_3)^{\frac{\de_2+\de_3-\de_1}{2}}},
\ee
where
\be\label{eq:Dcoeff}
D_{\de_1,\de_2,\de_3}=\frac{\pi^{\frac{d-2}{2}}\G(\frac{\de_1+\de_2+\de_3-d+2}{2})}{2\G(\de_1)\G(\de_2)\G(\de_3)}\G(\tfrac{\de_1+\de_2-\de_3}{2})\G(\tfrac{\de_1+\de_3-\de_2}{2})\G(\tfrac{\de_2+\de_3-\de_1}{2}).
\ee
So, we have
\be
I^{(0)}_{\mathrm{strong}}(\de,0;\de')
&=\frac{8}{\pi^3} \frac{2^{d-2}\vol(\SO(d-2))}{B_{\cP_{\cE}\cP_{\cE}\cP_{\de,0}}B_{\tl{\cP}_{\de,0}\cP_{\cE}\cP_{\de'}}}C_{3,3;\tl{\de}}D_{\tl{\de},3,\de'}\x   \nn \\
&\quad\int \frac{D^{d-2}zD^{d-2}z' D^{d-2}z_3}{\vol(\SO(d-1,1))}\<\cP_{\de,0}(z)\cP_{\tl{\cE}}(z_3)\cP_{\tl{\de'}}(z')\>\<\cP_{\cE}(z_3)\cP_{\tl{\de}}(z)\cP_{\de'}(z')\> \nn \\
&=\frac{8}{\pi^3} \frac{2^{d-2}\vol(\SO(d-2))}{B_{\cP_{\cE}\cP_{\cE}\cP_{\de,0}}B_{\tl{\cP}_{\de,0}\cP_{\cE}\cP_{\de'}}}C_{3,3;\tl{\de}}D_{\tl{\de},3,\de'}\p{\<\cP_{\de}\cP_{\tl{\cE}}\cP_{\tl{\de'}}\>,\<\cP_{\tl{\de}}\cP_{\cE}\cP_{\de'}\>}.
\ee
Plugging in the explicit expressions using \eqref{eq:bubblecoeff_definition}, \eqref{eq:C0_m1} and \eqref{eq:Dcoeff}, we finally obtain (after setting $d=4$)
\be
&I^{(0)}_{\mathrm{strong}}(\de,0;\de') \nn \\
&= \frac{(1-\de') \G \left(3-\frac{\de}{2}\right) \G \left(\frac{\de}{2}\right)^2 \G \left(\frac{\de+4}{2}\right)  \G \left(\tfrac{-\de-\de'+5}{2}\right) \G \left(\tfrac{-\de+\de'-1}{2}\right) \G \left(\tfrac{-\de+\de'+3}{2}\right) \G \left(\tfrac{\de+\de'+1}{2}\right)}{4\pi^5 \G (1-\de) \G (\de-1) \G (\de'-1)}.
\ee
Consequently, $C^{(0)}_{\mathrm{strong}}(\de,0;\de')$ is given by
\be\label{eq:Cfuncstrong_0thorder}
&C^{(0)}_{\mathrm{strong}}(\de,0;\de') \nn \\
&=\frac{\G \left(3-\frac{\de}{2}\right) \G \left(\frac{\de}{2}\right)^2 \G \left(\frac{\de+4}{2}\right) \G \left(\tfrac{\de-\de'+3}{2}\right) \G \left(\tfrac{-\de+\de'+3}{2}\right) \G \left(\tfrac{\de+\de'-3}{2}\right) \G \left(\tfrac{\de+\de'+1}{2}\right)}{4\pi^3 \G (\de-1) \G (\de) \G (\de'-1)}.
\ee

To find the coefficients for the celestial block expansion, we have to close the $\de$ and $\de'$ contours in \eqref{eq:EEEC_celestialblockexpansion0}. It turns out that the resulting celestial block expansion does not depend on the order of contour deformations, so let us close the $\de'$ contour first for simplicity. When $\de$ is on the principal series, the only poles of $\de'$ that are to the right of the principal series are at $\de'=\de+3+2k$, where $k$ is a nonnegative integer. We then find that the residues
\be
\Res_{\de'=\de+3+2k}C^{(0)}_{\mathrm{strong}}(\de,0;\de')
\ee
have poles at $\de=6+2n$, where $n$ is a nonnegative integer. Thus, the values of $\de$ and $\de'$ appearing in the celestial block expansion \eqref{eq:strongEEEC_celestialblockexpansion1} should be
\be
\de=6+2n, \de'=9+2n+2k.
\ee
This agrees with our previous predictions \eqref{eq:strongcoupling_deltaprediction} and \eqref{eq:strongcoupling_deltaprimeprediction} from the light-ray OPE. The coefficients $P_{\de,j;\de'}$ are given by
\be\label{eq:Pcoeff_leading}
&P^{(0)}_{6+2n,0;9+2n+2k} \nn \\
&=\frac{ (k+1) (k+2) (-1)^{k+n} \G (n+3)^2 \G (n+5) \G (k+2 n+6) \G (k+2 n+8)}{\pi^3 \G (n+1) \G (2 n+5) \G (2 n+6) \G (2 (k+n+4))}.
\ee
As a consistency check, we can expand $\cF^{(0)}_{\mathrm{strong}}$ in the collinear limit and find the celestial block expansion order by order using the expansion of the celestial block $G^c$ in the collinear limit given by \eqref{eq:firstexpressionforcelestialbock} (or \eqref{eq:threeptcelestial_result_derivation1}). We have checked up to $O(\z_{13}^6)$ that the coefficients obtained in this way agree with \eqref{eq:Pcoeff_leading}.

\subsection{$O(1/\l)$ correction}
Let us now consider the $O(1/\l)$ term of \eqref{eq:EEEC_strong1},
\be
&\cF^{(1)}_{\mathrm{strong}}(z_1,z_2,z_3,p)\nn \\
&=\frac{48}{\pi(-2p\.z_1)^3(-2p\.z_2)^3(-2p\.z_3)^3} \p{2-4\p{\z_{12}+\z_{13}+\z_{23}}+4\p{\z_{12}^2+\z_{13}^2+\z_{23}^2}}.
\ee
Plugging this into the celestial inversion formula, we get
\be\label{eq:inversionformula_subleaing1}
&I^{(1)}_{\mathrm{strong}}(\de,j;\de') \nn \\
&=\frac{48}{\pi} \frac{2^{d-2}\vol(\SO(d-2))}{B_{\cP_{\cE}\cP_{\cE}\cP_{\de,j}}B_{\tl{\cP}_{\de,j}\cP_{\cE}\cP_{\de'}}}\x \int \frac{D^{d-2}zD^{d-2}z' D^{d-2}z_1 D^{d-2}z_2 D^{d-2}z_3 d^{d-1}_{\mathrm{AdS}}p}{\vol(\SO(d-1,1))}  \nn \\
&\quad\frac{\p{2-4\p{\z_{12}+\z_{13}+\z_{23}}+4\p{\z_{12}^2+\z_{13}^2+\z_{23}^2}}}{(-2z_1\.p)^3(-2z_2\.p)^3(-2z_3\.p)^3}\<\cP_{\tl{\cE}}(z_1)\cP_{\tl{\cE}}(z_2)\tl{\cP}_{\de,j}(z)\>\<\cP_{\de,j}(z)\cP_{\tl{\cE}}(z_3)\cP_{\tl{\de'}}(z')\>\frac{1}{(-2z'\.p)^{\de'}}.
\ee

There are two types of $z_1,z_2$ integrals to consider,\footnote{Note that in \eqref{eq:inversionformula_subleaing1} we implicitly contract the indices of $\tl{\cP}_{\de,j}(z)$ and $\cP_{\de,j}(z)$, and in \eqref{eq:z1z2integral_subleading} we contract the indices with a polarization vector $w$.}
\be\label{eq:z1z2integral_subleading}
& \int D^{d-2}z_1 D^{d-2}z_2 \frac{\z_{12}^k}{(-2z_1\.p)^3(-2z_2\.p)^3}\<\cP_{\tl{\cE}}(z_1)\cP_{\tl{\cE}}(z_2)\tl{\cP}_{\de,j}(z,w)\>, \nn \\
& \int D^{d-2}z_1 D^{d-2}z_2 \frac{\z_{13}^k}{(-2z_1\.p)^3(-2z_2\.p)^3}\<\cP_{\tl{\cE}}(z_1)\cP_{\tl{\cE}}(z_2)\tl{\cP}_{\de,j}(z,w)\>,
\ee
where $k=0,1,2$. Since the dependence of the integral on $w$ should be $([z,w]\.[\cdots])^j$, the first line is only nonzero when $j=0$ and the second one is nonzero when $j \leq k$. However, when $j=1$, the three-point structure $\<\cP_{\tl{\cE}}(z_1)\cP_{\tl{\cE}}(z_2)\tl{\cP}_{\de,j}(z,w)\>$ is antisymmetric in $1\leftrightarrow 2$, and the contribution from $\z_{13}^k$ and $\z_{23}^k$ will cancel. So, $I^{(1)}_{\mathrm{strong}}(\de,j;\de')$ is nonzero for $j=0$ and $j=2$.

\subsubsection{$j=0$}
We first consider the $j=0$ case. The integral containing $\z_{12}^k$ can be computed by applying the differential operator $\ptl_p\.\ptl_p$ $k$ times to \eqref{eq:j0_beforepair}. The result is
\be\label{eq:subleading_j0_z1z2integral_1}
 \int D^{d-2}z_1 D^{d-2}z_2 \frac{\z_{12}}{(-2z_1\.p)^3(-2z_2\.p)^3}\<\cP_{\tl{\cE}}(z_1)\cP_{\tl{\cE}}(z_2)\tl{\cP}_{\de}(z)\> 
&=\frac{(\de-6)(\tl{\de}-6)}{36}C^{(0)}_{3,3;\tl{\de}}(-2p\.z)^{-\tl{\de}},
\ee
and
\be\label{eq:subleading_j0_z1z2integral_2}
& \int D^{d-2}z_1 D^{d-2}z_2 \frac{\z_{12}^2}{(-2z_1\.p)^3(-2z_2\.p)^3}\<\cP_{\tl{\cE}}(z_1)\cP_{\tl{\cE}}(z_2)\tl{\cP}_{\de}(z)\> \nn \\
&=\frac{(\de-6)(\de-8)(\tl{\de}-6)(\tl{\de}-8)}{2304}C^{(0)}_{3,3;\tl{\de}}(-2p\.z)^{-\tl{\de}}.
\ee

To compute the second type of integral in \eqref{eq:z1z2integral_subleading}, which contains $\z_{13}^k$ or $\z_{23}^k$, we can first study
\be\label{eq:subleading_z1weightshifting}
z_1^{\mu}\<\cP_{\de_1}(z_1)\cP_{\de_2}(z_2)\cP_{\de}(z)\>.
\ee
The factor $z_1^{\mu}$ can be viewed as a weight-shifting operator \cite{Karateev:2017jgd} that decreases the scaling dimension of $\cP_{\de_1}(z_1)$ by $1$. One can performing crossing on \eqref{eq:subleading_z1weightshifting} to make the weight-shifting operators act on $\cP_{\de}(z)$. We find
\be\label{eq:subleading_z1crossing}
&(z_3\.z_1)\<\cP_{\de_1}(z_1)\cP_{\de_2}(z_2)\cP_{\de}(z)\>\nn \\
&=\frac{(\de-\de_1+\de_2)(4-d+\de-\de_1+\de_2)}{2(d-3-\de)(d-2-2\de)}(z_3\.z)\<\cP_{\de_1-1}(z_1)\cP_{\de_2}(z_2)\cP_{\de+1}(z)\> \nn \\
&+ \frac{1}{2-d+d\de-2\de^2}(z_3\.D_z)\<\cP_{\de_1-1}(z_1)\cP_{\de_2}(z_2)\cP_{\de-1}(z)\> \nn \\
&- \frac{\de-\de_1+\de_2}{2(d-4)(d-3-\de)(\de-1)}(z_3\.\cD^{0-}_{z,w})\<\cP_{\de_1-1}(z_1)\cP_{\de_2}(z_2)\cP_{\de,1}(z,w)\>,
\ee
where $D_z$ is the Todorov operator and $\cD^{0-}_{z,w}$ is the weight-shifting operator that decreases spin by 1 defined in \cite{Karateev:2017jgd}. As discussed below \eqref{eq:z1z2integral_subleading}, the $\cD^{0-}_{z,w}\<\cP_{\de_1-1}(z_1)\cP_{\de_2}(z_2)\cP_{\de,1}(z,w)\>$ term will vanish after integrating over $z_1,z_2$. 

After using \eqref{eq:subleading_z1crossing} and a similar relation for $\z_{23}$, we find
\be\label{eq:subleading_j0_z1z2integral_3}
& \int D^{d-2}z_1 D^{d-2}z_2 \frac{\z_{13}+\z_{23}}{(-2z_1\.p)^3(-2z_2\.p)^3}\<\cP_{\tl{\cE}}(z_1)\cP_{\tl{\cE}}(z_2)\tl{\cP}_{\de}(z)\> \nn \\
&= 2\int D^{d-2}z_1 D^{d-2}z_2 \frac{-2z_{3\mu}}{(-2z_1\.p)^4(-2z_2\.p)^3(-2z_3\.p)}\nn \\
&\qquad \p{\frac{\tl{\de}^2}{4(\tl{\de}-1)^2}z^{\mu}\<\cP_{\de_{\tl{\cE}}-1}(z_1)\cP_{\cE}(z_2)\cP_{\tl{\de}+1}(z)\> -\frac{1}{2(\tl{\de}-1)^2}D_z^{\mu}\<\cP_{\de_{\tl{\cE}}-1}(z_1)\cP_{\tl{\cE}}(z_2)\cP_{\tl{\de}-1}(z)\>} \nn \\
&=\frac{2}{-2z_3\.p}\p{\frac{\tl{\de}^2}{4(\tl{\de}-1)^2}(-2z_3\.z)C_{4,3;\tl{\de}+1}^{(0)}(-2p\.z)^{-\tl{\de}-1} -\frac{1}{2(\tl{\de}-1)^2}(-2z_3\.D_z)C_{4,3;\tl{\de}-1}^{(0)}(-2p\.z)^{-\tl{\de}+1} } \nn \\
&=-\frac{1}{3}C^{(0)}_{3,3;\tl{\de}}(-2p\.z)^{-\tl{\de}}\p{{\tfrac{\tl{\de}-6}{2}}-\tl{\de}\z_{03}}.
\ee
Similarly, for $\z_{13}^2+ \z_{23}^2$ we have
\be\label{eq:subleading_j0_z1z2integral_4}
& \int D^{d-2}z_1 D^{d-2}z_2 \frac{\z_{13}^2+\z_{23}^2}{(-2z_1\.p)^3(-2z_2\.p)^3}\<\cP_{\tl{\cE}}(z_1)\cP_{\tl{\cE}}(z_2)\tl{\cP}_{\de}(z)\> \nn \\
&=\frac{1}{96(\tl{\de}-3)}C^{(0)}_{3,3;\tl{\de}}(-2p\. z)^{-\tl{\de}}\left(\tl{\de} \left(\tl{\de}^2+6 (\tl{\de}-4) (\tl{\de}+2)\z_{03}^2-2 (\tl{\de}-6) (3 \tl{\de}-4)\z_{03}-18 \tl{\de}+104\right)-192\right).
\ee

Combining \eqref{eq:subleading_j0_z1z2integral_1}, \eqref{eq:subleading_j0_z1z2integral_2}, \eqref{eq:subleading_j0_z1z2integral_3}, \eqref{eq:subleading_j0_z1z2integral_4}, we obtain
\be
& \int D^{d-2}z_1 D^{d-2}z_2\ \cF^{(1)}_{\mathrm{strong}}(z_1,z_2,z_3,p)\<\cP_{\tl{\cE}}(z_1)\cP_{\tl{\cE}}(z_2)\tl{\cP}_{\de}(z)\> \nn \\
&=\frac{48}{\pi}\frac{(\tl{\de}-4) \tl{\de} (\tl{\de}+2)}{576 (\tl{\de}-3)}C^{(0)}_{3,3;\tl{\de}}(-2p\. z)^{-\tl{\de}}\left(\tl{\de}^2-5 \tl{\de}+30+144 (\z_{03}-1) \z_{03}\right).
\ee
Hence, the inversion formula \eqref{eq:inversionformula_subleaing1} for $j=0$ is now given by
\be
&I^{(1)}_{\mathrm{strong}}(\de,0;\de') \nn \\
&=\frac{48}{\pi} \frac{2^{d-2}\vol(\SO(d-2))}{B_{\cP_{\cE}\cP_{\cE}\cP_{\de,j}}B_{\tl{\cP}_{\de,j}\cP_{\cE}\cP_{\de'}}}\frac{(\tl{\de}-4) \tl{\de} (\tl{\de}+2)}{576 (\tl{\de}-3)}C^{(0)}_{3,3;\tl{\de}}\x \int \frac{D^{d-2}zD^{d-2}z'  D^{d-2}z_3 d^{d-1}_{\mathrm{AdS}}p}{\vol(\SO(d-1,1))}  \nn \\
&\quad\frac{1}{(-2p\. z)^{\tl{\de}}(-2p\.z_3)^{3}}\left(\tl{\de}^2-5 \tl{\de}+30+144 (\z_{03}-1) \z_{03}\right)\<\cP_{\de}(z)\cP_{\tl{\cE}}(z_3)\cP_{\tl{\de'}}(z')\>\frac{1}{(-2z'\.p)^{\de'}}.
\ee
The integral over $p$ can be evaluated using \eqref{eq:Witten_3pt}. The result is
\be
&I^{(1)}_{\mathrm{strong}}(\de,0;\de') \nn \\
&=6\pi^2 A \frac{2^{d-2}\vol(\SO(d-2))}{B_{\cP_{\cE}\cP_{\cE}\cP_{\de,j}}B_{\tl{\cP}_{\de,j}\cP_{\cE}\cP_{\de'}}}\frac{(\tl{\de}-4) \tl{\de} (\tl{\de}+2)}{576 (\tl{\de}-3)}C^{(0)}_{3,3;\tl{\de}}D_{\tl{\de},3,\de'}   \nn \\
&\x\frac{\frac{3}{4} (\tl{\de}-\de'+3) (\tl{\de}-\de'+5) (\tl{\de}+\de'+1) (\tl{\de}+\de'+3)+(\tl{\de}+1) ((\tl{\de}-18) \tl{\de} (\tl{\de}+1)+12 (\de'-3) (\de'+1))}{\tl{\de} (\tl{\de}+1)} \nn \\
&\x \p{\<\cP_{\de}\cP_{\tl{\cE}}\cP_{\tl{\de'}}\>,\<\cP_{\tl{\de}}\cP_{\cE}\cP_{\de'}\>},
\ee
which implies that
\be
&C^{(1)}_{\mathrm{strong}}(\de,0;\de') \nn \\
&= \left(\de^2 \left(-6 \de'^2+12 \de'+10\right)+7 \de^4-28 \de^3+12 \de \left(\de'^2-2 \de'+3\right)+3 (\de'-1)^2 \left(\de'^2-2 \de'-3\right)\right) \nn \\
&\x(\de-4) (\de-1) (\de+2) (\de'-1)\frac{\G \left(3-\frac{\de}{2}\right) \G \left(\frac{\de}{2}\right)^2 \G \left(\frac{\de+4}{2}\right) \G \left(\tfrac{\de-\de'+3}{2}\right) \G \left(\tfrac{-\de+\de'+3}{2}\right) \G \left(\tfrac{\de+\de'-3}{2}\right) \G \left(\tfrac{\de+\de'+1}{2}\right)}{1536 \pi  (\de-3) (\de+1) \G (\de)^2 \G (\de')}.
\ee

We see that the Gamma functions are the same as the leading order result $C^{(0)}_{\mathrm{strong}}$, and hence the locations of the poles are the same. In particular, we only get poles at
\be
\de=6+2n,\quad \de'=9+2n+2k,
\ee
where $n,k$ are nonnegative integers. The coefficients $P^{(1)}_{\de,0;\de'}$ are given by the residues,
\be
&P^{(1)}_{6+2n,0;9+2n+2k} \nn \\
&= \left(3 k^2 \left(4 n^2+38 n+83\right)+12 k^3 (n+4)+3 k^4+6 k \left(6 n^2+43 n+76\right)+4 n^4+40 n^3+169 n^2+381 n+378\right)\nn \\
&\x\frac{ (-1)^{k+n} (n+1) (n+4) (2 n+5)(k+n+4) \G (k+3) \G (n+3)^2 \G (n+5) \G (k+2 n+6) \G (k+2 n+8)}{3\pi (2 n+3) (2 n+7) \G (k+1) \G (n+1) \G (2 n+6)^2 \G (2 k+2 n+9)}.
\ee
We have checked that the coefficients obtained by expanding $\cF^{(1)}_{\mathrm{strong}}$ in the collinear limit up to $O(\z_{13}^6)$ agree with the above expression.

\subsubsection{$j=2$}
Now let us consider the $j=2$ case. From the discussion below \eqref{eq:z1z2integral_subleading}, we must only consider the $z_1,z_2$-integral
\be
& \int D^{d-2}z_1 D^{d-2}z_2 \frac{\z_{13}^2+\z_{23}^2}{(-2z_1\.p)^3(-2z_2\.p)^3}\<\cP_{\tl{\cE}}(z_1)\cP_{\tl{\cE}}(z_2)\tl{\cP}_{\de,j=2}(z,w)\>. 
\ee
We can again use crossing equations for weight-shifting operators. Only terms with a $j=0$ three-point structure will be nonvanishing after we integrate over $z_1$ and $z_2$. There is only one such term for $(z_3\.z_1)^2\<\cP_{\de_1}\cP_{\de_2}\cP_{\de,2}(z,w)\>$, and its $6j$ symbol is given by\footnote{Our convention here is $\<\cP_{\de_1}\cP_{\de_2}\cP_{\de,j}(z,w)\>= \frac{(2z\.z_1w\.z_2-2z\.z_2w\.z_1)^j}{(-2z_1\.z_2)^{\frac{\de_1+\de_2-\de+j}{2}}(-2z_1\.z)^{^{\frac{\de_1+\de-\de_2+j}{2}}}(-2z_2\.z)^{\frac{\de_2+\de-\de_1+j}{2}}}$.}
\be
&(z_3\.z_1)^2\<\cP_{\de_1}\cP_{\de_2}\cP_{\de,2}(z,w)\> \nn \\
&=\frac{(\de-\de_1+\de_2-d+2) (\de-\de_1+\de_2-d+4)}{2 (d-2) d \de (\de+1) (\de-d+1) (\de-d+2)}(z_3\.\cD^{0+}_{z,w})\<\cP_{\de_1-2}(z_1)\cP_{\de_2}(z_2)\cP_{\de}(z)\> + \ldots.
\ee
Our $z_1,z_2$ integral is thus given by
\be
& \int D^{d-2}z_1 D^{d-2}z_2 \frac{\z_{13}^2+\z_{23}^2}{(-2z_1\.p)^3(-2z_2\.p)^3}\<\cP_{\tl{\cE}}(z_1)\cP_{\tl{\cE}}(z_2)\tl{\cP}_{\de,2}(z,w)\> \nn \\
&=\frac{\tl{\de}}{2(\tl{\de}-3)}C^{(0)}_{5,3;\tl{\de}}\frac{(2z\.p w\.z_3-2z\.z_3 w\.p)^2}{(-2z\.p)^{\tl{\de}+2}(-2z_3\.p)^2}.
\ee
Plugging this into the inversion formula \eqref{eq:inversionformula_subleaing1} for $j=2$, we find
\be
&I^{(1)}_{\mathrm{strong}}(\de,2;\de') \nn \\
&=\frac{48}{\pi} \frac{2^{d-2}\vol(\SO(d-2))}{B_{\cP_{\cE}\cP_{\cE}\cP_{\de,2}}B_{\tl{\cP}_{\de,2}\cP_{\cE}\cP_{\de'}}}\frac{\tl{\de}}{2(\tl{\de}-3)}C^{(0)}_{5,3;\tl{\de}}\x \int \frac{D^{d-2}zD^{d-2}z'  D^{d-2}z_3 d^{d-1}_{\mathrm{AdS}}p}{\vol(\SO(d-1,1))}  \nn \\
&\quad\frac{(2z\.p)^2z_{3\mu}z_{3\nu}-2(2z\.p)(2z\.z_3)z_{3\mu}p_{\nu}+(2z\.z_3)^2p_{\mu}p_{\nu}}{(-2z\.p)^{\tl{\de}+2}(-2z_3\.p)^5}\<\cP^{\mu\nu}_{\de,2}(z)\cP_{\tl{\cE}}(z_3)\cP_{\tl{\de'}}(z')\>\frac{1}{(-2z'\.p)^{\de'}}.
\ee

To perform the $p$ integral, we can view $p_{\mu}$ as a linear combination of the AdS weight-shifting operators \cite{Costa:2018mcg} and then perform crossing. We find
\be
p_{\mu}(-2p\.z)^{-\de}=\frac{1}{(\de-1)(d-2-2\de)}D_{z\mu}(-2p\.z)^{-\de+1} + \frac{2\de}{2\de-d+2}z_{\mu}(-2p\.z)^{-\de-1}.
\ee
After using this relation, the integral over $p$ is elementary. The result is
\be
&I^{(1)}_{\mathrm{strong}}(\de,2;\de') \nn \\
&=\frac{48}{\pi} \frac{2^{d-2}\vol(\SO(d-2))}{B_{\cP_{\cE}\cP_{\cE}\cP_{\de,2}}B_{\tl{\cP}_{\de,2}\cP_{\cE}\cP_{\de'}}}\frac{(\tl{\de}+\de'-5)(\tl{\de}+\de'-3)}{8(\tl{\de}+1)(\tl{\de}-3)}C^{(0)}_{5,3;\tl{\de}}D_{\tl{\de},5,\de'}\p{\<\cP_{\cE}\cP_{\de'}\tl{\cP}_{\de,2}\>,\<\tl{\cP}_{\cE}\tl{\cP}_{\de'}\cP_{\de,2}\>}.
\ee
Collecting together all the factors, we finally obtain
\be
&C^{(1)}_{\mathrm{strong}}(\de,2;\de') \nn \\
&=\frac{\G \left(4-\frac{\de}{2}\right) \G \left(\frac{\de}{2}\right) \G \left(\frac{\de+2}{2}\right) \G \left(\frac{\de+6}{2}\right) \G \left(\tfrac{\de-\de'+5}{2}\right) \G \left(\tfrac{-\de+\de'+5}{2}\right) \G \left(\tfrac{\de+\de'-1}{2}\right) \G \left(\tfrac{\de+\de'+3}{2}\right)}{24 \pi  \G (\de) \G (\de+2) \G (\de'-1)}.
\ee

After contour deformations, the only poles that will contribute to the celestial block expansion are at
\be
\de=8+2n,\qquad \de'=\de+5+2k=13+2n+2k.
\ee
The corresponding coefficients are given by
\be
&P^{(1)}_{8+2n,2;13+2n+2k} \nn \\
&=\frac{(-1)^{k+n} \G (k+5) \G (n+5) \G (n+7) \G (n+4) \G (k+2 n+10) \G (k+2 n+12)}{6 \pi  \G (k+1) \G (n+1) \G (2 (n+5)) \G (2 n+8) \G (2 (k+n+6))}.
\ee
We have checked that this result agrees with the expansion of $\cF^{(1)}_{\mathrm{strong}}$ in the collinear limit up to $O(\z_{13}^6)$. The validity of these expressions for the celestial block expansion of the strong-coupling EEEC is a strong check on both our expressions for the collinear expansion of celestial blocks \eqref{eq:firstexpressionforcelestialbock}, and the celestial inversion formula (\ref{eq:inversionformula_1}).

\section{Discussion and future directions}\label{sec:discussion}

In this work, we studied aspects of the celestial block decomposition of the three-point energy correlator (EEEC). We found that, in both strong and weak coupling examples, the EEEC admits an expansion in a discrete sum of celestial blocks, corresponding to light-ray operators appearing in repeated OPEs of energy detectors. We derived useful formulas for 3-point celestial blocks in an expansion around the collinear limit. We then explored the celestial block expansion in the collinear limit using lightcone bootstrap techniques and the Lorentzian inversion formula, both in QCD and $\cN=4$ SYM. The symmetry structure of the celestial block expansion allowed us to make certain predictions for higher orders in perturbation theory. We also determined the leading and first subleading contact terms in the EEEC in $\cN=4$ SYM. Finally, using techniques from harmonic analysis, we studied the full celestial block decomposition of the strong-coupling EEEC in $\cN=4$ SYM, for generic configurations of detectors. Along the way, we encountered several puzzles and novel objects that we summarize below.

This celestial block expansion must be compatible with crossing symmetry, which leads to constraints similar to those studied in the traditional conformal bootstrap.  It would be interesting to fully characterize the consistency conditions that multi-point energy correlators should satisfy, in order to set up a direct bootstrap program for them (and more general event shapes). It would also be interesting to incorporate ``generalized detectors" \cite{Korchemsky:2021okt,Korchemsky:2021htm}.

In the collinear EEEC, the density matrix $|\Psi\>\<\Psi|$ in which we evaluate the event shape gets highly boosted, and essentially projected onto its component with boost eigenvalue $\de_*'$ (the dimension of the lowest-twist spin-4 operator). These highly-boosted density matrices, which we might call ``light-ray density matrices," are naturally dual to light-ray operators, via taking expectation values. It is interesting to ask whether they provide a useful basis for studying other physical observables. It may also be interesting to explore information-theoretic properties of light-ray density matrices.

Our lightcone bootstrap analysis of collinear event shapes reveals the existence of two important contributions to the light-ray OPE: double-twist operators built out of a pair of detectors $[\cP_\cO \cP_\cO]$, and double-twist objects $[\cP_\cO \cP_{\cO'_*}]$ that formally look like an OPE between a light-ray operator and the light-ray density matrix obtained from boosting $|\Psi\>\<\Psi|$. This suggests that it may be possible more generally to make sense of an OPE between light-ray operators and light-ray density matrices.

In our analysis of the EEEC in QCD, we used the fact that QCD admits a conformal point at a particular value of the coupling in $d=4-\e$ dimensions. The existence of this conformal point has implications for the structure of perturbation theory at each loop order, even away from the conformal point, allowing us to apply selection rules from conformal symmetry in our work. However, the existence of a celestial block expansion should require only Lorentz invariance. It would be interesting to characterize how the structure of the space of light-ray operators and light-ray OPEs changes in the presence of a nonzero $\beta$-function.

In applying the Lorentzian inversion formula to the collinear EEEC, we were led to continue the EEEC into a ``doubly-Lorentzian" regime where the cross-ratios on the celestial sphere become independent real numbers. (This same analytic continuation to (a cover of) the celestial torus is frequently used in studying celestial amplitudes \cite{Atanasov:2021oyu}.) Although this continuation is straightforward at the order in perturbation theory we studied, we have little understanding of whether it is admissible at higher orders or nonperturbatively. Another important question is how event shapes behave in the ``celestial Regge limit" where the celestial cross ratios undergo the analytic continuation usually studied in the context of the Regge limit in CFT \cite{Cornalba:2006xm}. The two examples we studied (weakly-coupled $\cN=4$ SYM and QCD) exhibited very different behavior in this regime, and it would be interesting to understand what the general nonperturbative behavior can be.

Evidently, Lorentzian QFT observables still hold many mysteries. To better understand their structure, it will be important to study more examples and collect more data both in perturbation theory and beyond.

\section*{Acknowledgements}
We thank Hao Chen, Lance Dixon, Murat Kologlu, Petr Kravchuk, Ian Moult, Joshua Sandor, Sasha Zhiboedov, and Hua Xing Zhu for discussions. We thank Hao Chen, Ian Moult, and Hua Xing Zhu for sharing their Mathematica notebook with results for the weak-coupling collinear EEEC with us. CHC and DSD are
supported by Simons Foundation grant 488657 (Simons Collaboration on the Nonperturbative
Bootstrap) and a DOE Early Career Award under grant no. DE-SC0019085.
\newpage

\appendix

\section{Alternative derivation of the three-point celestial block}\label{app:celestialblock}
In this section we give an alternative derivation of an expression for the three-point celestial block
\be
\cC_{12\cP_{\de,j}}(z_1,z_2,\ptl_{z_2},\ptl_{w_2})\cC_{\cP_{\de,j}3\cP_{\de'}}(z_2,w_2,z_3,\ptl_{z_3})(-2p\.z_3)^{-\de'}.
\ee
Lorentz invariance and homogeneity imply that\footnote{We define $[a,b]\.[c,d]=2[ (a\.c) (b\.d) -(a\.d) (b\.c)]$}
\be
&\cC_{\cP_{\de,j}3\cP_{\de'}}(z_2,z_3,\ptl_{z_3})(-2p\.z_3)^{-\de'}=\frac{(-[z_2,w_2]\.[z_3,p])^j(-p^2)^{\frac{\de+\de_3+j-\de'}{2}}}{(-2p\.z_2)^{\de+j}(-2p\.z_3)^{\de_3+j}}f'_{\de'}(\z_{23}).
\ee
From the definition of $\cC_{\cP_{\de,j}3\cP_{\de'}}$, the leading term of $f'_{\de'}(\z_{23})$ should be given by
\be
f'_{\de'}(\z_{23})=\z_{23}^{\frac{\de'-\de-j-\de_3}{2}}\p{1+O(\zeta_{23})}.
\ee
Solving the Casimir equation with this boundary condition, we find
\be\label{eq:2ptcelestialblock_spinj}
f'_{\de'}(\z_{23})=\z_{23}^{\frac{\de'-\de-j-\de_3}{2}}{}_2F_1\p{\tfrac{\de'-\de_3+\de+j}{2},\tfrac{\de'+\de_3-\de+j}{2},\de'+2-\tfrac{d}{2},\z_{23}}.
\ee
This is the two-point celestial block where one of the external operators has nonzero $j$, generalizing the result derived in \cite{Kologlu:2019mfz}.

Now, we must compute
\be\label{eq:3ptcelestialblock_C12P}
\cC_{12\cP_{\de,j}}(z_1,z_2,\ptl_{z_2},\ptl_{w_2})\p{\frac{(-[z_2,w_2]\.[z_3,p])^j(-p^2)^{\frac{\de+\de_3+j-\de'}{2}}}{(-2p\.z_2)^{\de+j}(-2p\.z_3)^{\de_3+j}}f'_{\de'}(\z_{23})}.
\ee
Expanding the two-point celestial block $f'_{\de'}(\z_{23})$ given in~\eqref{eq:2ptcelestialblock_spinj}, we can rewrite~\eqref{eq:3ptcelestialblock_C12P} as
\be
\sum_{n=0}^{\oo}\frac{\p{\tfrac{\de'-\de_3+\de+j}{2}}_n\p{\tfrac{\de'+\de_3-\de+j}{2}}_n}{\p{\de'+2-\tfrac{d}{2}}_n n!}(-p^2)^{n}\cC_{12\cP_{\de,j}}(z_1,z_2,\ptl_{z_2},\ptl_{w_2})\p{\<\cP_{\de,j}(z_2,w_2)\cP_{\de_3}(z_3)\cP_{\de'+2n}(p)\>},
\ee
where
\be
\<\cP_{\de,j}(z_2,w_2)\cP_{\de_3}(z_3)\cP_{\de'+2n}(p)\>=\frac{(-[z_2,w_2]\.[z_3,p])^j}{(-2p\.z_2)^{\frac{\de'+\de+j-\de_3}{2}+n}(-2p\.z_3)^{\frac{\de'+\de_3+j-\de}{2}+n}(-2z_2\.z_3)^{\frac{\de+j+\de_3-\de'}{2}-n}}.
\ee
Note that the function $\<\cP_{\de,j}(z_2,w_2)\cP_{\de_3}(z_3)\cP_{\de'+2n}(p)\>$ is not a conformally invariant three-point function since $p$ is not null, and therefore $\cC_{12\cP_{\de,j}}$ acting on it does not give a conformal block. To compute the action of $\cC_{12\cP_{\de,j}}$ on $\<\cP_{\de,j}(z_2,w_2)\cP_{\de_3}(z_3)\cP_{\de'+2n}(p)\>$, we would like to express it as an expansion in conformal three-point functions in the limit that $p$ becomes null.  To do so, we express $p$ as a linear combination of two null vectors $p=z_0+v$. In the limit $v\to 0$, $p$ approaches the point $z_0$. We define the null vectors by
\be
z_0^{\mu}=p^{\mu}-\frac{p^2}{2p\.z_1}z_1^{\mu},\quad v^{\mu}=\frac{p^2}{2p\.z_1}z_1^{\mu},
\ee
essentially using $z_1$ as a reference direction. Consequently, subsequent expressions will not be manifestly symmetric with respect to $1\leftrightarrow 2$, but the symmetry will be restored in the final answer.
Note that $z_0$ and $v$ satisfy
\be
z_0^2=v^2=0,~p=z_0+v,~ -p^2=-2z_0\.v.
\ee
The cross ratios can then be written
\be
\z_{12}=\frac{-2z_2\.v}{-2z_2\.(z_0+v)},\quad \z_{13}=\frac{-2z_3\.v}{-2z_3\.(z_0+v)}, \quad \z_{23}=\frac{(-2z_2\.z_3)(-2z_0\.v)}{(-2z_2\.(z_0+v))(-2z_3\.(z_0+v))}.
\ee

In these variables, the expansion around the collinear limit is an expansion in small $v$. We can write the three-point celestial block as
\be
\sum_{n=0}^{\oo}\frac{\p{\tfrac{\de'-\de_3+\de+j}{2}}_n\p{\tfrac{\de'+\de_3-\de+j}{2}}_n}{\p{\de'+2-\tfrac{d}{2}}_n n!}(-2z_0\.v)^{n}\cC_{12\cP_{\de,j}}(z_1,z_2,\ptl_{z_2},\ptl_{w_2})\p{\<\cP_{\de,j}(z_2,w_2)\cP_{\de_3}(z_3)\cP_{\de'+2n}(z_0+v)\>}.
\ee
We find that the function $\<\cP_{\de,j}(z_2,w_2)\cP_{\de_3}(z_3)\cP_{\de'+2n}(z_0+v)\>$ has the following expansion:
\be
&\<\cP_{\de,j}(z_2,w_2)\cP_{\de_3}(z_3)\cP_{\de'+2n}(z_0+v)\>\nn \\
&=\sum_{m=0}^{\oo}\sum_{k=0}^m c_{m,k}(-2v\.z_0)^k (v\.D_{z_0})^{m-k}\<\cP_{\de,j}(z_2,w_2)\cP_{\de_3}(z_3)\cP_{\de'+2n+2k}(z_0)\>,
\ee
where $D_{z_0}$ is the Todorov operator acting on $z_0$.  The coefficients $c_{m,k}$ can be determined by expanding both sides of the above equation order by order in small $v$:
\be
c_{m,k}=\frac{\p{\frac{j+2n+\de-\de_3+\de'}{2}}_k\p{\frac{j+2n-\de+\de_3+\de'}{2}}_k}{(m-k)!k!}\frac{\p{\frac{d}{2}-(2k+1+2n+\de')}\G\p{\frac{d}{2}-(m+k+1+2n+\de')}}{\G\p{\frac{d}{2}-(k+2n+\de')}}.
\ee

Since the Todorov operator $D_{z_0}$ commutes with $\cC_{12\cP_{\de,j}}$, the three-point celestial block is therefore given by
\be
&\cC_{12\cP_{\de,j}}(z_1,z_2,\ptl_{z_2},\ptl_{w_2})\p{\frac{(-[z_2,w_2]\.[z_3,p])^j(-p^2)^{\frac{\de+\de_3+j-\de'}{2}}}{(-2p\.z_2)^{\de+j}(-2p\.z_3)^{\de_3+j}}f'_{\de'}(\z_{23})} \nn \\
&=\sum_{n=0}^{\oo}\sum_{m=0}^{\oo}\sum_{k=0}^m\frac{\p{\tfrac{\de'-\de_3+\de+j}{2}}_n\p{\tfrac{\de'+\de_3-\de+j}{2}}_n}{\p{\de'+2-\tfrac{d}{2}}_n n!}c_{m,k}(-2v\.z_0)^{n+k} (v\.D_{z_0})^{m-k} g_{\de,j}^{(\de_1,\de_2,\de_3,\de'+2n+2k)}(z_1,z_2,z_3,z_0) \nn \\
&=\sum_{n=0}^{\oo}\sum_{m=0}^{\oo}\sum_{k=0}^m B_{n,m,k}(-2v\.z_0)^{n+k} (v\.D_{z_0})^{m-k} g_{\de,j}^{(\de_1,\de_2,\de_3,\de'+2n+2k)}(z_1,z_2,z_3,z_0),
\label{eq:horriblesum}
\ee
where $g_{\de,j}^{(\de_1,\de_2,\de_3,\de'+2n+2k)}(z_1,z_2,z_3,z_0)$ is a usual four-point conformal block, and
\be\label{eq:Bcoeff_definition}
B_{n,m,k}=\frac{\p{\frac{j+\de-\de_3+\de'}{2}}_{n+k}\p{\frac{j-\de+\de_3+\de'}{2}}_{n+k}}{\p{\de'+2-\tfrac{d}{2}}_nn!(m-k)!k!}\frac{\p{\frac{d}{2}-(2k+1+2n+\de')}\G\p{\frac{d}{2}-(m+k+1+2n+\de')}}{\G\p{\frac{d}{2}-(k+2n+\de')}}.
\ee
The expression (\ref{eq:horriblesum}) can be further simplified. To do so, we redefine $n'\equiv n+k, N\equiv n+m$, which gives
\be
&\sum_{n=0}^{\oo}\sum_{m=0}^{\oo}\sum_{k=0}^m B_{n,m,k}(-2v\.z_0)^{n+k} (v\.D_{z_0})^{m-k} g_{\de,j}^{(\de_1,\de_2,\de_3,\de'+2n+2k)}(z_1,z_2,z_3,z_0) \nn \\
=&\sum_{N=0}^{\oo}\sum_{n'=0}^{N}\sum_{k=0}^{n'}B_{n'-k,N-n'+k,k}(-2v\.z_0)^{n'}(v\.D_{z_0})^{N-n'} g_{\de,j}^{(\de_1,\de_2,\de_3,\de'+2n')}(z_1,z_2,z_3,z_0).
\ee
It turns out that the coefficients $B_{n,m,k}$ given in \eqref{eq:Bcoeff_definition} satisfy
\be
\sum_{k=0}^{n'}B_{n'-k,N-n'+k,k} =B_{0,N,0}\de_{n',0}.
\ee
Therefore, two of the sums in (\ref{eq:horriblesum}) collapse, and the three-point celestial block becomes
\be\label{eq:threeptcelestial_result_derivation1}
&\cC_{12\cP_{\de,j}}(z_1,z_2,\ptl_{z_2},\ptl_{w_2})\p{\frac{(-[z_2,w_2]\.[z_3,p])^j(-p^2)^{\frac{\de+\de_3+j-\de'}{2}}}{(-2p\.z_2)^{\de+j}(-2p\.z_3)^{\de_3+j}}f'_{\de'}(\z_{23})} \nn \\
&=\sum_{N=0}^{\oo}\frac{\G(\tfrac{d}{2}-1-\de'-N)}{\G(\tfrac{d}{2}-1-\de')N!}(v\.D_{z_{0}})^{N}g_{\de,j}^{(\de_1,\de_2,\de_3,\de')}(z_1,z_2,z_3,z_0)  \nn \\
&=\G(\de'+2-\tfrac{d}{2})(\sqrt{v\.D_{z_0}})^{\tfrac{d}{2}-\de'-1}J_{\de'+1-\tfrac{d}{2}}\p{2\sqrt{v\.D_{z_0}}}g_{\de,j}^{(\de_1,\de_2,\de_3,\de')}(z_1,z_2,z_3,z_0).
\ee
This expression is more manifestly Lorentz-invariant than \eqref{eq:firstexpressionforcelestialbock}, at the cost of singling out $z_1$ as special.

It is convenient to write the result in terms of conformally-invariant cross ratios. Before doing so, we relabel the vector $v \to v'$ to avoid confusion with the cross-ratio $v$. Factoring out the homogeneity factor in the conformal block, we find (up to linear order in $v'$, for simplicity)
\be
&\cC_{12\cP_{\de,j}}(z_1,z_2,\ptl_{z_2},\ptl_{w_2})\p{\frac{(-[z_2,w_2]\.[z_3,p])^j(-p^2)^{\frac{\de+\de_3+j-\de'}{2}}}{(-2p\.z_2)^{\de+j}(-2p\.z_3)^{\de_3+j}}f'_{\de'}(\z_{23})} \nn \\
&=\p{1+ \frac{1}{\tfrac{d}{2}-2-\de'}(v'\.D_{z_0})+\cdots}T_{123\de'}(z_1,z_2,z_3,z_0)g_{\de,j}^{(123\de')}\p{u,v},
\ee
where $u,v$ and $T_{123\de'}$ are defined in \eqref{eq:uandv_definition} and \eqref{eq:homogeneityfactor} (with $p$ replaced by $z_0$). Finally, we can replace $z_0 \to p-\tfrac{p^2}{2p\.z_1}z_1, v' \to \tfrac{p^2}{2p\.z_1}z_1$. After expanding in $(-p^2)$, we obtain \eqref{eq:threept_celestialblock_crossratios} in the main text.

We can also derive the same result from \eqref{eq:firstexpressionforcelestialbock}. To compute the celestial block using \eqref{eq:AdSstate_expansion}, we start with the conformal block
\be
T_{123\de'}(z_1,z_2,z_3,z_0)g_{\de,j}^{(123\de')}\p{\frac{(-2z_1\.z_2)(-2z_3\. z_0)}{(-2z_1\.z_3)(-2z_2\. z_0)},\frac{(-2z_2\.z_3)(-2z_1\. z_0)}{(-2z_1\.z_3)(-2z_2\. z_0)}},
\ee
and go to the frame where $z_0$ is near the origin and $z_{1}, z_2, z_3$ are near infinity. Explicitly, let us set
\be\label{eq:123infinityframe}
z_0 = (1,\vec y^2, \vec y), \quad z_i = (\l^2 \vec y_i^2, 1, \l \vec y_i),
\ee
where $i=1,2,3$, and $\l$ is an expansion parameter. The first two terms, for example, from expanding out the Bessel function in \eqref{eq:firstexpressionforcelestialbock} then become
\be
&\p{1-\frac{\ptl_{\vec y}^2}{4(\de'+2-\frac{d}{2})}}\left.\p{T_{123\de'}(z_1,z_2,z_3,z_0)g_{\de,j}^{(123\de')}\p{\frac{(-2z_1\.z_2)(-2z_3\. z_0)}{(-2z_1\.z_3)(-2z_2\. z_0)},\frac{(-2z_2\.z_3)(-2z_1\. z_0)}{(-2z_1\.z_3)(-2z_2\. z_0)}}}\right|_{\vec y_0\to 0}
\nn\\
&=\left.T_{123\de'}(z_1,z_2,z_3,p)\bigg( g_{\de,j}^{(123\de')}(u,v) +\frac{\z_{13}}{d-4-2\de'}\cD_{u,v}^{(1)}g_{\de,j}^{(123\de')}(u,v)\bigg)\right|_{\substack {p\to (1,1,0) \\ z_i = (\l^2 \vec y_i^2, 1, \l \vec y_i)}} + O(\l^4),
\ee
which agrees with  \eqref{eq:threept_celestialblock_crossratios}. More generally, when acting on the appropriate class of functions,  $v'\.D_{z_0}=\ptl_{\vec y}^2/4$, so \eqref{eq:firstexpressionforcelestialbock} and \eqref{eq:threeptcelestial_result_derivation1} are equivalent.

\section{$\<R^{(1)}_{\de,j}\>$ and $\<R^{(0)}_{\de,j}\g^{(1)}_{\de,j}\>$ from direct decomposition}\label{app:Rcoeff_directdecomposition}
In this appendix, we give the coefficients $\<R^{(1)}_{\de,j}\>$ and $\<R^{(0)}_{\de,j}\g^{(1)}_{\de,j}\>$ up to $\de=12$ for $\cN=4$ SYM, the QCD gluon jet, and the QCD quark jet obtained by expanding the collinear EEEC order-by-order in small $r$, as in section \ref{sec:Rcoeff_directdecomposition}.
\subsection{$\cN=4$ SYM}
\be\label{eq:Rcoeff_N4SYM_directdecomposition2}
&\<R^{(1)}_{10,0}\>=\tfrac{141301}{352800}-\tfrac{2 \pi ^2}{49},&& \<R^{(1)}_{10,2}\>= \tfrac{107129}{529200}-\tfrac{19 \pi ^2}{1260}, \nn \\
&\<R^{(1)}_{10,4}\>= \tfrac{33394601}{85377600}-\tfrac{5 \pi ^2}{132},&& \<R^{(1)}_{10,6}\>= \tfrac{189283}{1801800}-\tfrac{3 \pi ^2}{286}, \nn \\
&\<R^{(1)}_{12,0}\>= \tfrac{84401}{3175200}-\tfrac{\pi ^2}{378},&& \<R^{(1)}_{12,2}\>=\tfrac{547098707}{5378788800}-\tfrac{9 \pi ^2}{1232}, \nn \\
&\<R^{(1)}_{12,4}\>= \tfrac{2674437767}{81162081000}-\tfrac{16 \pi ^2}{6435},&& \<R^{(1)}_{12,6}\>= \tfrac{1220098669}{19675656000}-\tfrac{7 \pi ^2}{1144}, \nn \\
&\<R^{(1)}_{12,8}\>= \tfrac{1030567}{68612544}-\tfrac{\pi ^2}{663}, \nn \\
&\<R^{(0)}_{10,0}\g^{(1)}_{10,0}\> = \tfrac{89}{210},&& \<R^{(0)}_{10,2}\g^{(1)}_{10,2}\> = \tfrac{8}{105},&& \<R^{(0)}_{10,4}\g^{(1)}_{10,4}\> = -\tfrac{1}{154}, \nn \\
&\<R^{(0)}_{12,0}\g^{(1)}_{12,0}\> = \tfrac{253}{630} ,&& \<R^{(0)}_{12,2}\g^{(1)}_{12,2}\> = \tfrac{28247}{194040} ,&& \<R^{(0)}_{12,4}\g^{(1)}_{12,4}\> = \tfrac{893}{90090}, \nn \\
&\<R^{(0)}_{12,6}\g^{(1)}_{12,6}\> = -\tfrac{1}{1560}.
\ee
\subsection{QCD}
For the gluon jet, we find
\be\label{eq:Rcoeff_QCD_gluon_directdecomposition2}
&\<R^{(1)g}_{8,0}\>= \tfrac{-2 \left(56889000 \pi ^2-563610307\right) C_A^2+2 \left(39690000 \pi ^2-399089339\right) C_A n_f T_F-3108375 C_F n_f T_F}{317520000}, \nn \\
&\<R^{(1)g}_{8,2}\>= \tfrac{C_A \left(\left(241456351-24418800 \pi ^2\right) C_A+2 \left(11207700 \pi ^2-110698537\right) n_f T_F\right)}{50803200}, \nn \\
&\<R^{(1)g}_{8,4}\>= \tfrac{\left(1713863-173600 \pi ^2\right) C_A^2+2 \left(107800 \pi ^2-1063963\right) C_A n_f T_F+120 C_F n_f T_F}{403200}, \nn \\
&\<R^{(1)g}_{10,0}\>= \tfrac{\left(4851797956-487317600 \pi ^2\right) C_A^2+2 \left(251143200 \pi ^2-2512589293\right) C_A n_f T_F+7121499 C_F n_f T_F}{1303948800}, \nn \\
&\<R^{(1)g}_{10,2}\>= \tfrac{\left(104601961181-10560904200 \pi ^2\right) C_A^2}{23051952000}+\tfrac{\left(13728260700 \pi ^2-136102341667\right) C_A n_f T_F}{23051952000} \nn \\
&\qquad\qquad -\tfrac{389 C_F n_f T_F}{1411200}, \nn \\
&\<R^{(1)g}_{10,4}\>= \tfrac{C_A \left(\left(31701145719-3211362000 \pi ^2\right) C_A+14 \left(336659400 \pi ^2-3323123011\right) n_f T_F\right)}{4098124800}, \nn \\
&\<R^{(1)g}_{10,6}\>= \tfrac{\left(459012085-46506600 \pi ^2\right) C_A^2+\left(76381200 \pi ^2-753853073\right) C_A n_f T_F+2520 C_F n_f T_F}{86486400}, \nn \\
&\<R^{(1)g}_{12,0}\>= \tfrac{\left(4613713731326-363812248800 \pi ^2\right) C_A^2}{1558311955200}+\tfrac{\left(254307853800 \pi ^2-2498045171789\right) C_A n_f T_F}{779155977600}\nn \\
&\qquad\qquad +\tfrac{788981 C_F n_f T_F}{113836800}, \nn \\
&\<R^{(1)g}_{12,2}\>= -\tfrac{\left(885896411400 \pi ^2-9950914029409\right) C_A^2}{2727045921600}+\tfrac{\left(139361777100 \pi ^2-1376830794827\right) C_A n_f T_F}{283329446400}\nn \\
&\qquad\qquad +\tfrac{9197 C_F n_f T_F}{19514880}, \nn \\
&\<R^{(1)g}_{12,4}\>= -\tfrac{\left(56981204280 \pi ^2-577018051339\right) C_A^2}{129859329600}+\tfrac{\left(655611356400 \pi ^2-6475204191751\right) C_A n_f T_F}{865728864000} \nn \\
&\qquad\qquad -\tfrac{86819 C_F n_f T_F}{4122518400}, \nn \\
&\<R^{(1)g}_{12,6}\>= \tfrac{C_A \left(3 \left(372815843400 \pi ^2-3679625658343\right) n_f T_F-7 \left(91148557500 \pi ^2-899626801693\right) C_A\right)}{865728864000}, \nn \\
&\<R^{(1)g}_{12,8}\>= \tfrac{\left(3609756605-365743840 \pi ^2\right) C_A^2+16 \left(41815620 \pi ^2-412703677\right) C_A n_f T_F+2352 C_F n_f T_F}{914833920},
\ee
and
\be\label{eq:Rcoeff_QCD_gluon_directdecomposition3}
&\<R^{(0)g}_{8,0}\g^{(1)}_{8,0}\>= \tfrac{-62 C_A^2-110 C_A n_f T_F-135 C_F n_f T_F}{5040},&& \<R^{(0)g}_{8,2}\g^{(1)}_{8,2}\>= -\tfrac{C_A (73 C_A+16 n_f T_F)}{20160}, \nn \\
&\<R^{(0)g}_{10,0}\g^{(1)}_{10,0}\> = \tfrac{1986 C_A^2-6740 C_A n_f T_F-1407 C_F n_f T_F}{47040}, \nn \\
&\<R^{(0)g}_{10,2}\g^{(1)}_{10,2}\> = \tfrac{5743 C_A^2-11981 C_A n_f T_F-594 C_F n_f T_F}{332640},&& \<R^{(0)g}_{10,4}\g^{(1)}_{10,4}\> = -\tfrac{C_A (39 C_A+7 n_f T_F)}{73920}, \nn \\
&\<R^{(0)g}_{12,0}\g^{(1)}_{12,0}\> = \tfrac{491846 C_A^2-1164457 C_A n_f T_F-102141 C_F n_f T_F}{6486480}, \nn \\
&\<R^{(0)g}_{12,2}\g^{(1)}_{12,2}\> = \tfrac{999298 C_A^2-1763153 C_A n_f T_F-34398 C_F n_f T_F}{24216192}, \nn \\
&\<R^{(0)g}_{12,4}\g^{(1)}_{12,4}\> = \tfrac{3197 C_A^2-6135 C_A n_f T_F-126 C_F n_f T_F}{720720},&& \<R^{(0)g}_{12,6}\g^{(1)}_{12,6}\> = -\tfrac{C_A (37 C_A+6 n_f T_F)}{686400}.
\ee
For the quark jet, we find
\be\label{eq:Rcoeff_QCD_quark_directdecomposition2}
&\<R^{(1)q}_{8,0}\>= \tfrac{\left(27930000 \pi ^2-275120246\right) C_A C_F}{35280000}-\tfrac{\left(206976 \pi ^2-2041751\right) C_F^2}{282240} +\tfrac{\left(201264317-20580000 \pi ^2\right) C_F n_f T_F}{35280000}, \nn \\
&\<R^{(1)q}_{8,2}\>= \tfrac{\left(10224900 \pi ^2-100838071\right) C_A C_F}{16934400}-\tfrac{\left(1774500 \pi ^2-17530127\right) C_F^2}{2822400} +\tfrac{\left(39243247-3981600 \pi ^2\right) C_F n_f T_F}{8467200},  \nn \\
&\<R^{(1)q}_{8,4}\>= \tfrac{C_F\left(7 \left(15000 \pi ^2-148003\right) C_A+\left(1685981-170800 \pi ^2\right) C_F\right)}{403200}, \nn \\
&\<R^{(1)q}_{10,0}\>=\tfrac{\left(1709457750 \pi ^2-16812815347\right) C_A C_F}{1792929600}-\tfrac{\left(21218400 \pi ^2-209515081\right) C_F^2}{39513600} \nn \\
&\qquad\qquad +\tfrac{\left(168438023821-17142602400 \pi ^2\right) C_F n_f T_F}{14343436800}, \nn \\
&\<R^{(1)q}_{10,2}\>= \tfrac{\left(13877671500 \pi ^2-136733593943\right) C_A C_F}{15367968000}-\tfrac{\left(64499400 \pi ^2-636790073\right) C_F^2}{127008000}\nn \\
&\qquad\qquad +\tfrac{\left(2801569019-284592000 \pi ^2\right) C_F n_f T_F}{256132800}, \nn \\
&\<R^{(1)q}_{10,4}\>=  \tfrac{\left(6747741000 \pi ^2-66583998913\right) C_A C_F}{12294374400}+\tfrac{\left(46333633219-4693827600 \pi ^2\right) C_F^2}{12294374400} \nn \\
&\qquad\qquad -\tfrac{\left(999600 \pi ^2-9863251\right) C_F n_f T_F}{1330560}, \nn \\
&\<R^{(1)q}_{10,6}\>=  \tfrac{C_F \left(4 \left(9538200 \pi ^2-94135219\right) C_A+\left(540714493-54784800 \pi ^2\right) C_F\right)}{172972800}, \nn \\
&\<R^{(1)q}_{12,0}\>=\tfrac{\left(83057703300 \pi ^2-830250288473\right) C_A C_F}{111307996800}-\tfrac{\left(702332400 \pi ^2-7516807777\right) C_F^2}{3073593600}\nn \\
&\qquad\qquad +\tfrac{\left(2504233505507-261779918400 \pi ^2\right) C_F n_f T_F}{222615993600}, \nn \\
&\<R^{(1)q}_{12,2}\>=  \tfrac{\left(112390544366100 \pi ^2-1118604851318477\right) C_A C_F}{159986694067200}-\tfrac{\left(18487161000 \pi ^2-193326296189\right) C_F^2}{86060620800}\nn \\
&\qquad\qquad +\tfrac{\left(20817079707113-2158310154000 \pi ^2\right) C_F n_f T_F}{1904603500800}, \nn \\
&\<R^{(1)q}_{12,4}\>=  \tfrac{\left(3216440026800 \pi ^2-31809587301651\right) C_A C_F}{5194373184000}-\tfrac{\left(518049932400 \pi ^2-5194727659819\right) C_F^2}{2597186592000} \nn \\
&\qquad\qquad +\tfrac{\left(488517802327-49855085280 \pi ^2\right) C_F n_f T_F}{51943731840}, \nn \\
&\<R^{(1)q}_{12,6}\>= -\tfrac{\left(31815610577491-3223780560000 \pi ^2\right) C_A C_F}{10388746368000}-\tfrac{\left(177924146400 \pi ^2-1756123265947\right) C_F^2}{1484106624000}\nn \\
&\qquad\qquad -\tfrac{\left(9737280 \pi ^2-96097693\right) C_F n_f T_F}{17297280}, \nn \\
&\<R^{(1)q}_{12,8}\>= \tfrac{C_F \left(\left(4699648800 \pi ^2-46383418021\right) C_A+14 \left(4303513889-436035600 \pi ^2\right) C_F\right)}{41167526400},
\ee
and
\be\label{eq:Rcoeff_QCD_quark_directdecomposition3}
&\<R^{(0)q}_{8,0}\g^{(1)}_{8,0}\>= -\tfrac{C_F (24 C_A+4 C_F+67 n_f T_F)}{1680},&& \<R^{(0)q}_{8,2}\g^{(1)}_{8,2}\>= \tfrac{C_F (13 C_A-40 C_F)}{6720}, \nn \\
&\<R^{(0)q}_{10,0}\g^{(1)}_{10,0}\> = -\tfrac{C_F (-23799 C_A+3652 C_F+86219 n_f T_F)}{517440}, \nn \\
&\<R^{(0)q}_{10,2}\g^{(1)}_{10,2}\> = -\tfrac{C_F (-4825 C_A+682 C_F+9012 n_f T_F)}{221760},&& \<R^{(0)q}_{10,4}\g^{(1)}_{10,4}\> = \tfrac{C_F (113 C_A-368 C_F)}{443520}, \nn \\
&\<R^{(0)q}_{12,0}\g^{(1)}_{12,0}\> = -\tfrac{C_F (-343735 C_A+5382 C_F+865310 n_f T_F)}{4324320}, \nn \\
&\<R^{(0)q}_{12,2}\g^{(1)}_{12,2}\> = -\tfrac{C_F (-38990713 C_A+463606 C_F+69682564 n_f T_F)}{887927040}, \nn \\
&\<R^{(0)q}_{12,4}\g^{(1)}_{12,4}\> = -\tfrac{C_F (-7296 C_A+655 C_F+13016 n_f T_F)}{1441440},&& \<R^{(0)q}_{12,6}\g^{(1)}_{12,6}\> = \tfrac{C_F (13 C_A-43 C_F)}{514800}.
\ee

\section{More details on the Lorentzian inversion formula calculation}\label{app:inversionformula_details}

In this section, we compute in full detail the coefficients $\<R^{(1)}_{j+4,j}\>$ and $\<R^{(0)}_{j+\tau_c,j;\de'_*}\g^{(1)}_{j+\tau_c,j}\>$ for $\cN=4$ SYM using the Lorentzian inversion formula. We compute $\<R^{(1)}_{j+4,j}\>$ in the first subsection, and then we compute $\<R^{(0)}_{j+\tau_c,j;\de'_*}\g^{(1)}_{j+\tau_c,j}\>$ in the next subsection.
\subsection{$\<R^{(1)}_{j+4,j}\>$}
From \eqref{eq:Lorentzianinversionformula_3}, we have
\be\label{eq:R1coeff_inversionformula_1}
\<R^{(1)}_{j+4,j}\>=&2\kappa_{2j+6}\int_0^1\frac{d\bar{z}}{\bar{z}^2} k_{2j+6}^{0,1}(\bar z)\mathrm{dDisc}_t[\cG^{(1)}(z,\bar{z})]|_{z^2}.
\ee
Therefore, we first need $\cG^{(1)}(z,\bar z)|_{z^2}$ in order to compute $\<R^{(1)}_{j+4,j}\>$. We find
\be
\cG^{(1)}(z,\bar{z})|_{z^2}=\frac{1}{4(1-\bar{z})^2}&\left(2\bar{z}^4(\bar{z}-2)\z_2-\bar{z}(1-\bar{z})\log(1-\bar{z})(4+2\bar{z}(\bar{z}-2)+\bar{z}(1-\bar{z})^2\log(1-\bar{z}))\right. \nn \\
&\qquad\left.+2\bar{z}^3(3-3\bar{z}+\bar{z}^2)\mathrm{Li}_2(\bar{z})\right).
\ee
Using the double discontinuities
\be\label{eq:dDisc_identities}
\mathrm{dDisc}_t[(1-\bar{z})^{\a}]=&-2\sin^2(\pi\a)(1-\bar{z})^{\a}, \nn \\
\mathrm{dDisc}_t[(1-\bar{z})^{\a}\log(1-\bar{z})]=&-2\sin^2(\pi\a)(1-\bar{z})^{\a}\log(1-\bar{z})-2\pi\sin(2\pi\a)(1-\bar{z})^{\a}, \nn \\
\mathrm{dDisc}_t[(1-\bar{z})^{\a}\mathrm{Li}_2(\bar{z})]=&-2\sin^2(\pi\a)(1-\bar{z})^{\a}\mathrm{Li}_2(\bar{z})+2\pi\sin(2\pi\a)(1-\bar{z})^{\a}\log(\bar{z}), \nn \\
\mathrm{dDisc}_t[(1-\bar{z})^{\a}\log^2(1-\bar{z})]=&-2\sin^2(\pi\a)(1-\bar{z})^{\a}\log^2(1-\bar{z})-4\pi\sin(2\pi\a)(1-\bar z)^{\a}\log(1-\bar{z}) \nn \\
&-4\pi^2\cos(2\pi\a)(1-\bar z)^{\a},
\ee
we obtain (after introducing a small regulator $\e$)
\be\label{eq:dDisc_Gz2order}
&\mathrm{dDisc}_t[\cG^{(1)}(z,\bar{z})]|_{z^2}\nn \\
&=-2\sin^2(\pi\e)\cG^{(1)}(z,\bar{z})|_{z^2}+2\pi\sin(2\pi\e)\x\p{-\frac{\bar{z}^{1+\e}}{4(1-\bar{z})^{1+\e}}}(4+2\bar{z}^2-4\bar{z})\nn \\
&\quad-2\pi\sin(2\pi\e)\x\frac{\bar{z}^{3+\e}}{2(1-\bar{z})^{2+\e}}(3-3\bar{z}+\bar{z}^2)\log(\bar{z})\nn \\
&\quad+4\pi\sin(2\pi\e)\x\p{-\frac{\bar{z}^{2+\e}}{4(1-\bar{z})^{-1+\e}}}\log(1-\bar{z})\nn \\
&\quad-4\pi^2\cos(2\pi\e)\x\p{-\frac{\bar{z}^{2+\e}}{(1-\bar{z})^{-1+\e}}}.
\ee
We can then plug \eqref{eq:dDisc_Gz2order} into \eqref{eq:R1coeff_inversionformula_1} and integrate each term over $\bar z$. After taking the $\e\to 0 $ limit, we obtain \eqref{eq:oneloop_coeff_analytical}.

\subsection{$\<R^{(0)}_{j+\tau_c,j;\de'_*}\g^{(1)}_{j+\tau_c,j}\>$}
To do the calculation for general twist operators, we will need to compute dDisc of more complicated functions than in the leading twist case.  It is convenient to first define the linear functional
\be
I_{\b}[f]\equiv\lim_{\e\to 0}\kappa_\beta\int_0^1\frac{d\bar z}{\bar z^2}k_{\b}^{0,1}(\bar z)\mathrm{dDisc}_t\left[\frac{\bar z^\e}{(1-\bar z)^\e}f(\bar z)\right],
\ee
where $\kappa_{\beta}$ is defined in \eqref{eq:kappaandmudef}, and
\be
k_{\b}^{0,1}(\bar z)=\bar z^{\frac{\b}{2}}{}_2F_1\p{\tfrac{\b}{2}, \tfrac{\b}{2}+1,\b ;\bar z}.
\ee

To compute anomalous dimensions, we will need to apply this functional to functions with power-law divergences or logarithmic divergences near $\bar z=1$. Using \eqref{eq:dDisc_identities}, we can obtain
\be\label{eq:Ifunctional_powerlaw}
I_{\b}\left[\frac{\bar z^{b}}{(1-\bar z)^{a}}\right]=&\frac{\G(\tfrac{\b}{2})^2\G(\tfrac{\b}{2}+1)\G(-1+b+\frac{\b}{2})}{\G(a)\G(a+1)\G(\b-1)\G(1-a+\frac{\b}{2})\G(b-a+\frac{\b}{2})} \nn \\
&\qquad \x {}_3F_2\left(\begin{matrix} &1-a,\quad b-a,\quad 1+\tfrac{\b}{2} & \\ &1-a+\tfrac{\b}{2},\quad b-a+\tfrac{\b}{2}&\end{matrix};1\right),
\ee
where $a$ should be a positive integer (for zero or negative $a$ the right hand side is zero). Also, for logarithmic divergences we can take the derivative of \eqref{eq:Ifunctional_powerlaw} with respect to $a$. For $a=0$, we have
\be\label{eq:Ifunctional_log1}
I_{\b}\left[\bar z^b\log (1-\bar z)\right]=-\frac{\G(\tfrac{\b}{2})\G(\tfrac{\b}{2}-1)}{\G(\b-1)},
\ee
and for $a=1$
\be\label{eq:Ifunctional_log2}
&I_{\b}\left[\frac{\bar z^{b}}{(1-\bar z)}\log (1-\bar z)\right] \nn \\
&=\frac{\G(\tfrac{\b}{2})\G(\tfrac{\b}{2}+1)}{\G(\b-1)}\p{1-S_1(\tfrac{\b}{2}-1)-S_1(\tfrac{\b}{2}+b-2)+{}_3F_2^{(0,0,1),(0,0)}\left(\begin{matrix} &1-b,\quad 1+\tfrac{\b}{2},\quad 0 & \\ &-1+b+\tfrac{\b}{2},\quad \tfrac{\b}{2}&\end{matrix};1\right)},
\ee
where $S_1$ is the harmonic number, and
\be
{}_3F_2^{(0,0,1),(0,0)}\left(\begin{matrix} &a_1,\quad a_2,\quad a_3 & \\ & b_1,\quad b_2&\end{matrix};1\right)=\ptl_{a_3}\p{{}_3F_2\left(\begin{matrix} &a_1,\quad a_2,\quad a_3 & \\ & b_1,\quad b_2&\end{matrix};1\right)}.
\ee
\eqref{eq:Ifunctional_powerlaw}, \eqref{eq:Ifunctional_log1}, and \eqref{eq:Ifunctional_log2} are the main results we need for the functional $I_{\b}$ in order to do the Lorentzian inversion formula calculation for general twists. In particular, for \eqref{eq:Ifunctional_log2}, it suffices to consider the $b=1$ case, which is given by
\be
I_{\b}\left[\frac{\bar z}{(1-\bar z)}\log (1-\bar z)\right] =\frac{\G(\tfrac{\b}{2})\G(\tfrac{\b}{2}+1)\p{1-2S_1(\tfrac{\b}{2}-1)}}{\G(\b-1)}.
\ee

By \eqref{eq:Lorentzianinversionformula_3}, the anomalous dimension coefficient $\<R^{(0)}_{j+\tau_c,j}\g^{(1)}_{j+\tau_c,j}\>$ with celestial twist $\tau_c$ should be given by
\be
\<R^{(0)}_{j+\tau_c,j}\g^{(1)}_{j+\tau_c,j}\>=4\kappa_{2j+\tau_c }\left.\p{z^{\frac{\tau_c}{2}-1}\int_0^1\frac{d\bar z}{\bar z^2}g^{\tl{\de}_i}_{j+1,j+\tau_c-1}(z,\bar z)\mathrm{dDisc}_t[\cG(z,\bar z)]}\right|_{z^{\frac{\tau_c}{2}}\log z}.
\ee
The 2d conformal block $g^{\tl{\de}_i}_{j+1,j+\tau_c-1}(z,\bar z)$ is
\be
g^{\tl{\de}_i}_{j+1,j+\tau_c-1}(z,\bar z)=k^{0,1}_{2-\tau_c}(z)k^{0,1}_{2j+\tau_c}(\bar z) + (z\leftrightarrow \bar z).
\ee
Expanding the above expression in small $z$, we find (the $(z\leftrightarrow \bar z)$ term doesn't contribute)
\be\label{eq:Rgamma_IbetaG}
\<R^{(0)}_{j+\tau_c,j}\g^{(1)}_{j+\tau_c,j}\>=4\sum_{m=0}^{\frac{\tau_c}{2}-3}\frac{(1-\tfrac{\tau_c}{2})_m(2-\tfrac{\tau_c}{2})_m}{(2-\tau_c)_mm!}I_{2j+\tau_c}\left.\left[\cG(z,\bar z)\right]\right|_{z^{\tfrac{\tau_c}{2}-m}\log z},
\ee
where we have used the fact that $\cG(z,\bar z)$ starts at $z^3$ to restrict the range of the sum over $m$. Therefore, our task now is to find $I_{\b}\left.\left[\cG(z,\bar z)\right]\right|_{z^n\log z}$ for general $n$. Focusing on the singular terms near $\bar z=1$, we find
\be
&\cG(z,\bar z)|_{\log z} \nn \\
&= z^3 \x \bigg(-\frac{1+z+2z^2}{2z(1-z)^2}\frac{1}{1-\bar z}-\frac{1}{2(1-z)^2}\frac{\bar z^3}{(1-\bar z)^2}\p{1+\frac{z \bar z}{2}-2z\bar z\sum_{k=1}^{\oo}\frac{z^k\bar z^k}{k(k+2)}} \nn \\
&\qquad\qquad~ -\frac{(1+z)\log(1-z)}{2z(1-z)}\frac{\bar z^2}{(1-\bar z)^2}-\frac{(1-z+2z^3)\log(1-z)}{2z^2(1-z)^2}\frac{\bar z^2}{(1-\bar z)} \nn \\
&\qquad\qquad~ -\frac{\bar z^2(z \bar z-(1-z) (1-\bar z))(z \bar z(1-z) (1-\bar z))}{2z(1-z)(1-\bar z)}\sum_{k=1}^{\oo}\frac{z^k}{k}\p{1-\frac{1}{(1-\bar z)^k}} \nn \\
&\qquad\qquad~ -\frac{z}{2(1-z)}\frac{\bar z \log(1-\bar z)}{1-\bar z}-\frac{3(1+2z^2-z^3)}{2(1-z)^3}\bar z\log(1-\bar z)\bigg) + O((1-\bar z)^0).
\ee

We can then apply the functional $I_{\b}$ to the above expression and expand in small $z$. After several lines of algebra, we obtain that for $n\geq 1$
\be\label{eq:IbetaG_ngeq1}
&\left.I_{\b}\left[\cG(z,\bar z)\right]\right|_{z^{3+n}\log z} \nn \\
&=\p{-n-2+\frac{3-7n-4n^2+2n(n+2)S_1(n)+(n+2)S_1(n+1)}{2(n+2)}}I_{\b}\left[\frac{1}{1-\bar z}\right] \nn \\
&\quad +\frac{1}{2}\p{S_1(n+1)+S_1(n)}I_\b\left[\frac{\bar z^2}{(1-\bar z)^2}\right] \nn \\
&\quad -\frac{1}{2}\p{(n+1)I_\b\left[\frac{\bar z^3}{(1-\bar z)^2}\right]+\frac{n}{2}I_\b\left[\frac{\bar z^4}{(1-\bar z)^2}\right]-2\sum_{k=1}^{n-1}\frac{n-k}{k(k+2)}I_\b\left[\frac{\bar z^{k+4}}{(1-\bar z)^2}\right]}\nn \\
&\quad -\frac{1}{2}\p{\frac{I_\b\left[\frac{\bar z^2}{(1-\bar z)^n}\right]}{n+1}-\frac{I_\b\left[\frac{\bar z^2}{(1-\bar z)^{n-1}}\right]}{n}+S_1(n-1)I_\b\left[\frac{\bar z^4}{(1-\bar z)}\right]-\sum_{k=1}^{n-1}\frac{1}{k}I_\b\left[\frac{\bar z^4}{(1-\bar z)^{1+k}}\right]} \nn \\
&\quad -\frac{1}{2}I_\b\left[\frac{\bar z}{(1-\bar z)}\log(1-\bar z)\right]-\frac{3}{2}n(n+2)I_\b\left[\log(1-\bar z)\right].
\ee
For the leading order term $n=0$, we simply have
\be\label{eq:IbetaG_n0}
\left.I_{\b}\left[\cG(z,\bar z)\right]\right|_{z^{3}\log z}=&-\frac{1}{4}I_{\b}\left[\frac{1}{1-\bar z}\right]-\frac{3}{2}I_{\b}\left[\log(1-\bar z)\right] \nn \\
=&-\frac{(\b-6)(\b+4)\G(\tfrac{\b}{2})\G(\tfrac{\b}{2}-1)}{16\G(\b-1)}.
\ee

Combining \eqref{eq:Ifunctional_powerlaw}, \eqref{eq:Ifunctional_log1}, \eqref{eq:Ifunctional_log2}, \eqref{eq:Rgamma_IbetaG}, \eqref{eq:IbetaG_ngeq1}, and \eqref{eq:IbetaG_n0}, we can then determine $\<R^{(0)}_{j+\tau_c,j}\g^{(1)}_{j+\tau_c,j}\>$ for any twist $\tau_c=6, 8, 10, \dots$. For general $\tau_c$, we obtain
\be\label{eq:Rgamma_generaltauc}
&\<R^{(0)}_{j+\tau_c,j}\g^{(1)}_{j+\tau_c,j}\>\nn \\
&=\frac{(-1)^{\frac{\tau_c}{2}}(\tau_c^2-2\tau_c+8(-1)^{\frac{\tau_c}{2}}-16)\G(\tfrac{\tau_c}{2})\G(\tfrac{\tau_c}{2}+1)\G(j+\tfrac{\tau_c}{2})\G(j+\tfrac{\tau_c}{2}+1)S_1(j+\tfrac{\tau_c}{2}-1)}{2\G(\tau_c-1)\G(2j+\tau_c-1)} \nn \\
&\quad+ \frac{\G(j+\tfrac{\tau_c}{2})\G(j+\tfrac{\tau_c}{2}-1)}{\G(2j+\tau_c-1)}P_{\tau_c}(j),
\ee
where $P_{\tau_c}(j)$ is a polynomial in $j$. It is given by
\be
P_{\tau_c}(j)&=\frac{(-1)^{\frac{\tau_c}{2}}(\tfrac{\tau_c}{2}-2)\G(\tfrac{\tau_c}{2})\G(\tfrac{\tau_c}{2}+2)(j+\tfrac{\tau_c}{2}-3)(j+\tfrac{\tau_c}{2}+2)}{2\G(\tau_c-1)}+ \sum_{m=0}^{\tfrac{\tau_c}{2}-4}\frac{(1-\tfrac{\tau_c}{2})_m(2-\tfrac{\tau_c}{2})_m}{m!(2-\tau_c)_m}Q_{m,\tau_c}(j),
\ee
and
\be
&Q_{m,\tau_c}(j) \nn \\
=&-2 \tau_c \left(j^2-2 j (m+2)+4 m+6\right)+j^2 (4 m+6)+\tau_c^2 (-2 j+m+4)+\frac{4 (j+m+2) (j+m+3)}{-2 m+\tau_c-6} \nn \\
&+\frac{4 (j+m+1) (j+m+2)}{-2 m+\tau_c-4}+\frac{4 (j+m) (j+m+1)}{-2 m+\tau_c-2}+j (6-4 m)+6 (m+1) (m+4)-\frac{\tau_c^3}{2} \nn \\
&+2\G(j+\tfrac{\tau_c}{2}+1)^2\left(\frac{{}_3F_2\left(\begin{matrix} &5+m-\tfrac{\tau_c}{2},\quad 5+m-\tfrac{\tau_c}{2},\quad j+\tfrac{\tau_c}{2} & \\ & j+4+m,\quad j+6+m&\end{matrix};1\right)}{\G(j+4+m)\G(j+6+m)\G(\tfrac{\tau_c}{2}-m-2)\G(\tfrac{\tau_c}{2}-m-4)}\right. \nn \\
&\qquad\qquad\qquad\qquad\qquad\left.-\frac{{}_3F_2\left(\begin{matrix} &4+m-\tfrac{\tau_c}{2},\quad 4+m-\tfrac{\tau_c}{2},\quad j+\tfrac{\tau_c}{2} & \\ & j+3+m,\quad j+5+m&\end{matrix};1\right)}{\G(j+3+m)\G(j+5+m)\G(\tfrac{\tau_c}{2}-m-1)\G(\tfrac{\tau_c}{2}-m-3)}\right) \nn \\
&+2\sum_{k=1}^{\tfrac{\tau_c}{2}-4-m}\frac{\G(j+\tfrac{\tau_c}{2})\G(j+\tfrac{\tau_c}{2}+1)\G(j+\tfrac{\tau_c}{2}+3){}_3F_2\left(\begin{matrix} &3-k,\quad -k,\quad j+\tfrac{\tau_c}{2}+1 & \\ & j+\tfrac{\tau_c}{2}-k,\quad  j+\tfrac{\tau_c}{2}+3-k&\end{matrix};1\right)}{k\G(k+1)\G(k+2)\G(j+\tfrac{\tau_c}{2}-k)\G(j+\tfrac{\tau_c}{2}+3-k)\G(j+\tfrac{\tau_c}{2}-1)}.
\ee

\section{Lightcone bootstrap and large-$j$ behavior of $\<R_{j+6,j}^{(n+1)g/q}\>$}\label{app:lightcone_bootstrap_largej}
In this appendix, we describe how to obtain the large-$j$ behavior of the coefficients $\<R_{j+6,j}^{(n+1)g/q}\>$ using the double lightcone limit result \eqref{eq:Gtilde_doublelightcone} and the crossing equation \eqref{eq:collinearEEEC_crossing_leading}.
\subsection{Infinite sums of $\SL(2,\R)$ blocks}

We will use lightcone bootstrap techniques from \cite{Simmons-Duffin:2016wlq} to study the large-$j$ behavior of the coefficients. An important ingredient for the lightcone bootstrap are identities for infinite sums of $\SL(2,\R)$ blocks. One identity derived in \cite{Simmons-Duffin:2016wlq} is given in \eqref{eq:SL2Rblock_infinitesum}. For the calculation in this appendix, we can write it as (assuming $a$ is negative)
\be\label{eq:SL2Rblock_infinitesum2}
\sum_{\substack{h=j+ h_0 \\ j=0,2,\dots}}&S_{a}^{r,s}(h)k^{r,s}_{2h}(1-z)=\frac{1}{2}z^a +O(z^{a+1}), \nn \\
S^{r,s}_a(h)=&\frac{1}{\G(-a-r)\G(-a-s)}\frac{\G(\bar h-r)\G(\bar h-s)}{\G(2\bar h-1)}\frac{\G(\bar h-a-1)}{\G(\bar h+a+1)}.
\ee
For our calculation, we will also need to compute an infinite sum of derivatives of the $\SL(2,\R)$ blocks, such as $\ptl_{h}k^{r,s}_{2h}(1-z)$ or $\ptl_{s}k^{r,s}_{2h}(1-z)$. In particular, we want to compute
\be
&\sum_{j=0,2,\cdots} S_{a}^{0,-1}(j+3)\ptl_{\de}^n k_{\de+j}^{0,-1}(1-z)|_{\de\to j+6} \label{eq:SL2Rsum_deltaderiv}, \\
&\sum_{j=0,2,\cdots} S_{a}^{0,-1}(j+3)\ptl_{\de'_{*}}^n k_{2j+6}^{0,\frac{3-\de'_{*}}{2}}(1-z)|_{\de'_{*}\to 5} \label{eq:SL2Rsum_deltaprimederiv}.
\ee

Let us first consider \eqref{eq:SL2Rsum_deltaderiv} with $n=1$. One way of computing it is to follow the original argument for lightcone bootstrap given in \cite{Fitzpatrick:2012yx,Komargodski:2012ek}. In the limit $h\to \oo$, $\ptl_h k_{2h}(z)\sim \frac{(4 \r)^{h}}{\sqrt{1-\r^2}}\log(4\r)$, and therefore the dominant contribution of the sum in $h$ is at $h\sim\frac{1}{1-\r}\sim\frac{1}{\sqrt{1-z}}$. Thus, we should study the large $h$ limit of $k^{0,-1}_{2h}(z)$ with $h\sqrt{1-z}$ fixed. Using the Euler integral representation of $k_{2h}(z)$, we find that
\be\label{eq:SL2R_Besselapproximation}
k_{2h}^{0,-1}(z)\sim 2^{2h}\sqrt{\frac{y}{\pi h}}K_1(2\sqrt{y}),\quad y=h^2(1-z),
\ee
in the fixed $y$, $h\to \oo$ limit. Now, we can replace $1-z \to z$ in \eqref{eq:SL2R_Besselapproximation} and plug it into \eqref{eq:SL2Rsum_deltaderiv} with $n=1$. Focusing on the leading term in the small $z$ limit, we find
\be
&\sum_{h=0,2,\dots}S_{a}^{0,-1}(h+3)\frac{1}{2}\ptl_{h}k^{0,-1}_{2h+6}(1-z)\nn \\
&\approx \frac{1}{4}\int_{0}^{\oo}dh\frac{2^{-4-2h}h^{-2a-\frac{1}{2}}\sqrt{\pi}}{\G(1-a)\G(-a)}\ptl_{h}\p{2^{2h+6}\sqrt{\frac{h^2(1-\bar{z})}{\pi h}}K_1(2h\sqrt{z})} \nn \\
&= \frac{1}{2}z^{a}\p{\log2+\sqrt{z}\frac{(1+4a)\G(-\half-a)\G(\half-a)}{4\G(1-a)\G(-a)}} + O(z^{a+1}),
\ee
where $\approx$ means that the Casimir-singular terms are the same. We have checked numerically that this formula is correct in the $z \to 0$ limit. 

Interestingly, we find that if we simply move the derivative $\ptl_h$ of $\sum_{h=0,2,\dots}S_{a}^{0,-1}(h+3)\frac{1}{2}\ptl_{h}k^{0,-1}_{2h+6}(1-z)$ to the function $S_{a}^{0,-1}(h+3)$, we will also get the same result in the small $z$ limit. One way of understanding this is by following the discussion in \cite{Simmons-Duffin:2016wlq}. We can write \eqref{eq:SL2Rsum_deltaderiv} as a contour integral
\be
\int_{-\e-i\oo}^{\e+i\oo} dh\frac{\pi(1+e^{i\pi h})}{2\tan(\pi h)}S_{a}^{0,-1}(h+3)\frac{1}{2^n}\ptl^n_{h}k^{0,-1}_{2h+6}(1-z),
\ee
and then integrate by parts to move all the derivatives $\ptl_h$. The leading term at small $z$ should come from $\ptl_h^nS_{a}^{0,-1}(h+3)$. In conclusion, the infinite sum \eqref{eq:SL2Rsum_deltaderiv} in the small $z$ limit is given by
\be\label{eq:SL2Rsum_deltaderiv_answer}
&\sum_{j=0,2,\cdots} S_{a}^{0,-1}(j+3)\ptl_{\de}^n k_{\de+j}^{0,-1}(1-z)|_{\de\to j+6} = \p{\frac{\log 2}{2}}^n z^a + O(z^{a+\frac{1}{2}}).
\ee

We now consider \eqref{eq:SL2Rsum_deltaprimederiv}. We claim that it is given by
\be
&\sum_{j=0,2,\cdots} S_{a}^{0,-1}(j+3)\ptl_{\de'_{*}}^n k_{2j+6}^{0,\frac{3-\de'_{*}}{2}}(1-z)|_{\de'_{*}\to 5} = c_n z^a\log^{n}z + O(z^a\log^{n-1}z).
\ee
The coefficient $c_n$ can be determined by applying $\ptl_{\de'_{*}}$ to \eqref{eq:SL2Rblock_infinitesum2} $n$ times. Since \eqref{eq:SL2Rblock_infinitesum2} is independent of $\de'_{*}$, we should get zero. Hence, we have
\be
&\sum_{j=0,2,\cdots} S_{a}^{0,-1}(j+3)\ptl_{\de'_{*}}^n k_{2j+6}^{0,\frac{3-\de'_{*}}{2}}(1-z)|_{\de'_{*}\to 5} \nn \\
&= -\sum_{m=1}^{n} \sum_{j=0,2,\cdots} \frac{n!}{m!(n-m)!}\left(\ptl_{\de'_{*}}^{m}S_{a}^{0,\frac{3-\de'_{*}}{}}(j+3)|_{\de'_{*}\to 5}\right) \left(\ptl_{\de'_{*}}^{n-m} k_{2j+6}^{0,\frac{3-\de'_{*}}{2}}(1-z)|_{\de'_{*}\to 5}\right) \nn \\
&= -\sum_{m=1}^{n} \sum_{j=0,2,\cdots} \frac{n!}{m!(n-m)!} \frac{(-1)^m}{4^{m}}\ptl_a^{m} S_{a}^{0,-1}(j+3)\left(\ptl_{\de'_{*}}^{n-m} k_{2j+6}^{0,\frac{3-\de'_{*}}{2}}(1-z)|_{\de'_{*}\to 5}\right) + O(z^a\log^{n-1}z) \nn \\
&=-\sum_{m=1}^{n}\frac{n!}{m!(n-m)!} \frac{(-1)^m}{4^{m}}c_{n-m} z^a\log^n z + O(z^a\log^{n-1}z).
\ee
This leads to the recursion relation
\be
c_n=-\sum_{m=1}^{n} \frac{n!}{m!(n-m)!} \frac{(-1)^m}{4^{m}} c_{n-m}.
\ee
With the initial condition $c_0=\frac{1}{2}$, one can then determine that $c_n=2^{-1-2n}$. Thus,
\be\label{eq:SL2Rsum_deltaprimederiv_answer}
&\sum_{j=0,2,\cdots} S_{a}^{0,-1}(j+3)\ptl_{\de'_{*}}^n k_{2j+6}^{0,\frac{3-\de'_{*}}{2}}(1-z)|_{\de'_{*}\to 5} =\frac{1}{2^{1+2n}}z^a\log^{n}z + O(z^a\log^{n-1}z).
\ee

\subsection{$\<R_{j+6,j}^{(n+1)}\>$ in the large-$j$ limit}
We are now ready to compute the large-$j$ behavior of $\<R_{j+6,j}^{(n+1)g/q}\>$. As a warmup, we first consider the leading order ($n=0$) case of the double lightcone limit \eqref{eq:Gtilde_doublelightcone}. Since we know that $R^{(1)}_{j+2,j}$ is nonzero only when $j=2$, the $z (1-\bar z)^0$ term of $\tl{\cG}^{(1)}(z\ll 1-\bar z \ll 1)$ should be produced by the first subleading term in the $\SL(2,\R)$ expansion of the $\tau_c=4$ operator or the leading $\SL(2,\R)$ expansion of the $\tau_c=6$ operator. More explicitly, the crossing equation \eqref{eq:collinearEEEC_crossing_leading} at leading order in the double lightcone limit is given by (suppressing the $g/q$ superscript for brevity)
\be\label{eq:crossing_LO_doublelightcone}
&10 R^{(1)}_{4,2} z (1-\bar z)^{0} + \cdots \nn \\
&=\p{\frac{z\bar{z}}{(1-z)(1-\bar{z})}}^{3}\left[\sum_{j=0,2,\cdots}\<R^{(1)}_{j+4,j}\>\p{(1-\bar z)^2 k^{0,-1}_{2j+4}(1-z) + (1-\bar z)^3\sum_{n=-1}^{1} C_nk_{2j+4+2n}^{0,-1}(1-z) + \cdots} \right. \nn \\
&\qquad\qquad\qquad\qquad\quad  \left. + \sum_{j=0,2,\cdots}\<R^{(1)}_{j+6,j}\>(1-\bar z)^3 k^{0,-1}_{2j+6}(1-z) + \cdots \right],
\ee
where $\cdots$ are all subleading terms in the double lightcone limit. The coefficients $C_n$ are from the $\SL(2,\R)$ block expansion of the conformal block, but their expressions won't be important for our discussion. The reason is that the $\<R^{(1)}_{j+4,j}\>$ coefficient actually has no contribution to the leading term in the left hand side. There are two ways of seeing this. First, we can simply look at the expression for $\<R^{(1)}_{j+4,j}\>$ in \eqref{eq:R1tau4_QCD_gluon} and \eqref{eq:R1tau4_QCD_quark} we obtained from the Lorentzian inversion formula and take the large-$j$ limit.\footnote{When taking the large-$j$ limit of the ${}_3F_2$ functions in $\<R^{(1)}_{j+4,j}\>$, it is more convenient to first use the Euler representation of the ${}_3F_2$ function to write it as an integral of the ${}_2F_1$ function, and then study the large-$j$ limit of the integral numerically. For example,
\be
{}_3F_2\left(\begin{matrix} & 2, \quad 3,\quad j+1 & \\ & j+4 ,\quad j+5 &\end{matrix};1\right)= \frac{\G(j+5)}{\G(j+1)\G(4)}\int_0^1 dz\ z^j (1-z)^3 {}_2F_1\p{2,3,j+4,z} \sim 1+\frac{6}{j}-\frac{18}{j^2}
\ee
Sometimes the ${}_2F_1$ is even explicitly known, and one can study the large-$j$ limit after evaluating the integral.
}
For both the gluon jet and quark jet, we find
\be
\<R^{(1)}_{j+4,j}\> \sim 4^{-j} j^{\frac{3}{2}} \sim S_{a=-1}^{0,-1}(j),
\ee
and therefore
\be
z^3 \sum_{j=0,2,4,\cdots} \<R^{(1)}_{j+4,j}\>k_{2j+4+2n}^{0,-1}(1-z) \sim z^2,
\ee
which is indeed subleading in \eqref{eq:crossing_LO_doublelightcone}. The other argument is to consider the term 
\be
\p{\frac{z\bar{z}}{(1-z)(1-\bar{z})}}^{3}(1-\bar z)^2 \sum_j \<R^{(1)}_{j+4,j}\>k_{2j+4}^{0,-1}(1-z).
\ee
If $\<R^{(1)}_{j+4,j}\>$ grows like $S_{a=-2}^{0,-1}(j)$ or faster at large-$j$, we should expect to see $z (1-\bar z)^{-1}$ on the left hand side of \eqref{eq:crossing_LO_doublelightcone}. However, such terms don't exist, so $z^3\sum_j \<R^{(1)}_{j+4,j}\>k_{2j+4+2n}^{0,-1}(1-z)$ must give subleading contribution to $z$. 

Based on the above discussion, the crossing equation \eqref{eq:crossing_LO_doublelightcone} becomes
\be
&10 R^{(1)}_{4,2} z (1-\bar z)^{0} + \cdots =\p{\frac{z\bar{z}}{(1-z)(1-\bar{z})}}^{3}\sum_{j=0,2,\cdots}\<R^{(1)}_{j+6,j}\>(1-\bar z)^3 k^{0,-1}_{2j+6}(1-z) + \cdots.
\ee
Using \eqref{eq:SL2Rblock_infinitesum2}, we obtain
\be\label{eq:Rtwist6_largej_2}
\<R^{(1)}_{j+6,j}\> \sim 20 R^{(1)}_{4,2} S_{a=-2}^{0,-1}(j+3) \sim \frac{5\sqrt{\pi}}{8}R^{(1)}_{4,2} 4^{-j}j^{\frac{7}{2}}.
\ee
which is \eqref{eq:Rtwistc6_largej} in the main text.

Let us now consider $n=1$. In this case, we should allow nonzero $\<R^{(2)}_{j+\tau_c,j}\>$ for general $\tau_c$. At this order, the crossing equation in the double lightcone limit becomes
\be
&10 R^{(1)}_{4,2}\g^{(1)}_{4,2} z\log z (1-\bar z)^{0} + \cdots \nn \\
&=\p{\frac{z\bar{z}}{(1-z)(1-\bar{z})}}^{3}\left[
\sum_{\tau_c < 6}\sum_{j} \<R^{(2)}_{j+\tau_c,j}\> \sum_{n=0}^{\oo}(1-\bar z)^{\frac{\tau_c}{2}+n} \sum_{m=-n}^{n} C^{\tau_c}_{n,m}k_{2j+\tau_c+2m}^{0,-1}(1-z) \right. \nn \\
&\qquad\qquad\qquad\qquad\quad + \sum_{\tau_c < 6} \sum_{j} \<R^{(1)}_{j+\tau_c,j}\g^{(1)}_{j+\tau_c,j}\> \left(\sum_{n=0}^{\oo}\frac{1}{2}(1-\bar z)^{\frac{\tau_c}{2}+n}\log(1-\bar z) \sum_{m=-n}^{n} C^{\tau_c}_{n,m}k_{2j+\tau_c+2m}^{0,-1}(1-z) \right. \nn \\
& \qquad\qquad\qquad\qquad\qquad\qquad\qquad\qquad\qquad\qquad + \left. \sum_{n=0}^{\oo}(1-\bar z)^{\frac{\tau_c}{2}+n} \sum_{m=-n}^{n} C^{\tau_c}_{n,m}\ptl_{\de}k_{\de+j+2m}^{0,-1}(1-z)|_{\de\to j+\tau_c} \right) \nn \\
& \qquad\qquad\qquad\qquad\quad + \sum_{\tau_c < 6} \sum_{j} \<R^{(1)}_{j+\tau_c,j}\g^{\prime(1)}_{*}\>  \sum_{n=0}^{\oo}(1-\bar z)^{\frac{\tau_c}{2}+n} \sum_{m=-n}^{n} C^{\tau_c}_{n,m}\ptl_{\de'_{*}}k_{2j+\tau_c+2m}^{0,\frac{3-\de'_{*}}{2}}(1-z)|_{\de'_{*} \to 5} \nn \\
& \qquad\qquad\qquad\qquad\quad + \sum_{j} \<R^{(2)}_{j+6,j}\>(1-\bar z)^3 k_{2j+6}^{0,-1}(1-z) \nn \\
& \qquad\qquad\qquad\qquad\quad + \sum_{j} \<R^{(1)}_{j+6,j}\g^{(1)}_{j+6,j}\>\left(\frac{1}{2}(1-\bar z)^3\log(1-\bar z)k_{2j+6}^{0,-1}(1-z) + (1-\bar z)^3\ptl_{\de}k_{\de+j}^{0,-1}(1-z)|_{\de \to j+6}\right) \nn \\
& \qquad\qquad\qquad\qquad\quad \left. + \sum_{j} \<R^{(1)}_{j+6,j}\g^{\prime(1)}_{*}\> (1-\bar z)^3\ptl_{\de'_{*}}k_{2j+6}^{0,\frac{3-\de'_{*}}{2}}(1-z)|_{\de'_{*} \to 5} + \cdots  \right].
\ee
We can use \eqref{eq:SL2Rblock_infinitesum2} and \eqref{eq:SL2Rsum_deltaderiv_answer} to argue that for $\tau_c<6$, $\<R^{(2)}_{j+\tau_c,j}\>$ and $\<R^{(1)}_{j+\tau_c,j}\g^{(1)}_{j+\tau_c,j}\>$ cannot grow as fast as $\ptl_a S_{a}^{0,-1}|_{a\to -2}$ at large $j$, otherwise the $z\log z$ term in the left hand side will have the wrong leading behavior for $1-\bar z \ll1$. Similarly, $\<R^{(1)}_{j+6,j}\g^{(1)}_{j+6,j}\>$ should not give any leading contribution, or we will get $z\log z\log(1-\bar z)$. Thus, we have
\be\label{eq:Rtauc6_largej_twoloop_crossing}
&10 R^{(1)}_{4,2}\g^{(1)}_{4,2} z\log z + \cdots \nn \\
&=z^3\left( \sum_{j} \<R^{(2)}_{j+6,j}\>k_{2j+6}^{0,-1}(1-z) + \sum_{j} \<R^{(1)}_{j+6,j}\g^{\prime(1)}_{*}\> \ptl_{\de'_{*}}k_{2j+6}^{0,\frac{3-\de'_{*}}{2}}(1-z)|_{\de'_{*} \to 5} + \cdots  \right).
\ee

As discussed in section \ref{sec:allorderEEEC}, we also need to deal with the degeneracies coming from $\g'_{*}$. To do so, let us make an ansatz for the large-$j$ behavior of $R^{(n+1)}_{j+6,j}$. We will assume
\be\label{eq:Rtauc6_largej_ansatz}
R^{(n+1)g}_{j+6,j}=\p{c^{(0)g}_1 A^{(n+1)}_1 + c^{(0)g}_2A^{(n+1)}_2}\ptl_{a}^nS_{a}^{0,-1}(j+3)|_{a\to -2} + O(4^{-j}j^{\frac{7}{2}}\log^{n-1}j), \nn \\
R^{(n+1)q}_{j+6,j}=\p{c^{(0)q}_1 A^{(n+1)}_1 + c^{(0)q}_2A^{(n+1)}_2}\ptl_{a}^nS_{a}^{0,-1}(j+3)|_{a\to -2} + O(4^{-j}j^{\frac{7}{2}}\log^{n-1}j), 
\ee
where the coefficients $c^{(0)g/q}_{i}$ are given in \eqref{eq:c0coeff_solution}, and our goal is to determine $A^{(n+1)}_1$ and $A^{(n+1)}_{2}$. In general, the coefficient $c^{\cO}$ at subleading order can also contribute, but we expect them to appear as $c^{(k)\cO}_i A^{(n+1-k)}_i \ptl_a^{n-k}S_a|_{a\to -2}$. Since $\ptl_a^{n-k}S_a|_{a\to -2}$ is subleading at large-$j$, we can just consider $c^{(0)\cO}_i$ for the leading large-$j$ behavior. Note that the ansatz \eqref{eq:Rtauc6_largej_ansatz} implies that for any nonnegative integers $n$ and $k$,
\be\label{eq:Rtwist6_degeneracy_relation}
\<R^{(n+1)g/q}_{j+6,j} (\g'_{*})^k\>=&\p{c^{(0)g/q}_1 A^{(n+1)}_1 (\g'_{*1})^k + c^{(0)g/q}_2A^{(n+1)}_2 (\g'_{*2})^k}\ptl_{a}^nS_{a}^{0,-1}(j+3)|_{a\to -2} \nn \\
&+ O(4^{-j}j^{\frac{7}{2}}\log^{n-1}j),
\ee
where $\g'_{*1}$ and $\g'_{*2}$ are given by \eqref{eq:gammaprimestar_1}. Comparing \eqref{eq:Rtauc6_largej_ansatz} to the $n=0$ result \eqref{eq:Rtwist6_largej_2}, we find
\be
A^{(1)}_1=&\frac{1280\pi}{\a_{+}-\a_{-}}(R^{(1)q}_{4,2} - \a_{-}R^{(1)g}_{4,2}), \nn \\
A^{(1)}_2=&\frac{1280\pi}{\a_{+}-\a_{-}}\a_{-}(\a_{+}R^{(1)g}_{4,2} -R^{(1)q}_{4,2}).
\ee

For the $n=1$ crossing equation \eqref{eq:Rtauc6_largej_twoloop_crossing}, we can also plug in the ansatz \eqref{eq:Rtauc6_largej_ansatz}, and use \eqref{eq:SL2Rblock_infinitesum2} and \eqref{eq:SL2Rsum_deltaprimederiv_answer} to compute the infinite sums. After taking care of the degeneracies, we obtain
\be
A^{(2)}_1=&\frac{320\pi}{\a_{+}-\a_{-}}(R^{(1)q}_{4,2} - \a_{-}R^{(1)g}_{4,2})(4\g^{(1)}_{4,2}-\g'_{*1}), \nn \\
A^{(2)}_2=&\frac{320\pi}{\a_{+}-\a_{-}}\a_{-}(\a_{+}R^{(1)g}_{4,2} - R^{(1)q}_{4,2})(4\g^{(1)}_{4,2}-\g'_{*2}).
\ee
For $n>1$, one can repeat the above argument and determine $A^{(n+1)}_1, A^{(n+1)}_2$. In particular, terms involving $R^{(n+1-p)}_{j+\tau_c,j}$ with $\tau_c < 6$ should not contribute to the $z\log^nz (1-\bar z)^0$ term. Moreover, only $\<R^{(n+1)}_{j+6,j}\>$ and $\<R^{(n+1-m)}_{j+6,j}(\g^{\prime(1)}_{*})^m\>$ can produce $z\log^nz$. So, we have
\be
&10 R^{(1)}_{4,2}\p{\g^{(1)}_{4,2}}^n z\log^n z + \cdots \nn \\
&=z^3\left( \sum_{j} \<R^{(n+1)}_{j+6,j}\>k_{2j+6}^{0,-1}(1-z) + \sum_{m=1}^{n} \sum_{j} \<R^{(n+1-m)}_{j+6,j}(\g^{\prime(1)}_{*})^m\> \ptl^m_{\de'_{*}}k_{2j+6}^{0,\frac{3-\de'_{*}}{2}}(1-z)|_{\de'_{*} \to 5} + \cdots  \right).
\ee
After using the infinite sum formula \eqref{eq:SL2Rblock_infinitesum2}, \eqref{eq:SL2Rsum_deltaprimederiv_answer} and the relation \eqref{eq:Rtwist6_degeneracy_relation}, we find that the solution takes a suprisingly simple form for $n\geq 1$:
\be
A^{(n+1)}_1=&\frac{320\pi(R^{(1)q}_{4,2} - \a_{-}R^{(1)g}_{4,2})(\g^{(1)}_{4,2})^{n-1}(4\g^{(1)}_{4,2}-\g'_{*1})}{\a_{+}-\a_{-}}, \nn \\
A^{(n+1)}_2=&\frac{320\pi \a_{-}(\a_{+}R^{(1)g}_{4,2} - R^{(1)q}_{4,2})(\g^{(1)}_{4,2})^{n-1}(4\g^{(1)}_{4,2}-\g'_{*2})}{\a_{+}-\a_{-}}.
\ee

It would be interesting to use some other methods to verify our results for the large-$j$ behavior of $\<R^{(n+1)}_{j+6,j}\>$. For example, $\<R^{(1)}_{j+6,j}\>$ can be obtained by performing a higher-twist version of the Lorentzian inversion formula calculation described in section \ref{sec:Rcoeff_Lorentzianinversionformula}. If the collinear EEEC at subleading order is known, one should also be able to find $\<R^{(n+1)}_{j+6,j}\>$ at higher values of $n$.

\section{Tree-level EEEC}\label{app:EEEC_Wardidentities}
In this appendix, we give the details of the calculation of tree-level EEEC in section \ref{sec:Wardidentities_contactterm}. In the first section, we compute the functions $\cF_0$ and $\cF_1$ in \eqref{eq:EEECprime_g2_exprgeneral}. In the second section, we derive the relation \eqref{eq:crossing_a0b0condition} using crossing symmetry.
\subsection{Computing $\cF_0$ and $\cF_1$}\label{app:treeEEEC_F0F1}
We first calculate the squared amplitude for the initial state created by $\Tr F^2$. We focus on processes with three out-going particles since $\cF_0$ and $\cF_1$ only get contributions from those processes. At tree-level, there are three possible processes with three outgoing particles. The first one includes three gluons, and its amplitude squared is given by (we use mostly positive metric)
\be
|\cM_{g+g+g}|^2=&-64g^2N_c(N_c^2-1)\x\nn \\
\bigg(&6(p_1\.p_2+p_1\.p_3+p_2\.p_3)\nn \\
&+2\p{\frac{(p_1\.p_2)^2+(p_2\.p_3)^2}{p_1\.p_3}+\frac{(p_1\.p_3)^2+(p_2\.p_3)^2}{p_1\.p_2}+\frac{(p_1\.p_2)^2+(p_1\.p_3)^2}{p_2\.p_3}} \nn \\
&+\frac{(p_1\.p_2)^3}{p_1\.p_3p_2\.p_3}+\frac{(p_1\.p_3)^3}{p_1\.p_2p_2\.p_3}+\frac{(p_2\.p_3)^3}{p_1\.p_2p_1\.p_3} \nn \\
&+3\p{\frac{p_1\.p_3 p_2\.p_3}{p_1\.p_2}+\frac{p_1\.p_2 p_2\.p_3}{p_1\.p_3}+\frac{p_1\.p_2 p_1\.p_3}{p_2\.p_3}}\bigg).
\ee
The second process has one gluon and two Weyl spinors
\be
|\cM_{g(p_1)+\l(p_2)+\bar \l(p_3)}|^2=-16g^2N_c(N_c^2-1)\frac{(p_1\.p_2)^2+(p_1\.p_3)^2}{p_2\.p_3}.
\ee
Finally, the third process has one gluon and two scalars
\be
|\cM_{g(p_1)+\f(p_2)+\f(p_3)}|^2=-32g^2N_c(N_c^2-1)\frac{p_1\.p_2p_1\.p_3}{p_2\.p_3}.
\ee

We can then define the total amplitude squared $|\cM|^2$ as
\be\label{eq:Msquared_total}
|\cM|^2=\frac{1}{3!}|\cM_{g+g+g}|^2 + 4|\cM_{g(p_1)+\l(p_2)+\bar \l(p_3)}|^2 + \frac{6}{2!}|\cM_{g(p_1)+\f(p_2)+\f(p_3)}|^2.
\ee
Note that due to the three identical gluons and two identical scalars in the final states, we should include symmetry factors $\frac{1}{3!}$ and $\frac{1}{2!}$ in the phase space measure for the corresponding final state. But here we choose to include those factors in $|\cM|^2$, so that we can just use the same phase space measure for all the final states. 

Comparing \eqref{eq:eeecnparam} and \eqref{eq:EEECprime_g2_exprgeneral}, one can show that the function $\cF_1$ is defined as
\be\label{eq:F1_def}
&\cF_1(\vec n_1,\vec n_2)\nn \\
&=\sum_{i,j}\int d\s\frac{E_i^2 E_j}{Q^3}\de\p{\vec n_1,\frac{\vec p_i}{E_i}}\de\p{\vec n_2,\frac{\vec p_j}{E_j}} \nn \\
&=\frac{1}{\s_{\mathrm{tot}}}\sum_{\{i,j\} \subset \{a,b,c\}}\int\frac{d^3\vec p_a}{(2\pi)^3} \frac{d^3\vec p_b}{(2\pi)^3} \frac{d^3\vec p_c}{(2\pi)^3} \frac{1}{2E_a} \frac{1}{2E_b} \frac{1}{2E_c}|\cM|^2(2\pi)^4\de(Q-E_a-E_b-E_c)\de^{(3)}\p{\vec p_a+\vec p_b +\vec p_c} \nn \\
&\qquad\qquad\qquad\qquad\x\frac{E_i^2 E_j}{Q^3}\de\p{\vec n_1,\frac{\vec p_i}{E_i}}\de\p{\vec n_2,\frac{\vec p_j}{E_j}},
\ee
and $\cF_0$ is defined as
\be\label{eq:F0_def}
&\cF_0(\vec n_1,\vec n_2,\vec n_3)\de((\vec n_1\x\vec n_2)\.\vec n_3)\nn \\
&=\frac{1}{\s_{\mathrm{tot}}}\sum_{\{i,j,k\} = \{a,b,c\}}\int\frac{d^3\vec p_a}{(2\pi)^3} \frac{d^3\vec p_b}{(2\pi)^3} \frac{d^3\vec p_c}{(2\pi)^3} \frac{1}{2E_a} \frac{1}{2E_b} \frac{1}{2E_c}|\cM|^2(2\pi)^4\de(Q-E_a-E_b-E_c)\de^{(3)}\p{\vec p_a+\vec p_b +\vec p_c} \nn \\
&\qquad\qquad\qquad\qquad\x\frac{E_i E_j E_k}{Q^3}\de\p{\vec n_1,\frac{\vec p_i}{E_i}}\de\p{\vec n_2,\frac{\vec p_j}{E_j}}\de\p{\vec n_3,\frac{\vec p_k}{E_k}}. 
\ee
We can then plug in \eqref{eq:Msquared_total} for the squared amplitude $|\cM|^2$. Also, the total cross section is
\be
\s_{\mathrm{tot}}=\frac{N_c^2-1}{2\pi}Q^4.
\ee
Performing the integral in \eqref{eq:F1_def}, we then obtain that $\cF_1$ is given by \eqref{eq:F1_fullexpr}. For \eqref{eq:F0_def}, the delta function $\de((\vec n_1\x\vec n_2)\. \vec n_3)$ on the left-hand side will be canceled by one of the delta functions in $\de^{(3)}\p{\vec p_a+\vec p_b +\vec p_c}$ on the right-hand side. After performing the calculation, we then find that $\cF_0$ is given by \eqref{eq:F0_intildeF0} and \eqref{eq:F0tilde_fullexpr}. The step function $\th(\z_1+\z_2-1)$ in \eqref{eq:F0_intildeF0} comes from the condition $\vec p_a+\vec p_b+ \vec p_c=0$. If $\z_1+\z_2<1$, one can easily draw a line such that all three momenta $\vec p_{a,b,c}$ lie on the same side of the line, and therefore there are no solutions to $\vec p_a+\vec p_b+ \vec p_c=0$.

\subsection{Crossing symmetry of $\cF_0$}\label{app:treeEEEC_F0_crossing}
We now consider crossing symmetry of $\cF_0(\vec n_1,\vec n_2,\vec n_3)$. If $\vec n_1,\vec n_2,\vec n_3$ are all different from each other, it is not too hard to check that the function $\cF_0(\vec n_1,\vec n_2,\vec n_3)$ given by \eqref{eq:F0_intildeF0} and \eqref{eq:F0tilde_fullexpr} is crossing symmetric. So we want to focus on the delta functions and show that they are also crossing symmetric. In particular, we will consider $2\leftrightarrow 3$. Note that \eqref{eq:F0_intildeF0} can also be written as
\be
\cF_0(\vec n_1,\vec n_2,\vec n_3)=\frac{1}{16}\sqrt{(\vec n_1-\vec n_3)^2(\vec n_1+ \vec n_3)^2(\vec n_2-\vec n_3)^2(\vec n_2+ \vec n_3)^2}\tl{\cF}_0(\vec n_1,\vec n_2,\vec n_3),
\ee
and the singular part of $\tl{\cF}_0$ is given by \eqref{eq:F0tilde_singular_expr}.

We now preform the crossing $2\leftrightarrow 3$ and first look at the contact term at $r_2=0$. Naively, we expect that the contact term looks like
\be\label{eq:r2zero_contactterm}
\frac{\de(r'_2)}{r'_2}\frac{1}{16}\sqrt{(\vec n_1-\vec n_2)^2(\vec n_1+ \vec n_2)^2(\vec n_2-\vec n_3)^2(\vec n_2+ \vec n_3)^2}\de\p{(\vec n_1\x \vec n_2)\.\vec n_3},
\ee
where the new variable $r'_2$ is given by
\be
r'_2&=\frac{1}{4}\sqrt{(\vec n_1+\vec n_2)^4 + (\vec n_2 - \vec n_3)^4}=h_{2}(\th_2)r_2 + O(r_2^2),
\ee
and the function $h_{2}(\th_2)$ is
\be
h_{2}(\th_2)=\sqrt{\cos^2\th_2+(\sqrt{\cos\th_2}-\sqrt{\sin\th_2})^4}.
\ee

If we integrate \eqref{eq:r2zero_contactterm} against a test function $F(\vec n_1,\vec n_2,\vec n_3)$, we find
\be
&\int d\Omega_{\vec n_1} d\Omega_{\vec n_2} d\Omega_{\vec n_3} F(\vec n_1,\vec n_2,\vec n_3) \frac{\de(r'_2)}{r'_2}\frac{1}{16}\sqrt{(\vec n_1-\vec n_2)^2(\vec n_1+ \vec n_2)^2(\vec n_2-\vec n_3)^2(\vec n_2+ \vec n_3)^2}\de\p{(\vec n_1\x \vec n_2)\.\vec n_3} \nn \\
&=8\pi^2\int_0^{\frac{\pi}{4}}d\th_2 \frac{1}{h_2(\th_2)^2}\frac{\sqrt{\cos\th_2}-\sqrt{\sin\th_2}}{\sqrt{\sin\th_2}}\int dr_2 \de(r_2) F(\vec n_1,\vec n_2,\vec n_3) \nn \\
&=2\pi^3 F(-\vec n_3, \vec n_3,\vec n_3),
\ee
which agrees with the original contact term before crossing. However, there is actually another delta function coming from the $[\cdots]_{0}$ distribution. This is due to the relation
\be
\left[\frac{1}{ax}\right]_{0}=\frac{1}{a}\left[\frac{1}{x}\right]_{0} + \frac{\log a}{a}\de(x).
\ee
Therefore, from the $[\cdots]_{0}$ distribution we have
\be
\frac{f_0(\th'_2)}{r'_2}\left[\frac{1}{r'_2}\right]_{0} \to \frac{f_0(\th'_2)}{h_{2}(\th_2)r_2} \frac{\log h_{2}(\th_2)}{h_{2}(\th_2)}\de(r_2),
\ee
where
\be
\th'_2=\mathrm{tan}^{-1}\p{\frac{(\vec n_1 + \vec n_2)^2}{(\vec n_2-\vec n_3)^2}} = \mathrm{tan}^{-1}\p{(1-\sqrt{\mathrm{tan}\th_2})^2} + O(r_2).
\ee
Therefore, we should also consider
\be
8\pi^2\int_0^{\frac{\pi}{4}}d\th_2 \frac{f_0(\th_2')\log h_2(\th_2)}{h_2(\th_2)^2}\frac{\sqrt{\cos\th_2}-\sqrt{\sin\th_2}}{\sqrt{\sin\th_2}}\int dr_2 \de(r_2) F(\vec n_1,\vec n_2,\vec n_3).
\ee
It turns out that the $\th_2$ integral actually vanishes. So, for the $r_2=0$ contact term, the $[\cdots]_{0}$ distribution doesn't produce new delta function after crossing $2\leftrightarrow 3$.

Now we consider the other two contact terms. After $2\leftrightarrow 3$, $r_1$ and $r_3$ become
\be
r'_1&=\frac{1}{4}\sqrt{(\vec n_1-\vec n_2)^4 + (\vec n_2 + \vec n_3)^4}= h_{1\to3}(\th_3)r_3 + O(r_3^2), \nn \\
r'_3&=\frac{1}{4}\sqrt{(\vec n_1+\vec n_2)^4 + (\vec n_2 + \vec n_3)^4}= h_{3\to1}(\th_1)r_1 + O(r_1^2),
\ee
where
\be
h_{1\to3}(\th_3)&=\sqrt{\sin^2\th_3+\p{\sqrt{\cos\th_3}+\sqrt{\sin\th_3}}^4}, \nn \\
h_{3\to1}(\th_1)&=\sqrt{\sin^2\th_1+\p{\sqrt{\cos\th_1}-\sqrt{\sin\th_1}}^4}.
\ee
Also, $\th_1$ and $\th_3$ become
\be
\th'_1=&\mathrm{tan}^{-1}\p{\frac{(\vec n_2 + \vec n_3)^2}{(\vec n_1-\vec n_2)^2}} = \mathrm{tan}^{-1}\p{\frac{\sin\th_3}{(\sqrt{\cos\th_3}+\sqrt{\sin\th_3})^2}} + O(r_3), \nn \\
\th'_3=&\mathrm{tan}^{-1}\p{\frac{(\vec n_2 + \vec n_3)^2}{(\vec n_1+\vec n_2)^2}} =  \mathrm{tan}^{-1}\p{\frac{\sin\th_1}{(\sqrt{\cos\th_1}-\sqrt{\sin\th_1})^2}} + O(r_1).
\ee

Integrating the $\de(r_1')$ term, we get
\be
&\int d\Omega_{\vec n_1} d\Omega_{\vec n_2} d\Omega_{\vec n_3} F(\vec n_1,\vec n_2,\vec n_3) \frac{\de(r'_1)}{r'_1}\frac{1}{16}\sqrt{(\vec n_1-\vec n_2)^2(\vec n_1+ \vec n_2)^2(\vec n_2-\vec n_3)^2(\vec n_2+ \vec n_3)^2}\de\p{(\vec n_1\x \vec n_2)\.\vec n_3} \nn \\
&=8\pi^2\int_0^{\frac{\pi}{2}}d\th_3 \frac{1}{h_{1\to3}(\th_3)^2}\frac{\sqrt{\cos\th_3}+\sqrt{\sin\th_3}}{\sqrt{\cos\th_3}}\int dr_3 \de(r_3) F(\vec n_1,\vec n_2,\vec n_3) \nn \\
&=2\pi^3 F(-\vec n_3, -\vec n_3,\vec n_3).
\ee
We should also include the contribution from the $\left[\frac{1}{r_1'}\right]_{0}$ distribution. This term will give
\be
&8\pi^2\int_0^{\frac{\pi}{2}}d\th_3 \frac{f_0(\th_1')\log h_{1\to3}(\th_3)}{h_{1\to3}(\th_3)^2}\frac{\sqrt{\cos\th_3}+\sqrt{\sin\th_3}}{\sqrt{\cos\th_3}}\int dr_3 \de(r_3) F(\vec n_1,\vec n_2,\vec n_3) \nn \\
&=8\pi^2\x\frac{\pi-\sqrt{2}\pi+2\log 2+\sqrt{2}\log(3-2\sqrt{2})}{512\pi^4} F(-\vec n_3, -\vec n_3,\vec n_3).
\ee
Comparing this result with the contact terms before crossing, we find that for the delta functions to be crossing-symmetric, we must have \eqref{eq:crossing_a0b0condition}. Also, if we consider the contact term at $r_1=0$, we will get the same condition.

\section{More details on the celestial inversion formula}\label{app:strongcoupling}
In this appendix, we give the derivation for the orthogonality relation of celestial partial waves \eqref{eq:celestialpartialwave_orthogonality} and the integral identity \eqref{eq:j0_beforepair}. We also show that the contributions at infinity of the contour deformations of \eqref{eq:EEEC_celestialblockexpansion0} vanish.
\subsection{Orthogonality of celestial partial waves}
To derive the orthogonality relation, let us consider a natural pairing
\be
\int D^{d-2}z_1 D^{d-2}z_2 D^{d-2}z_3 d^{d-1}_{\mathrm{AdS}}p\ \Psi^{c}_{\de_5,j_5;\de'_5}(z_1,z_2,z_3,p)\Psi^{c(\tl{\de}_i)}_{\tl{\de}_6,j_6;\tl{\de}'_6}(z_1,z_2,z_3,p),
\ee
where $d_{AdS}^{d-1}p=2d^dp\de(p^2+1)\theta(p^0)$ is an integral over the AdS space defined by $p^2=-1$. Plugging in the definition of the celestial partial wave, we get
\be\label{eq:orthorelation_Psi_pairing}
&\int D^{d-2}z_1 D^{d-2}z_2 D^{d-2}z_3 d^{d-1}_{\mathrm{AdS}}p D^{d-2}z D^{d-2}z' D^{d-2}z'' D^{d-2}z''' \nn \\
& \<\cP_{\de_1}(z_1)\cP_{\de_2}(z_2)\cP_{\de_5,j_5}(z)\>\<\tl{\cP}_{\de_5,j_5}(z)\cP_{\de_3}(z_3)\cP_{\de'_5}(z')\>\frac{1}{(-2z'\.p)^{\tl{\de'_5}}} \nn \\
&\x \<\tl{\cP}_{\de_1}(z_1)\tl{\cP}_{\de_2}(z_2)\tl{\cP}_{\de_6,j_6}(z'')\>\<\cP_{\de_6,j_6}(z'')\tl{\cP}_{\de_3}(z_3)\tl{\cP}_{\de'_6}(z''')\>\frac{1}{(-2z'''\.p)^{\de'_6}} \nn \\
&=B_{12\cP_{\de_5,j_5}}\de_{\cP_5\cP_6}\int  D^{d-2}z_3 d^{d-1}_{\mathrm{AdS}}p D^{d-2}z D^{d-2}z' D^{d-2}z''' \nn \\
&\<\tl{\cP}_{\de_5,j_5}(z)\cP_{\de_3}(z_3)\cP_{\de'_5}(z')\>\frac{1}{(-2z'\.p)^{\tl{\de'_5}}}\<\cP_{\de_5,j_5}(z)\tl{\cP}_{\de_3}(z_3)\tl{\cP}_{\de'_6}(z''')\>\frac{1}{(-2z'''\.p)^{\de'_6}} \nn \\
&=B_{12\cP_{\de_5,j_5}}B_{\tl{\cP}_{\de_5,j_5}3\cP_{\de'_{5}}}\de_{\cP_5\cP_6}\de_{\cP'_{5}\cP'_{6}} \nn \\
&\x \int  d^{d-1}_{\mathrm{AdS}}p D^{d-2}z' \frac{1}{(-2z'\.p)^{\tl{\de'_5}}} \frac{1}{(-2z'\.p)^{\de'_5}}.
\ee
In the first and the second equality above, we have used the bubble formula (eq. (2.32) in \cite{Karateev:2018oml}), and $B_{12\cP_{\de_5,j_5}}$, $B_{\tl{\cP}_{\de_5,j_5}3\cP_{\de'_{5}}}$ are bubble coefficients given by \eqref{eq:bubblecoeff_definition}. Moreover, the integral in the last line of \eqref{eq:orthorelation_Psi_pairing} is given by
\be\label{eq:bulktoboundary_2ptpairing}
\int \frac{D^{d-2}z d_{\mathrm{AdS}}^{d-1}p}{\vol(\SO(d-1,1))}(-2p\. z)^{-\de}(-2p\. z)^{-\tl{\de}}=\frac{1}{2^{d-2}\vol(\SO(d-2))},
\ee
where we use the conformal group to gauge fix $p=(1,0,0,\dots,0)$ and $z=(1,1,0,\dots,0)$. The stabilizer group after the gauge-fixing is $\SO(d-2)$, and the Fadeev-Popov determinant for the gauge-fixing is $1$.

Therefore, the orthogonality relation for the celestial partial wave is
\be
&\int\frac{D^{d-2}z_1 D^{d-2}z_2 D^{d-2}z_3 d^{d-1}_{\mathrm{AdS}}p}{\vol(\SO(d-1,1))}\ \Psi^{c}_{\de_5,j_5;\de'_5}(z_1,z_2,z_3,p)\Psi^{c(\tl{\de}_i)}_{\tl{\de}_6,j_6;\tl{\de'}_6}(z_1,z_2,z_3,p) \nn \\
&=\frac{1}{2^{d-2}\vol(\SO(d-2))}B_{12\cP_{\de_5,j_5}}B_{\tl{\cP}_{\de_5,j_5}3\cP_{\de'_{5}}}\de_{\cP_5\cP_6}\de_{\cP'_5 \cP'_{6}},
\ee
where
\be
\de_{\cP_5\cP_6}=2\pi\de(s_5-s_6)\de_{j_5,j_6},
\ee
for $\de_5=\frac{d-2}{2}+is_5,\de_6=\frac{d-2}{2}+is_6$ with $s_5,s_6>0$.

\subsection{Derivation of \eqref{eq:j0_beforepair}}
We now consider the identity
\be
&\int D^{d-2}z_1D^{d-2}z_2\<\tl{\cP}_{\de_1}(z_1)\tl{\cP}_{\de_2}(z_2)\cP_{\de,j=0}(z)\>(-2p\. z_1)^{-\de_1}(-2p\. z_2)^{-\de_2}\nn \\
&=C_{\de_1,\de_2;\de}(-p^2)^{\frac{\de-\de_1-\de_2}{2}}(-2p\. z)^{-\de}.
\ee
By Lorentz symmetry and homogeneity of $p$ and $z$, the right-hand side must be proportional to $(-p^2)^{\frac{\de-\de_1-\de_2}{2}}(-2p\. z)^{-\de}$. Thus, our goal here is showing that the coefficient $C_{\de_1,\de_2;\de}$ is given by \eqref{eq:C0_m1}. Our strategy is to first fix $p^2=-1$, and integrate both sides of \eqref{eq:j0_beforepair} against $(-2p\. z)^{-\tl{\de}}$ over $z$ and $p$. More precisely, for the right-hand side, we have
\be\label{eq:AppF_integral_LHS}
\int \frac{D^{d-2}z d_{\mathrm{AdS}}^{d-1}p}{\vol(\SO(d-1,1))} C_{\de_1,\de_2;\de}(-2p\. z)^{-\de}(-2p\. z)^{-\tl{\de}} = C_{\de_1,\de_2;\de}\frac{1}{2^{d-2}\vol(\SO(d-2))},
\ee
which follows from \eqref{eq:bulktoboundary_2ptpairing}. For the left-hand side, we want to compute
\be
\int \frac{D^{d-2}z_1D^{d-2}z_2 D^{d-2}z d_{\mathrm{AdS}}^{d-1}p}{\vol(\SO(d-1,1))} \<\tl{\cP}_{\de_1}(z_1)\tl{\cP}_{\de_2}(z_2)\cP_{\de,0}(z)\>(-2p\. z_1)^{-\de_1}(-2p\. z_2)^{-\de_2}(-2p\. z)^{-\tl{\de}}.
\ee
One can immediately recognize that the $p$ integral is a three-point Witten diagram, and can be evaluated using \eqref{eq:Witten_3pt}. Furthermore, the remaining integral over $z_1,z_2,z$ is a conformally-invariant three-point pairing. Therefore, the left-hand side is given by
\be\label{eq:AppF_integral_RHS}
&\int \frac{D^{d-2}z_1D^{d-2}z_2 D^{d-2}z d_{\mathrm{AdS}}^{d-1}p}{\vol(\SO(d-1,1))} \<\tl{\cP}_{\de_1}(z_1)\tl{\cP}_{\de_2}(z_2)\cP_{\de,0}(z)\>(-2p\. z_1)^{-\de_1}(-2p\. z_2)^{-\de_2}(-2p\. z)^{-\tl{\de}} \nn \\
&=D_{\de_1,\de_2,\tl{\de}}\p{\<\tl{\cP}_{\de_1}\tl{\cP}_{\de_2}\cP_{\de}\>,\<\cP_{\de_1}\cP_{\de_2}\tl{\cP}_{\de}\>} \nn \\
&=\frac{\pi^{\frac{d-2}{2}}\G(\frac{\de_1+\de_2+\tl{\de}-d+2}{2})\G(\frac{\de_1+\de_2-\tl{\de}}{2})\G(\frac{\de_1+\tl{\de}-\de_2}{2})\G(\frac{\de_2+\tl{\de}-\de_1}{2})}{2\G(\de_1)\G(\de_2)\G(\tl{\de})}\frac{1}{2^{d-2}\vol(\SO(d-3))}.
\ee

Finally, comparing \eqref{eq:AppF_integral_LHS} and \eqref{eq:AppF_integral_RHS}, we obtain
\be
C_{\de_1,\de_2;\de} = \frac{\pi^{\frac{d-2}{2}}\G(\frac{\de_1+\de_2+\tl{\de}-d+2}{2})\G(\frac{\de_1+\de_2-\tl{\de}}{2})\G(\frac{\de_1+\tl{\de}-\de_2}{2})\G(\frac{\de_2+\tl{\de}-\de_1}{2})}{2\G(\de_1)\G(\de_2)\G(\tl{\de})}\vol(S^{d-3}),
\ee
which agrees with \eqref{eq:C0_m1}.

\subsection{Celestial block at large $\de$ and $\de'$}
In this section, we study \eqref{eq:EEEC_celestialblockexpansion0},
\be
\cF(z_1,z_2,z_3,p)=&\sum_{j}\int_{\frac{d-2}{2}-i\oo}^{\frac{d-2}{2}+i\oo}\frac{d\de}{2\pi i}\int_{\frac{d-2}{2}-i\oo}^{\frac{d-2}{2}+i\oo}\frac{d\de'}{2\pi i} C(\de,j;\de') G^{c}_{\de,j;\de'}(z_1,z_2,z_3,p),
\ee
and make sure that the contributions at infinity vanish when doing the contour deformations. For concreteness, we  consider the leading order strong-coupling EEEC, so $C(\de,j;\de')$ is given by \eqref{eq:Cfuncstrong_0thorder}. When we first close the $\de'$ contour to the right, the locations of the poles are at $\de'=\de+3+2k$. Furthermore, \eqref{eq:Cfuncstrong_0thorder} in the large $\de'$ limit behaves like
\be
C^{(0)}_{\mathrm{strong}}(\de,j;\de' \to \oo) \sim 2^{-\de'}\p{\cdots},
\ee
where $\p{\cdots}$ grows sub-exponentially at large $\de'$. Therefore, a sufficient condition for the contribution at infinity of the $\de'$ contour to vanish is
\be\label{eq:dep_infinity_condition}
\lim_{\mathrm{Re}(\de')\to \oo} 2^{-\de'}\p{\textrm{sub-exponential}}G^c_{\de,j;\de'}(z_1,z_2,z_3,p) = 0.
\ee

After closing the $\de'$ contour, \eqref{eq:EEEC_celestialblockexpansion0} becomes
\be
\cF(z_1,z_2,z_3,p)=&\sum_{j}\sum_{k=0}^{\oo}\int_{\frac{d-2}{2}-i\oo}^{\frac{d-2}{2}+i\oo}\frac{d\de}{2\pi i} \p{\mathrm{Res}_{\de'=\de+3+2k}C^{(0)}_{\mathrm{strong}}(\de,j;\de')}G^{c}_{\de,j;\de'=\de+3+2k}(z_1,z_2,z_3,p).
\ee
On the $\de$-plane, $\mathrm{Res}_{\de'=\de+3+2k}C^{(0)}_{\mathrm{strong}}$ has poles at $\de=6+2n$. The contributions from these poles will reproduce the celestial block coefficients \eqref{eq:Pcoeff_leading}. Note that at large $\de$,
\be
\mathrm{Res}_{\de'=\de+3+2k}C^{(0)}_{\mathrm{strong}} \sim 2^{-\de}\p{\cdots},
\ee
where $\p{\cdots}$ grows sub-exponentially at large $\de$. Thus, a sufficient condition for the contribution at infinity of the $\de$ contour to vanish is
\be\label{eq:de_infinity_condition}
\lim_{\mathrm{Re}(\de)\to \oo} 2^{-\de}\p{\textrm{sub-exponential}}G^c_{\de,j;\de+3+2k}(z_1,z_2,z_3,p) = 0.
\ee
If the two conditions \eqref{eq:dep_infinity_condition} and \eqref{eq:de_infinity_condition} are true, the celestial block expansion \eqref{eq:strongEEEC_celestialblockexpansion1} can be obtained from \eqref{eq:EEEC_celestialblockexpansion0} by contour deformation.

To show that \eqref{eq:dep_infinity_condition} and \eqref{eq:de_infinity_condition} are true, we will need to understand the behavior of the celestial block $G^c$ at large $\de$ or $\de'$. For four-point conformal blocks, one can determine their large $\De$ behavior by studying the Casimir equation in the limit $\De\to \oo$ \cite{Kos:2013tga, Kos:2014bka, Penedones:2015aga, Kravchuk:2017dzd, Erramilli:2019njx}. For the celestial block, the analogous Casimir equations are
\be\label{eq:Gc_Casimir}
&\bigg(-\frac{1}{2}L^{(12)}_{\mu\nu}L^{(12)\mu\nu} - \de(\de-d+2) -j(j+d-4)\bigg)G^{c}_{\de,j;\de'}=0,\nn \\
&\bigg(-\frac{1}{2}L^{(123)}_{\mu\nu}L^{(123)\mu\nu} - \de'(\de'-d+2)\bigg)G^{c}_{\de,j;\de'}=0,
\ee
where
\be
&L^{(12)}_{\mu\nu} = \sum_{i=1}^{2}z_{i\mu}\frac{\ptl}{\ptl z^{\nu}_i}-z_{i\nu}\frac{\ptl}{\ptl z^{\mu}_i}, \nn \\
&L^{(123)}_{\mu\nu} = \sum_{i=1}^{3}z_{i\mu}\frac{\ptl}{\ptl z^{\nu}_i}-z_{i\nu}\frac{\ptl}{\ptl z^{\mu}_i}.
\ee
If we take the limit given by \eqref{eq:dep_infinity_condition}, the second line of \eqref{eq:Gc_Casimir} will give a differential equation that the leading behavior of $G^c$ in this limit must satisfy. Similarly, for the limit given by \eqref{eq:de_infinity_condition}, \eqref{eq:Gc_Casimir} will give two differential equations. However, since $G^c$ can depend nontrivially on three cross ratios, $\z_{12},\z_{13},\z_{23}$, these differential equations are not as simple as as the conformal block case, and it is difficult to solve them directly.

Thus, we are led to consider an alternative method for studying $G^c$ at large $\de$ or $\de'$. We find that this can be achieved by writing down an integral representation with finite integration range for the celestial block. When we consider the limits given by \eqref{eq:dep_infinity_condition} and \eqref{eq:de_infinity_condition}, the integral will be dominated by a saddle point and the behavior of $G^c$ can be determined.

\subsubsection{Warmup: conformal blocks at large $\De$ revisited}

The method also applies to the conformal block, so let us first consider this simpler case. They key idea is to use the {\it Lorentzian} shadow representation of the block \cite{Polyakov:1974gs}. Here, we follow the notation of \cite{Kravchuk:2018htv}, where the block can be written
\be
G_{\De,J}(x_i) \sim \int_{1>x_0>2} d^dx_0D^{d-2}z|T_{d-\De,2-d-J}(x_1,x_2,x_0,z)|T_{\De,J}(x_3,x_4,x_0,z),
\ee
where the causality configuration is $1>2, 3>4$, and all other points are spacelike. Here, $\sim$ means that the two sides can differ by a factor independent of the positions $x_i$, and
\be
T_{\De,J}(x_1,x_2,x_0,z)=\frac{(2z\.x_{20}x_{10}^2-2z\.x_{10}x_{20}^2)^J}{(-x_{12}^2)^{\frac{\De_1+\De_2-\De+J}{2}}(x_{10}^2)^{\frac{\De_1+\De-\De_2+J}{2}}(x_{20}^2)^{\frac{\De_2+\De-\De_1+J}{2}}}
\ee
is a conformal three-point structure.
Since we are interested in the large $\De$ limit, we focus on the $\De$-dependence of the integrand,
\be
\p{\frac{(-x_{10}^2)(-x_{20}^2)(-x_{34}^2)}{(-x_{12}^2)(-x_{30}^2)(-x_{40}^2)}}^{\frac{\De}{2}}.
\ee

Let us choose lightcone coordinates $x=(u,v,x_\perp)$, where $x^2=uv+x_{\perp}^2$, and set $x_1=(u_1,v_1,0),x_2=(0,0,0),x_3=(1,1,0),x_4=\oo$. The conditions $1>2$ and $1\approx 3$ become $v_1<0,0<u_1<1$. Our integral becomes
\be
(-u_1v_1)^{-\frac{\De}{2}}\int dudvd^{d-2}x_{\perp}\p{(-uv-r^2)(-(u-u_1)(v-v_1)-r^2)}^{\frac{\De}{2}}\p{(u-1)(v-1)+r^2}^{-\frac{\De}{2}}\p{\cdots},
\ee
where $\p{\cdots}$ are independent of $\De$. In the large $\De$ limit, a saddle point appears at $u=1-\sqrt{1-u_1}, v=1-\sqrt{1-v_1},r=0$. Therefore, the leading large $\De$ behavior is given by
\be
\cG_{\De,J} \sim (-u_1v_1)^{-\frac{\De}{2}}\p{(1-\sqrt{1-u_1})(\sqrt{1-v_1}-1)}^{\De}\p{\cdots}.
\ee

Going back to the more familiar cross ratios $\r,\bar \r$ \cite{Hogervorst:2013sma}, we find
\be
\frac{-u_1v_1}{(1-\sqrt{1-u_1})^2(\sqrt{1-v_1}-1)^2}=\frac{1}{\r \bar \r}.
\ee
Therefore, in the large $\De$ limit, $G_{\De,J}$ is proportional to $(\r\bar \r)^{\frac{\De}{2}}\p{\cdots}$. One can fix the position-independent factor by considering the OPE limit of $G_{\De,J}$. Eventually, we obtain
\be
G_{\De,J} \sim 4^{\De}(\r\bar \r)^{\frac{\De}{2}}\p{\cdots},
\ee 
which agrees with the known result. To match the full result in \cite{Kos:2013tga, Kos:2014bka, Penedones:2015aga, Kravchuk:2017dzd, Erramilli:2019njx}, we could additionally include the 1-loop determinant around the saddle point. However, that such subleading terms will not be important in our analysis.

\subsubsection{Lorentzian integrals for the celestial block}

For the three-point celestial block $G^c$, we can write down a similar integral representation with finite integration range by continuing to ``double Lorentzian" signature, as in section~\ref{sec:analytic_bootstrap}. That is, we must analytically continue the celestial sphere to a Lorentzian signature space, so that the full spacetime has signature $(2,d-2)$. We then have
\be\label{eq:Gc_integralrepresentation}
G^c_{\de,j;\de'}(z_i,p) \sim \int_{\substack{1>0>2 \\ 0>0'>3}} D^{d-2}z_0 D^{d-2}z'_0 \frac{1}{(-2p\.z'_0)^{\de'}}\<\tl{\cP}_{\de'}(z'_0)\cP_3\cP_{\de,j}(z_0)\>\<\tl{\cP}_{\de,j}(z_0)\cP_1\cP_2\>
\ee
where $z_i,p$ are in $(2,d-2)$ signature. We can view $z_i$ as the embedding space coordinates of $\R^{1,d-3}$, and the causality constraints $1>0>2$ and $0>0'>3$ should be understood in this space. Since the right-hand side is a solution of the Casimir equations \eqref{eq:Gc_Casimir} by construction, one can show that it is proportional to $G^c$ (up to a factor independent of $z_i$ and $p$) by considering its various OPE limits. To study the integral more explicitly, we will pick the frame  $z_i=(1, \vec y_i^2, \vec y_i), p=(p_{+},1,\vec y_p)$, where
\be\label{eq:doubleLorentzianframe}
\vec y_1=(1,-1,0),\quad \vec y_2=(0,0,0), \quad \vec y_3=(-1,1,0),\quad \vec y_p=(y_p^{+},y_p^{-},0),
\ee
where the coordinate for $\vec y$ is $(y^{+},y^{-},\vec y_{\perp})$, and $\vec y^2=y^{+}y^{-}+\vec y_{\perp}^2$. This satisfies the causality constraint $1>2>3$. Also, the relation between $p_{+},y_p^{+},y_p^{-}$ and $\z_{12},\z_{13},\z_{23}$ can be obtained straightforwardly using the definition of the cross ratios.

Let us now consider the limit $\de'\to \oo$ corresponding to \eqref{eq:dep_infinity_condition}. For this limit, it is more convenient to study
\be
\int_{1>2>0'>3} D^{d-2}z'_0 \frac{1}{(-2p\.z'_0)^{\de'}} g^{(\de_1,\de_2,\de_3,\tl{\de'})}_{\de,j}(z_1,z_2,z_3,z'_0).
\ee
In the limit $\de' \to \oo$, the conformal block $g^{(\de_1,\de_2,\de_3,\tl{\de'})}_{\de,j}(z_1,z_2,z_3,z'_0)$ behaves like \cite{Behan:2014dxa}
\be
g^{(\de_1,\de_2,\de_3,\tl{\de'})}_{\de,j}(z_1,z_2,z_3,z'_0) \sim \p{\frac{(-2z_2\.z_3)}{(-2z_2\.z'_0)(-2z_3\.z'_0)}}^{-\frac{\de'}{2}}\p{\cdots},
\ee
where $\p{\cdots}$ grows sub-exponentially. Therefore, we must consider the integral
\be
\int_{1>2>0'>3} D^{d-2}z'_0 \frac{1}{(-2p\.z'_0)^{\de'}}\p{\frac{(-2z_2\.z_3)}{(-2z_2\.z'_0)(-2z_3\.z'_0)}}^{-\frac{\de'}{2}}.
\ee

After considering the integral in the frame $z_i=(1, \vec y_i^2, \vec y_i), p=(p_{+},1,\vec y_p)$ given by \eqref{eq:doubleLorentzianframe} and solving for its saddle point in the large $\de'$ limit, we can find the behavior of the integral at large $\de'$. We can further determine the position-independent factor in \eqref{eq:Gc_integralrepresentation} by matching the integral with the collinear limit of $G^c$ given by \eqref{eq:threept_celestialblock_crossratios}.\footnote{When comparing the saddle point result to the collinear limit of the celestial block, one should note that in the main text we always set $-p^2=1$. However, in the frame \eqref{eq:doubleLorentzianframe}, $-p^2$ depends on the cross ratios and is not equal to $1$. Therefore, we should first factor out the homogeneity factors on both sides of \eqref{eq:Gc_integralrepresentation} and just compare the remaining functions that depend on the cross ratios, which are independent of the choice of conformal frame.}${}^{,}$\footnote{From \eqref{eq:threept_celestialblock_crossratios}, one can show that if we first take the collinear limit and then the $\de' \to \oo$ limit, the leading behavior of $G^c$ at large $\de'$ is $\p{\mathrm{max}\p{\z_{13},\z_{23}}}^\frac{\de'}{2}\x(\textrm{sub-exponential})$.} We find that at large $\de'$,
\be\label{eq:Gc_largedep_leading}
G^c_{\de,j;\de' \to \oo} \sim\ & T_{123\de'}(z_1,z_2,z_3,p)\p{\frac{2(1-\sqrt{1-\z_{13}})}{\z_{13}}}^{\de'}(\textrm{sub-exponential}) \nn \\
&+ T_{123\de'}(z_1,z_2,z_3,p)\p{\frac{2(1-\sqrt{1-\z_{23}})}{\sqrt{\z_{13}\z_{23}}}}^{\de'}(\textrm{sub-exponential}),
\ee
where $T_{123\de'}$ is the homogeneity factor defined in \eqref{eq:homogeneityfactor}. We have checked that this result indeed solves the Casimir equations \eqref{eq:Gc_Casimir} in the $\de'$ limit. Furthermore, if $z_i$'s are on the celestial sphere, all the cross ratios should satisfy $\z_{ij}\in (0,1)$. Using \eqref{eq:Gc_largedep_leading}, we find that for $\z_{ij}\in (0,1)$, $2^{-\de'}G^c_{\de,j;\de'}$ is always decaying exponentially at large $\de'$, and thus the condition \eqref{eq:dep_infinity_condition} is true.

Finally, we consider the limit corresponding to \eqref{eq:de_infinity_condition}, where we set $\de'=\de+3+2k$ and take $\de\to \oo$. For this limit, we must consider the integral
\be
\int_{\substack{1>0>2 \\ 0>0'>3}} D^{d-2}z_0 D^{d-2}z'_0 \frac{1}{(-2p\.z'_0)^{\de}}&\frac{1}{(-2z'_0\.z_3)^{-\de}(-2z_0\.z_3)^{\de}} \nn \\
&\x \frac{1}{(-2z_0\.z_1)^{-\frac{\de}{2}}(-2z_0\.z_2)^{-\frac{\de}{2}}(-2z_1\.z_2)^{\frac{\de}{2}}}.
\ee
In the large $\de$ limit, it turns out that the dominant contribution of the $z'_0$ integral comes from the top of the diamond $0>0'>3$. Hence, we should set $z'_0=z_0$ and solve for the saddle point of the $z_0$ integral. After comparing the saddle point result to the collinear limit of the celestial block,\footnote{If we first take the collinear limit and then the limit corresponding to \eqref{eq:de_infinity_condition}, the leading behavior of $G^c$ is $\z_{12}^{\frac{\de}{2}}\x(\textrm{sub-exponential})$} we obtain
\be\label{eq:Gc_largededep_leading}
G^c_{\de \to \oo,j;\de'=\de+3+2k \to \oo} \sim T_{123\de'}(z_1,z_2,z_3,p)\p{\frac{2(1-\sqrt{1-\z_{12}})}{\sqrt{\z_{13}\z_{12}}}}^{\de}(\textrm{sub-exponential}) 
\ee
The result is indeed a solution to the Casimir equations \eqref{eq:Gc_Casimir} in the corresponding limit. Moreover, \eqref{eq:Gc_largededep_leading} implies that the condition \eqref{eq:de_infinity_condition} holds for $\z_{ij}\in (0,1)$.

\bibliographystyle{JHEP}
\bibliography{refs}

\end{document}